\documentclass[10pt,preprint]{aastex}

\pdfoutput=1

\usepackage{url}
\usepackage{grffile}

\newcounter{column_number}
\newcommand{\numberthecolumn}{\colhead{(\arabic{column_number})}\stepcounter{column_number}}

\newcommand{\hii}{H{\scriptsize II} }
\newcommand{\etacar}{$\eta$~Car }
\newcommand{\Chandra}{{\em Chandra} }

\shorttitle{CCCP Diffuse Emission}
\shortauthors{Townsley et al.} 

\begin{document}

\title{THE {\em CHANDRA} CARINA COMPLEX PROJECT:  DECIPHERING THE ENIGMA OF CARINA'S DIFFUSE X-RAY EMISSION}

\author{ Leisa K. Townsley\altaffilmark{*}\altaffilmark{1}, 
Patrick S. Broos\altaffilmark{1},  
You-Hua Chu\altaffilmark{2},
Marc Gagn{\'e}\altaffilmark{3},
Gordon P. Garmire\altaffilmark{1},
Robert A. Gruendl\altaffilmark{2},
Kenji Hamaguchi\altaffilmark{4},
Mordecai-Mark Mac Low\altaffilmark{5},
Thierry Montmerle\altaffilmark{6},
Ya{\"e}l Naz{\'e}\altaffilmark{7},
M. S. Oey\altaffilmark{8},
Sangwook Park\altaffilmark{9},
Robert Petre\altaffilmark{10},
Julian M. Pittard\altaffilmark{11}
}

\altaffiltext{*}{townsley@astro.psu.edu} 

\altaffiltext{1}{Department of Astronomy \& Astrophysics, Pennsylvania State University, 525 Davey Laboratory, University Park, PA 16802, USA}

\altaffiltext{2}{Department of Astronomy, University of Illinois at Urbana-Champaign, 1002 West Green Street, Urbana, IL 61801, USA}

\altaffiltext{3}{Department of Geology and Astronomy, West Chester University, West Chester, PA 19383, USA}

\altaffiltext{4}{CRESST and X-ray Astrophysics Laboratory NASA/GSFC, Greenbelt, MD 20771, USA; Department of Physics, University of Maryland, Baltimore County, 1000 Hilltop Circle, Baltimore, MD 21250, USA}

\altaffiltext{5}{Department of Astrophysics, American Museum of Natural History, New York, NY 10024-5192, USA}

\altaffiltext{6}{Institut d'Astrophysique de Paris, 98bis, Bd Arago, 75014 Paris, France}

\altaffiltext{7}{GAPHE, D{\'e}partement AGO, Universit{\'e} de Li{\`e}ge, All{\'e}e du 6 Ao{\^u}t 17, Bat. B5C, B4000-Li{\`e}ge, Belgium}

\altaffiltext{8}{Department of Astronomy, University of Michigan, 830 Dennison Building, Ann Arbor, MI   48109-1042, USA}

\altaffiltext{9}{Department of Physics, University of Texas at Arlington, Arlington, TX 76019, USA}

\altaffiltext{10}{Astrophysics Science Division, NASA Goddard Space Flight Center, Greenbelt, MD 20771, USA}

\altaffiltext{11}{School of Physics and Astronomy, The University of Leeds, Leeds, UK}


\begin{abstract}
We present a 1.42 square degree mosaic of diffuse X-ray emission in the Great Nebula in Carina from the {\em Chandra X-ray Observatory} Advanced CCD Imaging Spectrometer camera.  After removing $>$14,000 X-ray point sources from the field, we smooth the remaining unresolved emission, tessellate it into segments of similar apparent surface brightness, and perform X-ray spectral fitting on those tessellates to infer the intrinsic properties of the X-ray-emitting plasma.  By modeling faint resolved point sources, we estimate the contribution to the extended X-ray emission from unresolved point sources and show that the vast majority of Carina's unresolved X-ray emission is truly diffuse.  Line-like correlated residuals in the X-ray spectral fits suggest that substantial X-ray emission is generated by charge exchange at the interfaces between Carina's hot, rarefied plasma and its many cold neutral pillars, ridges, and clumps.
\end{abstract}




\section{INTRODUCTION \label{sec:intro}}

The Carina star-forming complex is one of the most impressive sites of massive star formation and feedback in the nearby Galaxy \citep[D=2.3~kpc,][]{SmithN06a}, with a well-documented population of $>$70 massive stars \citep{SmithN06b}.  With the recent X-ray discovery of a neutron star in the Carina complex \citep{Hamaguchi09,Pires09}, there is renewed interest in the possibility of past supernova activity in the complex.  This paper is the 15th in a series of 16 papers chronicling the findings of the \Chandra Carina Complex Project (CCCP), a wide, shallow X-ray survey of the Great Nebula in Carina using the {\it Chandra X-ray Observatory} and the imaging CCD array of its Advanced CCD Imaging Spectrometer camera \citep[ACIS-I,][]{Garmire03}.  An overview and introduction to the CCCP is given in \citet{Townsley11a}; multiwavelength images that help to situate Carina's extended X-ray emission with respect to more familiar morphological features of the Nebula are also shown there.  We recommend that readers not familiar with X-ray studies of the Carina Nebula first peruse that paper and its references, as some familiarity with those efforts is assumed here.    
    
One of the major motivations for the CCCP survey was to explore in more detail the extensive soft X-ray emission that {\it Einstein} \citep{Seward82} and {\it ROSAT} \citep{Corcoran95} imaged in Carina.  Figure 1 in the CCCP introductory paper \citep{Townsley11a} and Figure~\ref{fig:patsmooth} below show that this soft emission appears to fill the lower lobe of Carina's bipolar superbubble structure \citep{SmithN00}, outlined in Figure~\ref{fig:patsmooth}a by mid-infrared PAH emission; this morphology suggests that the unresolved X-ray emission is truly diffuse and traces hot plasma filling the superbubble, supplied by Carina's massive star winds, one or more cavity supernova explosions inside the superbubble, or both.
More recent work with {\it XMM-Newton} and {\it Suzaku} show enhanced abundances of some elements, including iron, and complex X-ray spectra with many line features \citep{Hamaguchi07,Ezoe08}.  These papers suggest that the diffuse emission is not strongly corrupted by unresolved point source emission, but the actual spatial distribution and spectral characteristics of Carina's vast point source population was not well-known from the data available at the time.

Although the CCCP is a very shallow survey, it covers most of the young stellar clusters in Carina and 
resolves out $>$14,000 X-ray point sources \citep{Broos11a}.  One of the main results of the CCCP is that, despite identifying this large number of point sources, substantial unresolved X-ray emission remains.  We will show in this paper that the diffuse nature of most of this emission is confirmed; anticipating this result, we will interchange the terms `extended', `unresolved', and `diffuse' for the remainder of this paper.

As described in \citet{Townsley11a}, the CCCP survey layout (repeated in Figure~\ref{fig:patsmooth} below) was designed in part to capture regions of the Carina Nebula that showed soft unresolved emission in the {\it ROSAT} data.  While much of the {\it ROSAT} field was covered, the CCCP survey was not large enough to image fully either of Carina's bipolar superbubble lobes.  In particular, the CCCP only sampled the northern edge of the southern, larger superbubble.  {\it ROSAT} showed that soft X-ray emission extends past the limits of the CCCP; it may fill the entire southern superbubble \citep[see Figure 1 in][]{Townsley11a}.  Thus estimates of the total luminosity of Carina's soft diffuse X-ray emission from the CCCP must be lower limits.


\section{MORPHOLOGY \label{sec:morph}}

\begin{figure}[htb] 
\begin{center}
\includegraphics[width=0.525\textwidth]{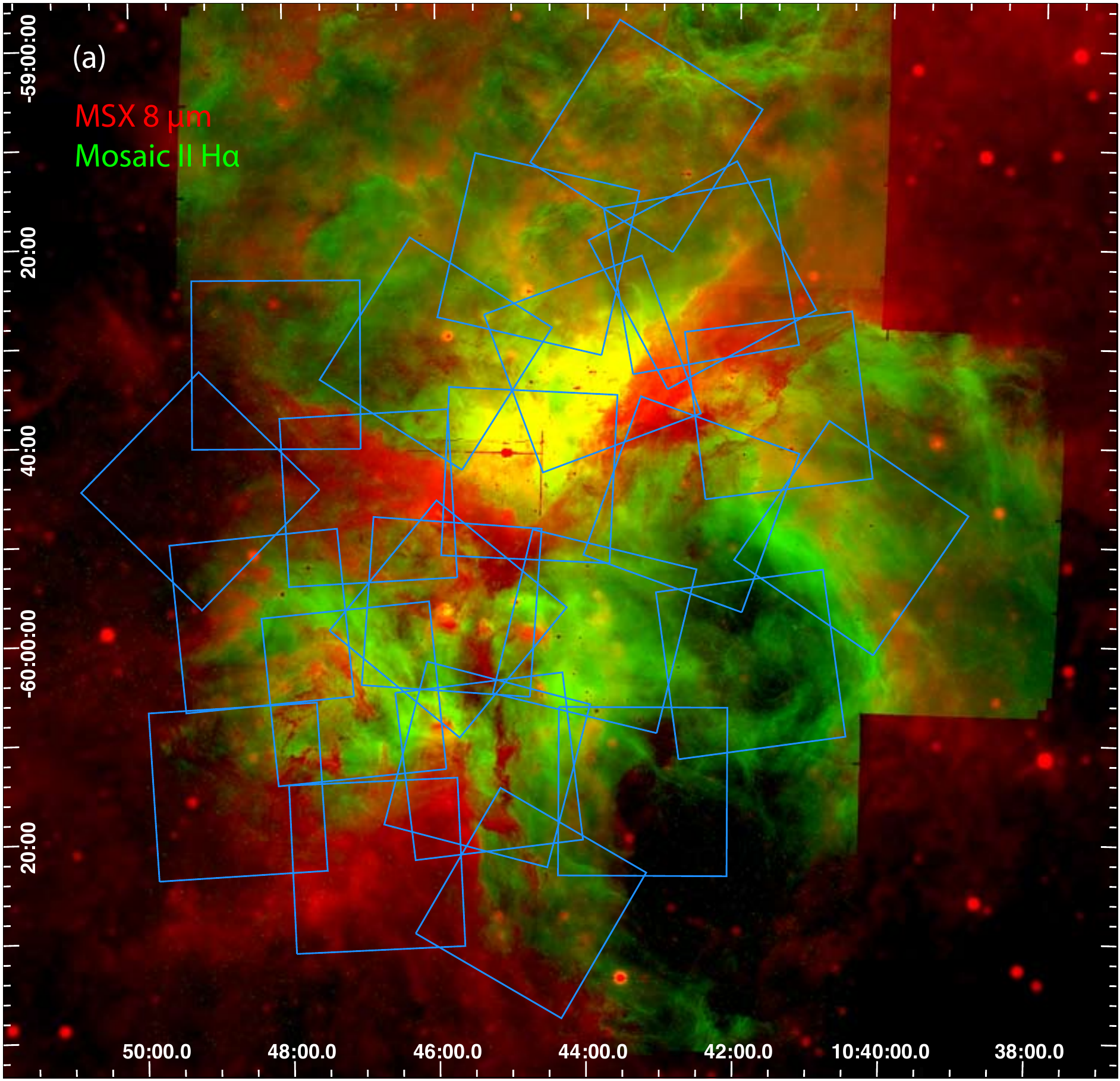}
\includegraphics[width=0.435\textwidth]{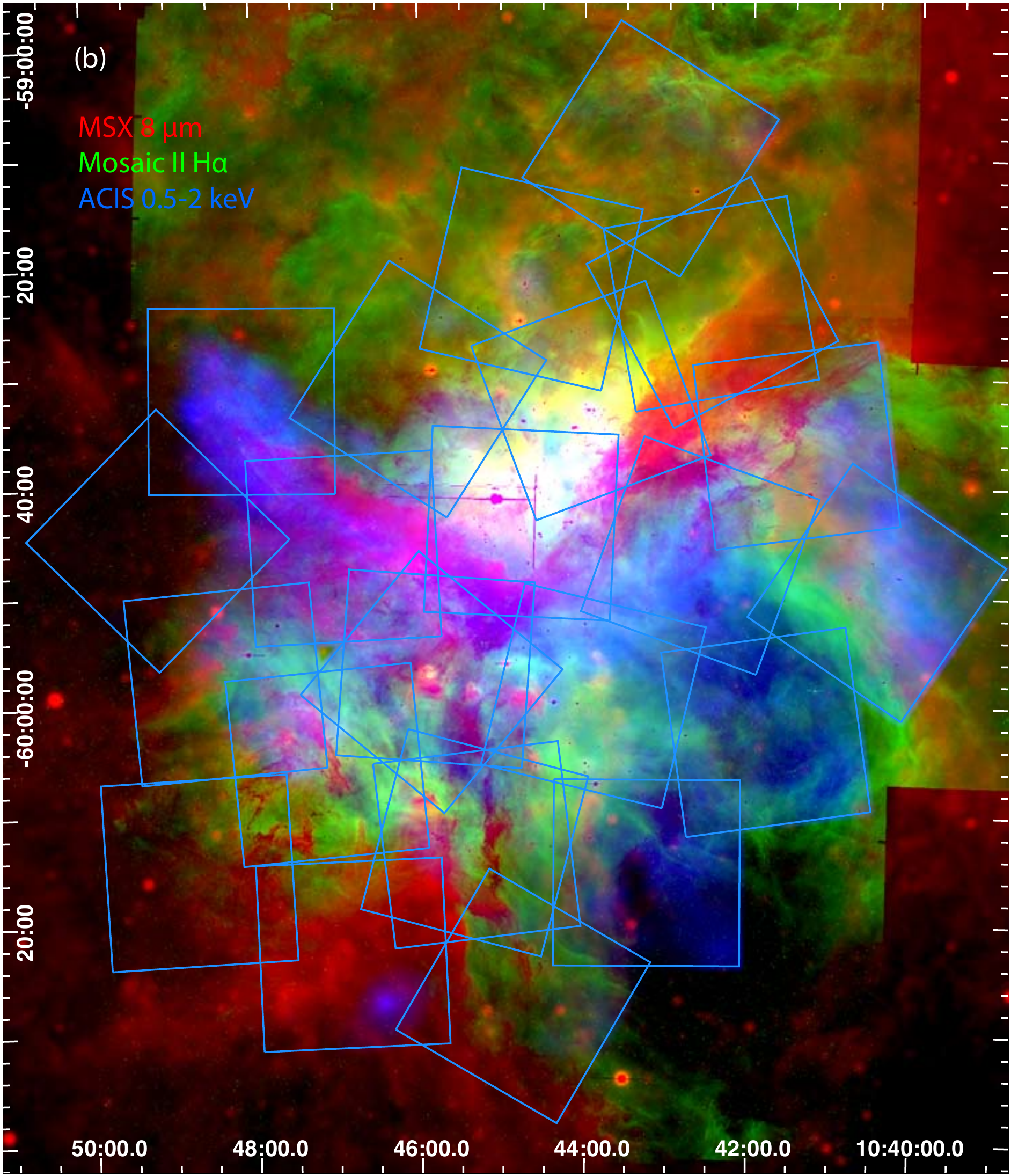}
\includegraphics[width=0.48\textwidth]{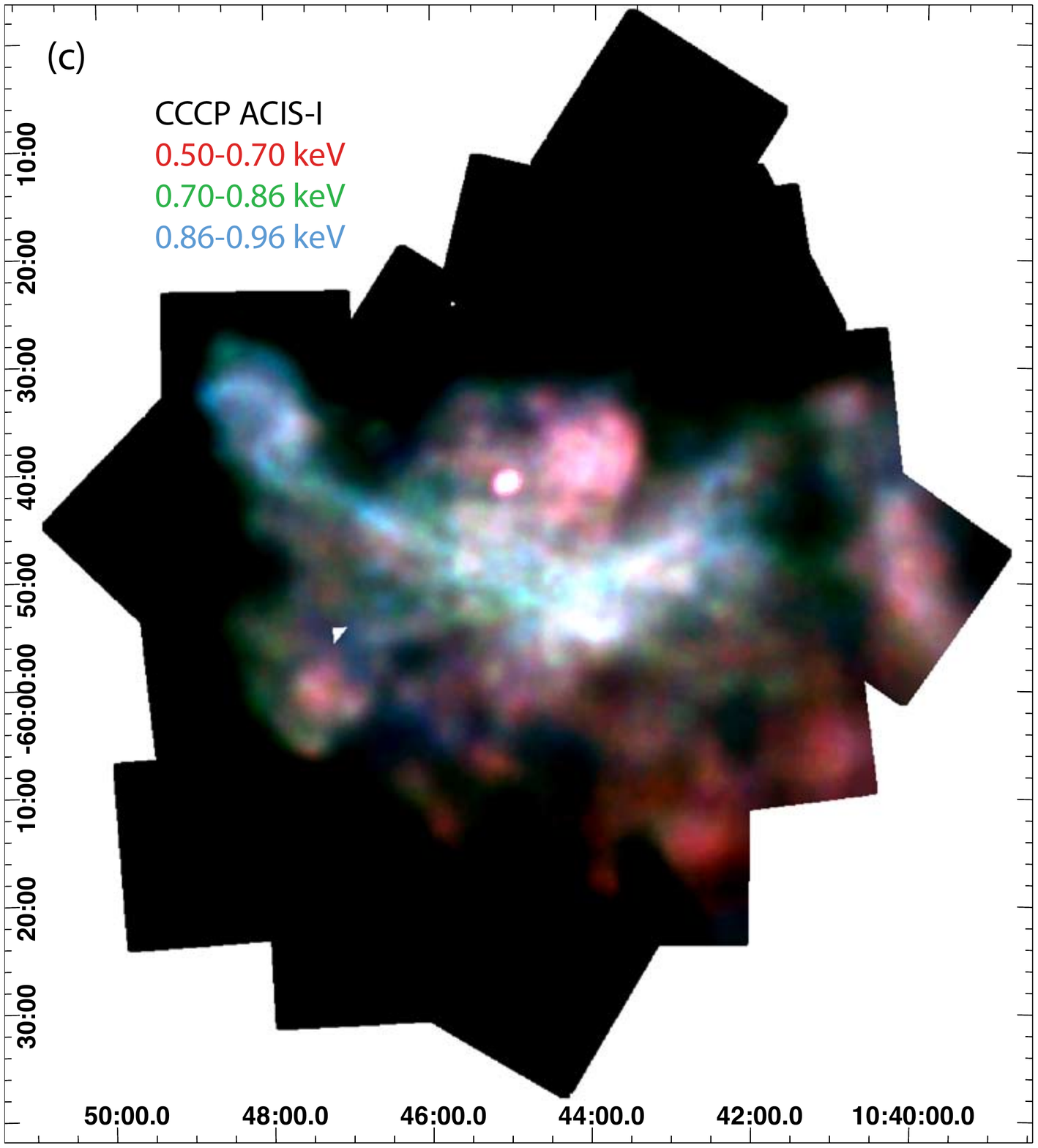}
\includegraphics[width=0.48\textwidth]{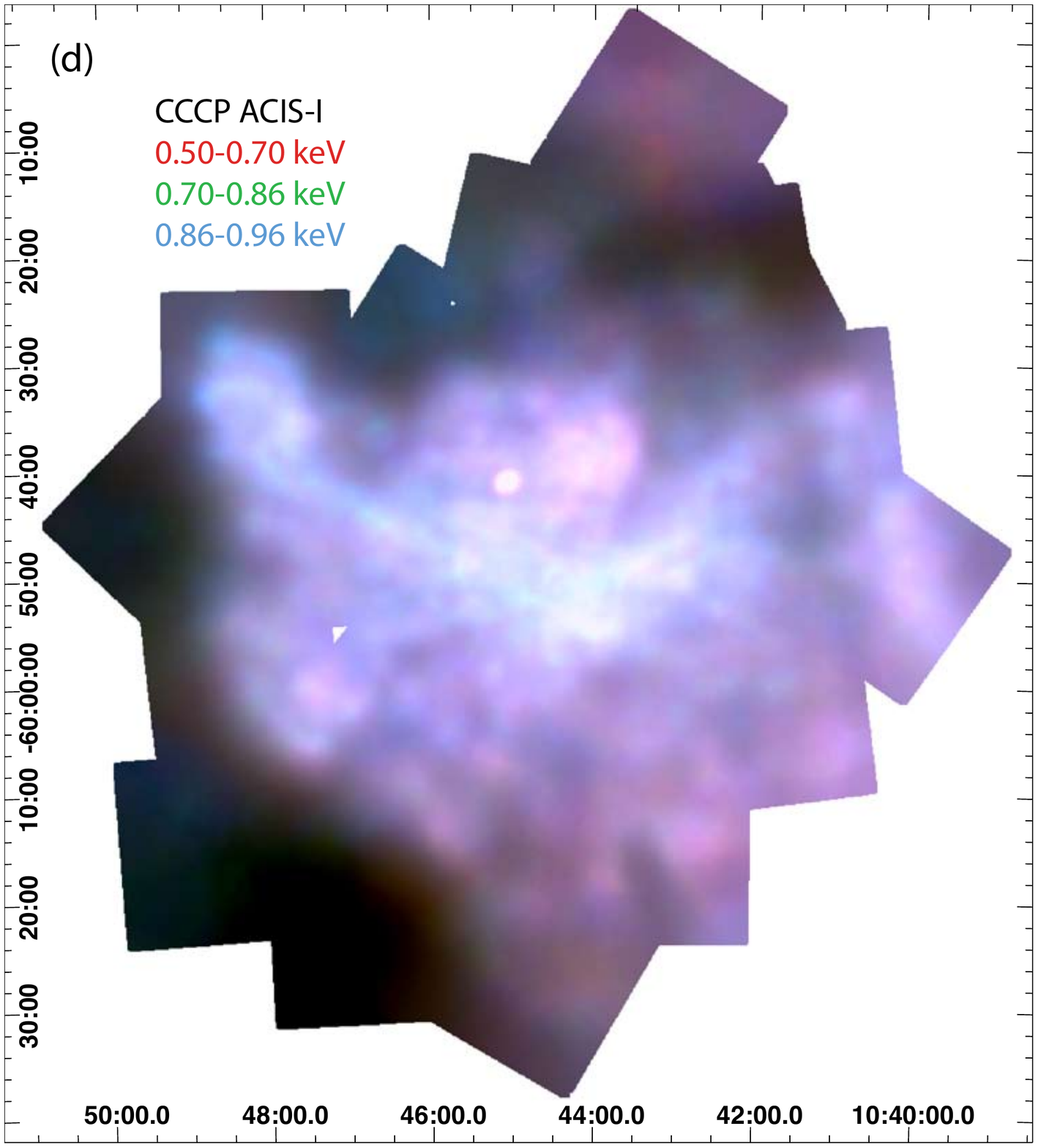}
\caption{
(a) The multiwavelength context of the CCCP.  {\em MSX} 8~$\mu$m data in red and H$\alpha$ from the MOSAIC~II camera (CTIO) in green, with the locations of ACIS-I pointings for the CCCP mosaic outlined by blue squares.
(b) The same visual and infrared images as (a), now zoomed slightly and with an adaptively-smoothed CCCP soft-band (0.5--2~keV) apparent surface brightness image (point sources excised before smoothing) added in blue.
(c) A smoothed apparent surface brightness image of Carina's soft extended X-ray emission (point sources excised) from the CCCP, with red = 500--700~eV, green = 700--860~eV, blue = 860--960~eV.
(d) The same image as (c), scaled now to bring out the faint diffuse structures.
At a distance of 2.3~kpc, $10\arcmin \sim 6.7$~pc.  Here and throughout this paper, coordinates on images are celestial J2000.
} 
\label{fig:patsmooth}
\end{center}
\end{figure}

The morphology of Carina's diffuse X-ray emission can be examined by removing the $>$14,000 X-ray point sources revealed by \Chandra \citep{Broos11a}.  Figure~\ref{fig:patsmooth} shows multiwavelength images to put Carina's diffuse X-ray emission in context, plus images of Carina's soft diffuse X-ray emission, made using our adaptive-kernel smoothing code \citep{Broos10,Townsley03} on the X-ray events remaining after the point sources have been excised.  In these X-ray images, the small holes where point sources were removed have been smoothed over; a few larger holes remain, where individual CCCP pointings didn't quite overlap and where bright point sources have been deleted.  Figure~\ref{fig:patsmooth}a shows PAH emission in red from the {\em MSX} 8~$\mu$m data and dense ionized gas in green using an H$\alpha$ image taken by the MOSAIC~II camera \citep{Muller98} on the Cerro Tololo Inter-American Observatory (CTIO) Blanco 4m Telescope.  Figure~\ref{fig:patsmooth}b repeats this image, adding in the soft-band (0.5--2~keV) CCCP diffuse X-ray image.  Figures~\ref{fig:patsmooth}c and d show just the soft X-ray emission, now in 3 narrower soft bands.  The large white elliptical structure slightly northeast of field center in these panels is $\eta$~Car's X-ray nebula (heavily smoothed).  The image color-coding in these two panels is designed to highlight the soft diffuse X-ray structures.

A first impression upon comparing Figure~\ref{fig:patsmooth} with the {\it ROSAT} data in \citet{Townsley11a} is that Carina's diffuse X-ray emission is much more complicated than the {\it ROSAT} image would suggest.  The brightest region stretches across the center of the field, exhibiting linear structures tens of parsecs long that appear to radiate from a central region made up of short arcs.  The eastern linear structure abruptly bends northwestward, then arcs back to the east, making a prominent hook shape.  A similar hook may terminate the bright linear structure to the west, but this ``arm'' appears to be foreshortened, making its terminal structure harder to distinguish.  All of these bright linear structures that cross the center of the field appear harder (bluer) than the surrounding emission.

A bright region of soft (red) diffuse emission lies west of $\eta$~Car; this emission suffuses the southern half of the Trumpler~14 (Tr14) star cluster (Townsley et al.\ in prep.).  It is sharply cut off at its western edge by the ``Northern'' molecular cloud and Carina I photodissociation region \citep{Rathborne04,Preibisch11a}.  North of this structure the diffuse emission is much fainter; no bright diffuse emission is apparent in the upper superbubble lobe north of Tr14 and Trumpler~16 \citep[Tr16,][]{Wolk11}.  In particular, no bright diffuse emission is associated with the Trumpler~15 \citep[Tr15,][]{Wang11} stellar cluster.  

No star cluster accompanies the region of brightest diffuse emission at the field center.  This bright emission lies south of the famous V-shaped dust lanes that distinguish visual images of Carina (Figures~\ref{fig:patsmooth}a and b).  The soft diffuse emission filling the lower superbubble lobe in the {\it ROSAT} image is also apparent in the \Chandra images, but it has more spatial structure, with prominent dark regions separating areas of brighter emission.  A bright linear structure also runs down the western edge of the survey field, only partially captured by our ACIS pointings.  This structure is fainter, broader, and softer than the linear features that criss-cross the field center, running east and west.  

Figure~\ref{fig:patsmooth}d is scaled to bring out fainter diffuse X-ray structures at the periphery of the mosaic.  Even with this extreme image stretch, it is clear that Carina's diffuse emission falls off precipitously to the east and south, perhaps indicating that molecular material in the South Pillars either shadows soft diffuse X-ray emission behind it or that this cold material forms a barrier to the hot plasma.  Another dark region crosses the northern part of the field west of Tr15, north of Tr14, and south of the cluster Bochum~10.  It is not clear without further analysis whether this is due to shadowing or to an absence of hot plasma in this region.  

All of these diffuse X-ray structures make more sense when placed in a multiwavelength context; see multiwavelength images in \citet[][e.g., Figure~18]{Townsley11a} to understand the relative locations of the hot and cold components of Carina's interstellar medium (ISM).  From Figure~\ref{fig:patsmooth} and these images it is clear that the eastern arm of diffuse X-ray emission is aligned with the eastern arm of the V-shaped dust lane prominent in visual images of Carina.  It is often anticoincident with the 8~$\mu$m emission there, appearing to shine through ``holes'' in the heated dust.  The brightest diffuse X-ray emission is located south of the western arm of the V-shaped dust lane, though, and appears superposed on visual emission from dense ionized gas.  The bright X-ray ``hook'' at the eastern edge of the mosaic is not associated with any prominent visual or infrared emission.

The spatial resolution in these diffuse images is limited by the shallowness of the CCCP survey; we are not taking full advantage of \Chandra's sharp imaging capabilities for diffuse structures due to limited photon statistics.  Nevertheless, these images hint that Carina's diffuse soft X-rays trace a highly complex network of hot plasmas suffusing its star clusters and southern superbubble.  Color changes indicate varying obscuration, plasma temperatures, or both.  Sharp linear features may indicate shocks.  These structures do not give the impression that Carina's superbubbles are uniformly filled with hot gas from its massive stellar winds, but these apparent surface brightness images can be deceiving, perhaps hiding simple physics behind a complex obscuring screen that complicates our view.  In the work that follows, we will attempt to disentangle the mix of emitting and absorbing components that confound these images through spatial segmentation followed by spectral analysis, with the goal of characterizing the hot plasmas that generate Carina's diffuse X-ray emission.


        
\section{DATA ANALYSIS \label{sec:analysis}}

One challenge in studying scenes of spatially-complex diffuse emission is finding a way to parse the image into segments for quantitative analysis.  This can be done ``by eye,'' extracting regions using simple geometric shapes such as circles, ovals, or squares.  Another approach is to make a contour plot of the apparent surface brightness of the scene, then extract regions to study based on the contour lines.  Both of these approaches are adequate for fairly simple surface brightness distributions, but they fall short when the scene becomes highly complex, as we see in Carina.  Thus we chose to parse Carina's diffuse emission using tessellates (space-filling polygons); the publicly-available software package {\it WVT Binning} \citep{Diehl06} proved to be very useful and effective for this exercise.

\subsection{WVT Binning \label{sec:wvt}}

\begin{figure}[htb] 
\begin{center}  
\includegraphics[width=1.0\textwidth]{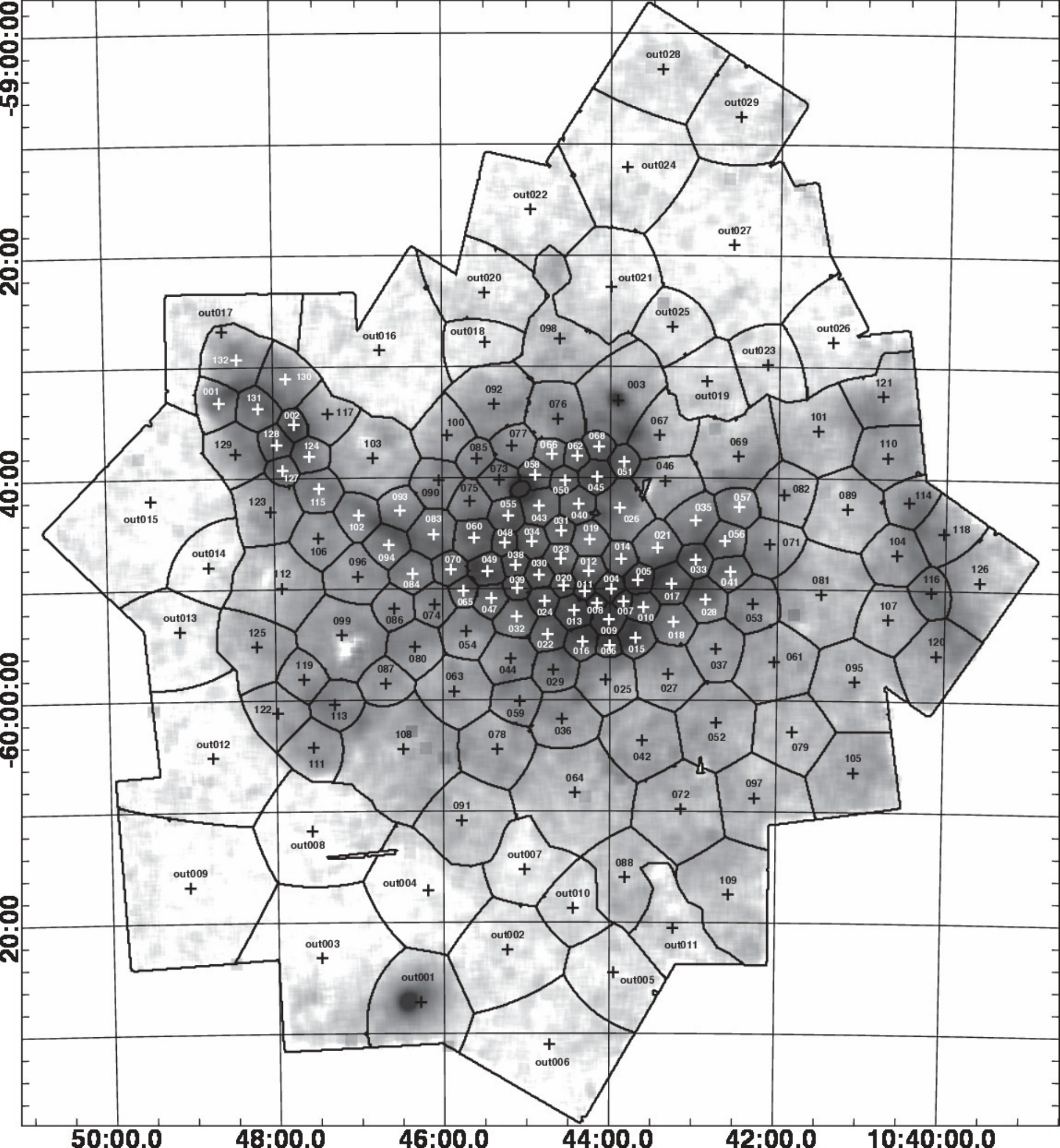}
\caption{
An apparent surface brightness image of Carina's soft-band (0.5--2~keV) diffuse X-ray emission; point sources have been excised and regions with high surface brightness appear dark.  Overlaid are tessellates from Diehl and Statler's {\it WVT Binning} code.  Tessellate centers are marked with +'s; coordinates are given in Table~\ref{tbl:tess_properties}.  Outside tessellates are labeled ``out\#''; inside tessellates are labeled just by their number.  Small unlabeled regions mark areas of missing data where the ACIS-I pointings failed to overlap.
} 
\label{fig:tessellates}
\end{center}
\end{figure}

Figure~\ref{fig:tessellates} shows an apparent surface brightness image of Carina's soft (0.5--2~keV) diffuse emission (point sources removed, smoothed with a boxcar of 5 image pixels for display purposes).  The bright X-ray nebula around \etacar (at about 10:45 -59:41) has been masked for the following analysis.  

This surface brightness image was supplied to the {\it WVT Binning} code (weighted Voronoi tessellation) by \citet{Diehl06}.  This code tessellates the image to achieve a specified signal-to-noise ratio.  We started by simply tessellating the whole image; results were not ideal because some tessellates were placed across regions where the apparent diffuse surface brightness was changing substantially.  So we contoured the diffuse emission using {\it SAOImage ds9} \citep{Joye03} to define ``inside'' and ``outside'' regions, where ``inside'' has high surface brightness 
and ``outside'' has lower surface brightness.  We saved this contour as a {\it ds9} ``region file,'' then used it as a mask for tessellating the ``inside'' regions with a S/N goal of 70 and the ``outside'' regions with a S/N goal of 40 (so the regions are not too large).  In Figure~\ref{fig:tessellates}, outside tessellates are labeled ``out\#''; inside tessellates are labeled just by their number.  Some inside tessellate numbers are shown in white to make them easier to see against the dark regions of high surface brightness in the underlying image. 
    
Table~\ref{tbl:tess_properties} gives tessellate details.  Included there are notes describing interesting features in the Carina Nebula (e.g., known young stellar clusters) that certain tessellates encompass.  While tessellation provides an objective means of segmenting the X-ray surface brightness for quantitative analysis, it does have the adverse effect of averaging over high-spatial-resolution structures, e.g., the sharp linear features seen in Figure~\ref{fig:patsmooth}.  

\begin{deluxetable}{@{}rccrrrl@{}}
\centering  \tabletypesize{\tiny} \tablewidth{0pt}

\tablecaption{Tessellate Properties \label{tbl:tess_properties}}

\tablehead{
\colhead{Name} &
\multicolumn{2}{c}{Position} &
\multicolumn{3}{c}{Extraction} & 
\colhead{Notes} \\

\colhead{} &
\multicolumn{2}{c}{\hrulefill} &
\multicolumn{3}{c}{\hrulefill} &
 \colhead{} \\

\colhead{} &
\colhead{$\alpha$ (J2000.0)} & \colhead{$\delta$ (J2000.0)} & 
\colhead{area} & \colhead{$C_{t,net}$} & \colhead{photon surface flux} & \colhead{} \\

\colhead{} &                                           
\colhead{(\arcdeg)} & \colhead{(\arcdeg)} & 
\colhead{(arcmin$^2$)} & \colhead{(counts)}  & \colhead{($10^{-6}$photon~s$^{-1}$~cm$^{-2}$~arcmin$^{-2}$)} & \colhead{} \\

\numberthecolumn & 
\numberthecolumn & \numberthecolumn & 
\numberthecolumn & \numberthecolumn & \numberthecolumn & \numberthecolumn
\setcounter{column_number}{1}
}

\startdata
inside001  &162.161180 & -59.553155 &    14.8 &  5138  & 16.4 &   \\
inside002  &161.940064 & -59.586033 &     5.8 &  4630  & 38.1 &   \\
inside003  &160.977518 & -59.551157 &    47.4 & 13620  & 12.7 & Tr14   \\
inside004  &160.998395 & -59.833481 &     5.0 &  8923  & 46.5 &   \\
inside005  &160.919355 & -59.820423 &     6.0 &  7556  & 47.7 &   \\
\enddata

\tablecomments{Table~\ref{tbl:tess_properties} is available in its entirety in the electronic edition of the Journal.  The first few lines are shown here for guidance regarding its form and content.
\\Col.\ (1):  Diffuse region label.
\\Cols.\ (2) and (3):  Mean J2000 right ascension and declination for events in the tessellate.
\\Col.\ (4):  Geometric area of the tessellate in square arcminutes, irrespective of point source masking.
\\Col.\ (5):  Counts extracted in the total energy band (0.5--7~keV), less counts expected from instrumental background.
\\Col.\ (6):  A quantity that accounts for calibration details such that it can be used to compare the apparent diffuse emission between tessellates; larger values imply brighter tessellates.
}
\end{deluxetable} 

\subsection{Spectral Fitting \label{sec:fitting}}

\begin{figure}[htb] 
\begin{center}  
\includegraphics[width=0.48\textwidth]{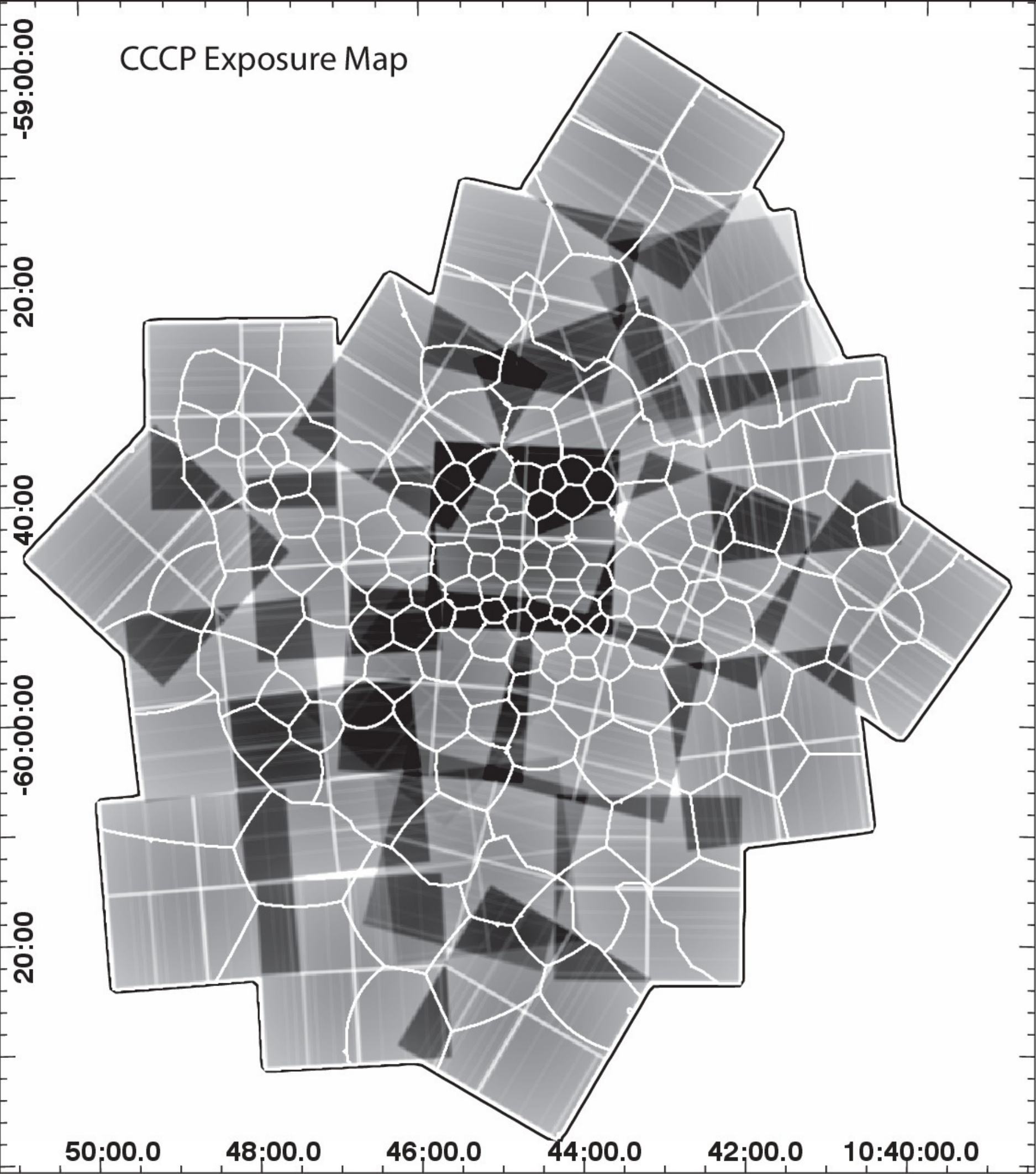}
\includegraphics[width=0.48\textwidth]{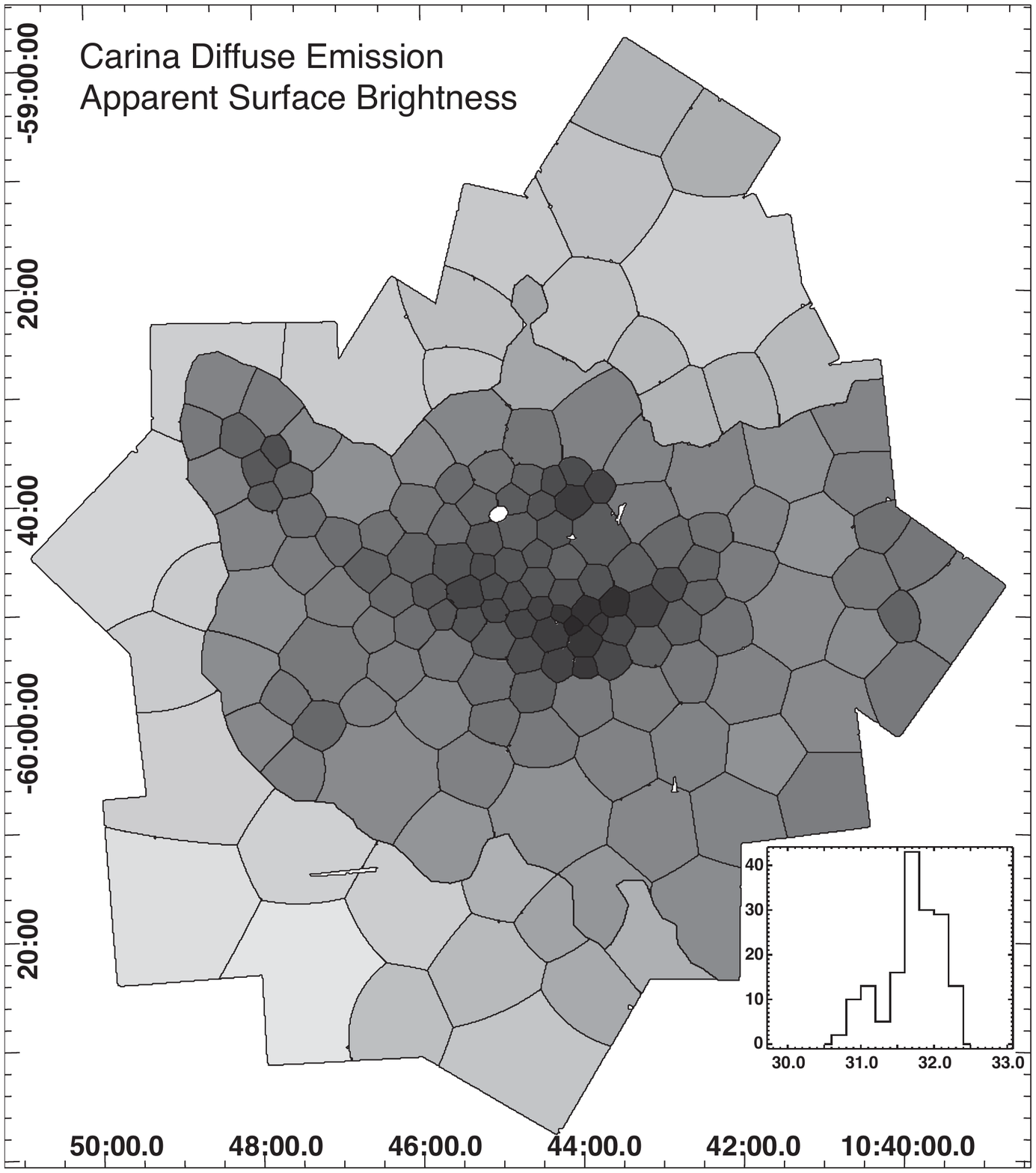}
\caption{
(a) The CCCP exposure map overlaid with our tessellates from {\it WVT Binning}.  The survey consists of 22 ACIS-I pointings and is composed of 38 individual ObsID's.  See \citet{Townsley11a} for details.
(b)  An idealized version of Carina's apparent surface brightness obtained by performing X-ray spectral fitting on the tessellates.  Our spectral fitting is described below; this soft-band (0.5--2~keV) tessellate image is presented here to facilitate comparison with the smoothed data shown in Figures~\ref{fig:patsmooth} and \ref{fig:tessellates}, and to familiarize the reader with the look of tessellate maps, as these will be used extensively throughout this paper.  All tessellate maps include inset histograms of the parameter value being mapped; in this case the histogram shows the log of the apparent surface brightness.  Here and for all tessellate maps, darker tones represent larger values of the quantity being mapped.  As for images, tessellate map coordinates are celestial J2000.
} 
\label{fig:emap}
\end{center}
\end{figure}

We were able to perform many experimental spectral fits on each Carina tessellate because {\it ACIS Extract} ({\it AE}), our publicly-available custom software for {\it Chandra}/ACIS data analysis, accomodates diffuse as well as pointlike sources \citep{Broos03,Broos10}.  We translated the tessellates generated by {\it WVT Binning} into {\it ds9} region files; these regions were used as input to {\it AE}'s diffuse spectral fitting code.  As for point sources, {\it AE} can take a diffuse patch of sky defined by a {\it ds9} region file (a tessellate) and extract the spectrum of the X-rays contained within that tessellate, computing appropriately-weighted calibration files (the so-called ARFs and RMFs) to account for the many partially-overlapping ObsID's that contribute to each tessellate.  To illustrate the value of this capability for a complex project such as the CCCP, Figure~\ref{fig:emap}a shows the Carina tessellates in white overlaid on the CCCP exposure map.

{\it ACIS Extract} can then perform automated {\it XSPEC} \citep{Arnaud96} spectral fits, including background subtraction and parameter error estimation, on these extracted tessellate spectra.  Again this automated capability is essential for a wide survey such as the CCCP; in this analysis we are working with 161 tessellates covering 5112 square arcminutes (1.42 square degrees) on the sky.  Our Carina tessellate spectra typically have $\sim$6000 counts or more and cover a few to many tens of square arcminutes.

Figure~\ref{fig:emap}b shows the tessellated version of the apparent surface brightness.  The bright \etacar nebula has been masked here and in all following tessellation work; its mask appears as a white oval at RA$= 10^{h}45^{m}04^{s}$, Dec$= -59^{\circ}41\arcmin00\arcsec$.  Comparing this image to Figures~\ref{fig:patsmooth} and \ref{fig:tessellates}, it is clear that tessellate maps provide at best a highly stylized representation of diffuse X-ray emission properties.  Larger tessellates of course represent more averaged physical quantities as they sample wide swaths of the Nebula.  We caution readers to consider the tessellate maps derived from spectral fitting in the following sections to be ``cartoon'' abstractions, not to be taken too literally or overinterpreted.  

Despite their abstract nature, tessellate maps of spectral fit parameters will prove helpful in sorting out Carina's diffuse emission.  First we must consider the possible constituent emission components as we attempt to build a spectral model to characterize that emission.

\subsubsection{Background Components \label{sec:bkgd}}

Instrumental background is estimated using ACIS stowed data \citep{Hickox06} and is subtracted from the tessellate spectra before diffuse analysis begins \citep{Broos10}, thus we assume that the background components remaining in the tessellate spectra are only celestial in origin.  One of the most dissatisfying aspects of our first attempts at {\it Chandra}/ACIS diffuse X-ray analysis \citep{Townsley03,Townsley06} was trying to choose a universal celestial background spectrum to apply to the whole field.  For Carina, we attempted to define a model using simultaneous fitting of 3 outside tessellates, then froze all parameters (including normalizations) of the resulting model, under the assumption that the background was constant across the field.  This simply never worked well -- sometimes it appeared that the background was oversubtracted, sometimes undersubtracted.  For this analysis, we have given up on the hope of a single, spatially flat background model, aside from one hard thermal plasma component that we include in every fit (with its $N_H$, kT, and normalization frozen) to represent a spatially invariant hard background, presumably composed of the cosmic X-ray background (largely unresolved active galactic nuclei, or AGN) and the Galactic Ridge emission \citep{Hamaguchi07}.  The normalization for this component was determined by taking the median normalization in fits to all outside tessellates.  Another, more flexible hard thermal plasma component (with $N_H$, kT, and normalization allowed to vary) takes up any additional hard background.

For the CCCP, we also have the unusual distinction of having discovered a cluster of galaxies superposed on the South Pillars, providing a bright, obscured, diffuse, redshifted hard thermal plasma component to the background.  It was described in \citet{Townsley11a}.  Happily this structure only affects a few outside tessellates (mainly Outside001), primarily in the hard band (2--7~keV).

\subsubsection{Foreground Components \label{sec:frgd}}  
  
A 2004 {\it FUSE/ROSAT} study by \citet{Andersson04} describes diffuse X-ray emission surrounding the Southern Coalsack, a dark cloud that lies roughly in the direction of Carina but is very much in the foreground (D$\sim$200~pc).  Figure 6 in that paper shows an X-ray halo around the Coalsack; their interpretation is that this is due to the interaction of this cold cloud with the hot plasma in the Upper Centaurus-Lupus Superbubble.  Close examination of this figure shows that other neighboring dark clouds have similar X-ray halos.  Noting that Carina is located at l$\sim288^{\circ}$, one can imagine that such an X-ray halo from a foreground cloud crosses in front of the Carina Nebula (sitting just off the right edge of the figure).  This foreground X-ray halo may contribute to the soft counts that we see in the CCCP data, but there is no obvious signature of it in the {\it ROSAT} image of Carina \citep[Figure 1 in][]{Townsley11a}.

\citet{Hamaguchi07} note that the Local Hot Bubble (LHB) contributes spatially uniform soft emission in the direction of Carina and model it with a thermal plasma with kT = 0.1~keV.  They quote a surface brightness for this emission of $4 \times 10^{-4}$~counts~s$^{-1}$~arcmin$^{-2}$ from \citet{Snowden98}.  Using the {\it PIMMS} tool by Koji Mukai\footnote{\url{http://heasarc.gsfc.nasa.gov/docs/software/tools/pimms.html}}, this translates to a total-band (0.5--7~keV) ACIS-I surface brightness of $1.08 \times 10^{-5}$~counts~s$^{-1}$~arcmin$^{-2}$.  For a typical integration time of 60~ks, then, we should expect a spatially uniform distribution of LHB counts with a surface density of 0.65~counts~arcmin$^{-2}$.  From Table 1, the largest tessellate (outside027) has an area of 188.4~arcmin$^{2}$, so this could have 122 LHB events (out of 10346 counts).  The tessellate with the smallest number of counts is outside017 (with 3328 counts); its area is 64.7~arcmin$^{2}$, so it could have 42 LHB events.  In both of these extreme cases, the LHB could be contributing $\sim$1\% of the tessellate counts.

Another possible source of unobscured soft X-ray emission is solar wind charge exchange \citep[SWCX, e.g.,][]{Snowden09}; this is a time-dependent phenomenon associated with space weather.  Although we filtered our data to remove times of high background \citep{Broos11a}, any residual SWCX emission is not easily removed in our composite dataset, made up of 38 individual observations obtained over a period of several years (see Figure~\ref{fig:emap}).  Strong SWCX emission should leave the imprint of the observation in which it occurred on the soft-band images; we should see enhanced line emission (SWCX has no continuum component) primarily from O{\scriptsize VII} and O{\scriptsize VIII} \citep{Snowden04} in the ACIS-I 0.5--0.7~keV image in Figure~\ref{fig:patsmooth} (panels c and d), filling the square pattern of the ACIS-I CCD array, matching the roll angle of the observation in question.  We see no such pattern, thus we assume that SWCX is not strongly contaminating any of our observations.

While the emission that we attribute below to Carina itself could include a contribution from these foreground emission sources, we don't think it dominates, because there are clear anticorrelations between Carina's diffuse emission and known colder structures in the Carina complex traced by, e.g., dense ionized gas in visual Digitized Sky Survey (DSS) or H$\alpha$ images, or PAH emission seen in {\it MSX} or {\it Spitzer} images \citep{Townsley11a}.  This implies that Carina's soft diffuse X-rays are either shadowed or displaced by its colder structures; in either case, it suggests that the soft diffuse emission that we see in this direction is mainly generated in the Carina Nebula, not in front of it.

Another concern is unresolved foreground stars; although field stars are generally X-ray-faint, the space cone subtended by the CCCP encompasses many foreground stars, given our assumed distance of 2.3~kpc to Carina \citep{SmithN06a}.  The composite spectrum of individually-detected foreground stars in the CCCP shows soft thermal plasma emission, with components at kT = 0.2 and 0.6~keV \citep{Broos11b}; unresolved foreground stars are likely to have a similar spectral shape that could easily masquerade as soft diffuse emission.  As part of an investigation into CCCP contaminating point source populations, \citet{Getman11} performed extensive Monte Carlo simulations of the foreground star population and its X-ray emission; those authors estimate that the $\sim$200,000 foreground stars in the space cone to Carina that remain unresolved in the CCCP study could contribute $\sim$10$^{4}$ soft-band counts to the diffuse emission.  Again, since we have over $10^{6}$ counts in our composite diffuse spectrum, foreground stars are contributing only at roughly the 1\% level.

\subsubsection{Carina Components \label{sec:carcomps}}
  
Even within the Carina Nebula, there are several different sources of unresolved X-ray emission that can contribute to the tessellate spectra.  There should be one or more truly diffuse emission components; from experience with other massive star-forming regions \citep{Townsley03,Townsley06}, the brightest of these tend to be soft thermal plasmas with $0.1 < kT < 1$~keV, although harder thermal plasmas and/or non-thermal (synchrotron) emission have been seen, perhaps in conjunction with cavity supernovae \citep[e.g.,][]{Wolk02,Fujita09,Townsley09,Townsley11b}.  We cannot rule out the possibility that we might see these plasmas still in a state of non-equilibrium ionization.  They could be generated by stellar winds, by cavity supernovae, or by both. 

Another source of unresolved emission, especially in a survey as shallow as the CCCP, is the pre-Main Sequence (pre-MS) star population.  There are many tens of thousands of unresolved young stars contributing to Carina's ``diffuse'' emission \citep{Feigelson11,Preibisch11b,Povich11}.  Based on the spectra of the young stars that we do resolve (Section~\ref{sec:softptcomps} below) and on more sensitive studies of pre-MS stars \citep[e.g.,][]{Preibisch05}, we expect this unresolved stellar component to be represented by a composite thermal plasma with both soft and hard components and that its normalization should be spatially variable, following the distribution of known clusters.  More obscured sources are harder to detect in a shallow survey and would tend to lose their soft components due to obscuration, so we might suspect that the unresolved stellar population would appear somewhat harder than the resolved stellar population.

\etacar and its surrounding bright X-ray nebula dominate the emission in their immediate vicinity.  We have attempted to mask \etacar itself, its readout streak, and its surrounding nebula, but some residual emission could remain due to scattered light.  Tessellates that could be affected are those that surround the mask that excludes these sources, namely inside043, inside055, inside058, and inside073.

\subsubsection{The Spectral Model \label{sec:model}}

Choosing a spectral model to account for all of these components proved difficult, and we experimented with many combinations of thermal plasmas in collisional ionization equilibrium (CIE), thermal plasmas with non-equilibrium ionization (NEI), and power laws (to allow for the possibility of synchrotron emission).  In the end, we settled on a phenomenological model that combines several NEI and CIE thermal plasmas and provides an adequate fit for many tessellates.  In {\it XSPEC} parlance it takes this form: \\
{\it TBabs}1*{\it vpshock}1 + {\it TBabs}2*{\it vpshock}2 + {\it TBabs}3*{\it vpshock}3 + {\it TBabs}4*{\it apec}4 + {\it TBabs}5*{\it apec}5 + {\it TBabs}6*{\it apec}6 \\
where, from the {\it XSPEC} manual \citep{Arnaud96}, {\it TBabs} is the Tuebingen-Boulder ISM absorption model \citep{Wilms00}, {\it vpshock} is a variable-abundance plane-parallel shock (NEI) plasma model \citep{Borkowski01}, and {\it apec} is a CIE plasma model \citep{SmithR01}.  Table~\ref{tbl:components} details the components of this model; these six components are also referred to as kT1 -- kT6, respectively.  The {\it vpshock} model used was the version in {\it XSPEC}; we caution the reader that this model is known to be incomplete in its modeling of spectral lines, especially the Fe~L lines.

\begin{deluxetable}{@{}lrrl@{}}
\centering  \tabletypesize{\tiny} \tablewidth{0pt}

\tablecaption{Spectral Model Components \label{tbl:components}}

\tablehead{
\colhead{Name} &
\colhead{Lower Limit} &
\colhead{Upper Limit} &
\colhead{Purpose} \\

\numberthecolumn & 
\numberthecolumn & 
\numberthecolumn &
\numberthecolumn 
\setcounter{column_number}{1}
}

\startdata
{\it TBabs}1 ($N_{H1}$) & $0.016 \times 10^{22}$~cm$^{-2}$ & $0.8 \times 10^{22}$~cm$^{-2}$ & absorbing column for {\it vpshock}1 \\
{\it vpshock}1 (kT1) & 0.1~keV & 1.0~keV & the softer NEI plasma component, usually long timescale \\
{\it TBabs}2 ($N_{H2}$) & $0.016 \times 10^{22}$~cm$^{-2}$ & $0.8 \times 10^{22}$~cm$^{-2}$ & absorbing column for {\it vpshock}2 \\
{\it vpshock}2 (kT2) & 0.1~keV & 1.0~keV & the intermediate NEI plasma component, usually short timescale \\
{\it TBabs}3 ($N_{H3}$) & $0.016 \times 10^{22}$~cm$^{-2}$ & $1.0 \times 10^{22}$~cm$^{-2}$ & absorbing column for {\it vpshock}3 \\
{\it vpshock}3 (kT3) & 0.4~keV & 2.0~keV & the harder NEI plasma component, usually long timescale \\
{\it TBabs}4 ($N_{H4}$) & $0.16 \times 10^{22}$~cm$^{-2}$ & $10.0 \times 10^{22}$~cm$^{-2}$ & absorption associated mainly with unresolved young stars in Carina \\
{\it apec}4 (kT4) & 2.0~keV & 4.0~keV & a CIE plasma designed primarily to take up unresolved young star emission \\
{\it TBabs}5 ($N_{H5}$) & $0.16 \times 10^{22}$~cm$^{-2}$ & $10.0 \times 10^{22}$~cm$^{-2}$ & absorption associated with the hard thermal plasma kT5 \\
{\it apec}5 (kT5) & 4.0~keV & 15.0~keV & a hard CIE plasma to account for, e.g., the cluster of galaxies, background AGN, or \etacar \\
{\it TBabs}6 ($N_{H6}$) & N/A & N/A & absorption associated with the hard X-ray background, frozen at 2.0e22~cm$^{-2}$ \\
{\it apec}6 (kT6) & N/A & N/A & a hard CIE plasma to account for the cosmic X-ray background and Galactic Ridge emission, frozen at 10~keV \\
\enddata

\tablecomments{
\\Cols.\ (2) and (3):  Allowed range for parameter value in the spectral fit.
}
\end{deluxetable}

To represent Carina's diffuse emission, we chose three NEI plasmas, each suffering independent obscuration and with their electron-density-weighted ionization timescales \citep{SmithR10} allowed to vary.  For most tessellates, two of the NEI plasmas (kT1 and kT3) favored CIE (long timescales); while these could be replaced with a CIE model such as {\it vapec} \citep{SmithR01}, we preferred to allow for the possibility of NEI conditions and retained the {\it vpshock} model for these components.  The other NEI plasma (kT2) almost always ran to its lower timescale limit, implying a low-density, highly non-equilibrium plasma perhaps indicative of a recent shock \citep{SmithR10}.  We hesitate to adopt this interpretation completely, however, because this model component shows substantial emission measure in most tessellates, whereas we might suspect that recent shocks in Carina's complex ISM would be more localized.  

Alternatively, this component may simply be an indication of a need for soft ``continuum'' counts, i.e., a smooth model component that is not strongly line-dominated at low energies.  It is well-modeled by thermal bremsstrahlung emission, but we currently have no physical explanation for such a component so we did not use it.  We attempted to replace kT2 with a power law component that could give such soft counts and might represent synchrotron emission from a cavity supernova, but the resulting power law slope was always steeper than the $\Gamma$ = 2--3 that is typically seen in relativistic
electrons accelerated behind a supernova shock \citep[e.g.,][]{Bamba03}.  Although we lack an adequate physical interpretation for the kT2 component, we suspect that it represents different emission components in different tessellates; these components could include foreground emission (unresolved foreground stars and foreground diffuse components listed above), the soft thermal plasma component of Carina's unresolved pre-MS star population, synchrotron emission from a cavity supernova remnant, or some different physical emission mechanism that we have failed to identify.  Since the timescale in the {\it vpshock} model is weighted by electron density, short ``timescales'' may in fact be indicating low densities.  In cases where its emission measure is low, kT2 may not be needed at all; the other NEI components might take up the extra emission.  It is left in place for all tessellates primarily because it is easier to compare spectral fit parameters between tessellates if they all use the same model. 

Substantial improvement to the fits was seen when the abundances of the soft NEI plasmas were allowed to vary.  Importantly, in these fits the abundances Z were not allowed to go below solar values (our reasoning is described in Section~\ref{sec:maps}); their allowed range was $1.0 < Z < 5.0$ in solar units.  The upper limit of 5Z$_{\odot}$ is arbitrary; it was set at this value simply because the abundances are not well-constrained by our spectra and we didn't want them to become unreasonably large.  The abundances were linked between the three soft plasmas, under the (perhaps flawed) assumption that they all originate in the Carina complex.  

Through experimentation, we found that Si and Fe (with Ni linked to Fe) often required supersolar abundances to get the best fit.  The Fe abundance affects the Fe-L line complex at $\sim$0.8~keV; these plasmas are too soft to generate any evidence of Fe-K$\alpha$ emission at 6.7~keV.  Extrasolar abundances for O, Ne, Mg, and/or S were also occasionally found to improve the fits, although their actual values are not well-constrained.  These abundance variations only affect Components 1 and 3, the NEI plasmas that tend towards CIE, because they show strong line emission.  Again we advise the reader that a physical interpretion of supersolar abundances requires caution; enhancing an element's abundance often strengthens line features in the model, but adding another plasma component can sometimes have the same effect.

Below, we treat kT4 -- kT6 essentially as ``nuisance'' components, accounting for spectral features that we know must be present given the constituents of Carina and along its line of sight, but that likely do not comprise diffuse emission from Carina's massive star winds or cavity supernovae.  Component 4 has subsolar abundance (frozen at 0.3Z$_{\odot}$) and a restricted plasma temperature range; this is a coarse attempt to have it account for the hard component of pre-MS star emission.  Component 5 is a catch-all hard plasma that accounts for hard emission not adequately modeled by other components, regardless of its origin.  We use it to represent emission from the \etacar nebula in some tessellates and from the cluster of galaxies in others, for example.  

Component 6, along with a frozen $N_{H}$ and kT, has an {\it XSPEC} emission measure per square parsec frozen at $\sim 8.1 \times 10^{53}$~cm$^{-3}$~pc$^{-2}$, based on fits to the outside tessellates (the units involve a ``per square parsec'' to reflect the fact that we are working with surface brightnesses in our spectral fitting).  By freezing the model normalization, we presume that the surface brightness of this component is constant across the entire field.  For some tessellates, kT5 may take up any extra hard emission (presumably from unresolved AGN or Galactic Ridge emission) that deviates from this assumption.  Of course AGN are more appropriately modeled with a power law since their emission is nonthermal, but we retain the thermal plasma spectral shape for both kT5 and kT6 since they may represent more than just AGN emission.

\begin{figure}[htb] 
\begin{center}  
\includegraphics[width=0.48\textwidth]{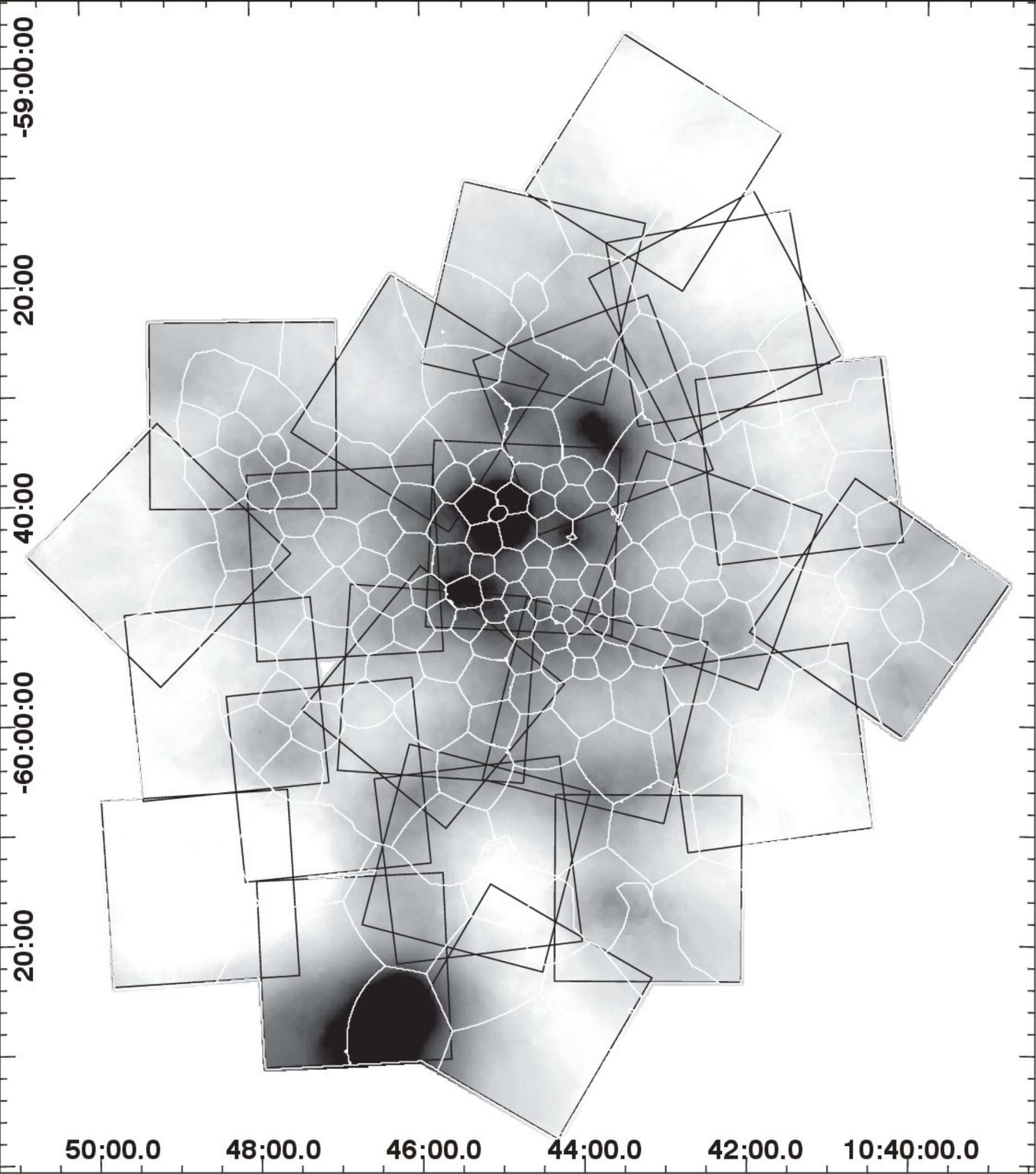}
\caption{A hard-band (2--7~keV) smoothed apparent surface brightness image of Carina's diffuse emission (point sources have been removed).  Diffuse emission tessellates are shown in white and the 38 ACIS-I observations are shown as black squares.  The \etacar X-ray nebula has {\em not} been masked in this image so it appears bright, as does the cluster of galaxies at the southern edge of the field and the wings of the piled-up massive star WR~25, located just west of the \etacar nebula in the same ACIS-I pointing.  Other bright structures are Tr14 and the small obscured cluster CCCP-Cl~14 \citep{Feigelson11}.  
} 
\label{fig:hardimg}
\end{center}
\end{figure}

We examined hard-band (2--7~keV) images of the CCCP mosaic using a variety of smoothing techniques and spatial scales, to search for a truly diffuse hard component; again such emission might come from synchrotron processes in the Carina complex and would be strong evidence for recent cavity supernovae.  An example is shown in Figure~\ref{fig:hardimg}.  We found no evidence for a hard diffuse component; rather the smoothed images showed enhancements around the edges of the individual ACIS-I pointings and at the centers of known stellar clusters, tracing the unresolved stellar population across the field.  Pointing edges are enhanced because the \Chandra PSF degrades strongly far off-axis, thus diminishing our point source detection sensitivity.

This model is not physical in the sense that a given model component can always be guaranteed to trace a given physical component of the emission; in fact we suspect that some model components represent a mix of physical components and that model components represent different physical components in different tessellates.  An example of modeling such complex diffuse X-ray emission with more physically-motivated models is the recent study of M101 by \citet{Kuntz10}.  

The fits that we present are also not unique; very different combinations of absorptions and plasma temperatures may also give reasonable results, since the components can trade off normalizations with each other.  This possible shifting of power from one fitted component to another is a general problem that can affect all spectral fitting of X-ray observations with multiple emission and absorption components.

While we found in our fitting that enhancing abundances worked better than adding more plasma components, we cannot guarantee that our model encompasses all the physics at play in the Carina complex; since some tessellates still have poor-quality spectral fits, we are rather guaranteed that our model, despite its complexity, is incomplete.  In Section~\ref{sec:charge} below, we propose an X-ray emission mechanism not usually considered for star-forming regions as a partial solution to the missing physics in our spectral model.

 
\section{RESULTS \label{sec:results}}

The results of our spectral fitting can be presented in many ways:  classic X-ray spectra showing the data, composite model, model components, and fit residuals for each tessellate; tables of parameter values; tessellate maps showing how a given model parameter varies across the field; tessellate maps of inferred quantities such as intrinsic surface brightnesses.  Below we give examples of all of these analysis products.

\subsection{Spectra \label{sec:spectra}}

\begin{figure}[htb] 
\begin{center}  
\includegraphics[width=0.32\textwidth]{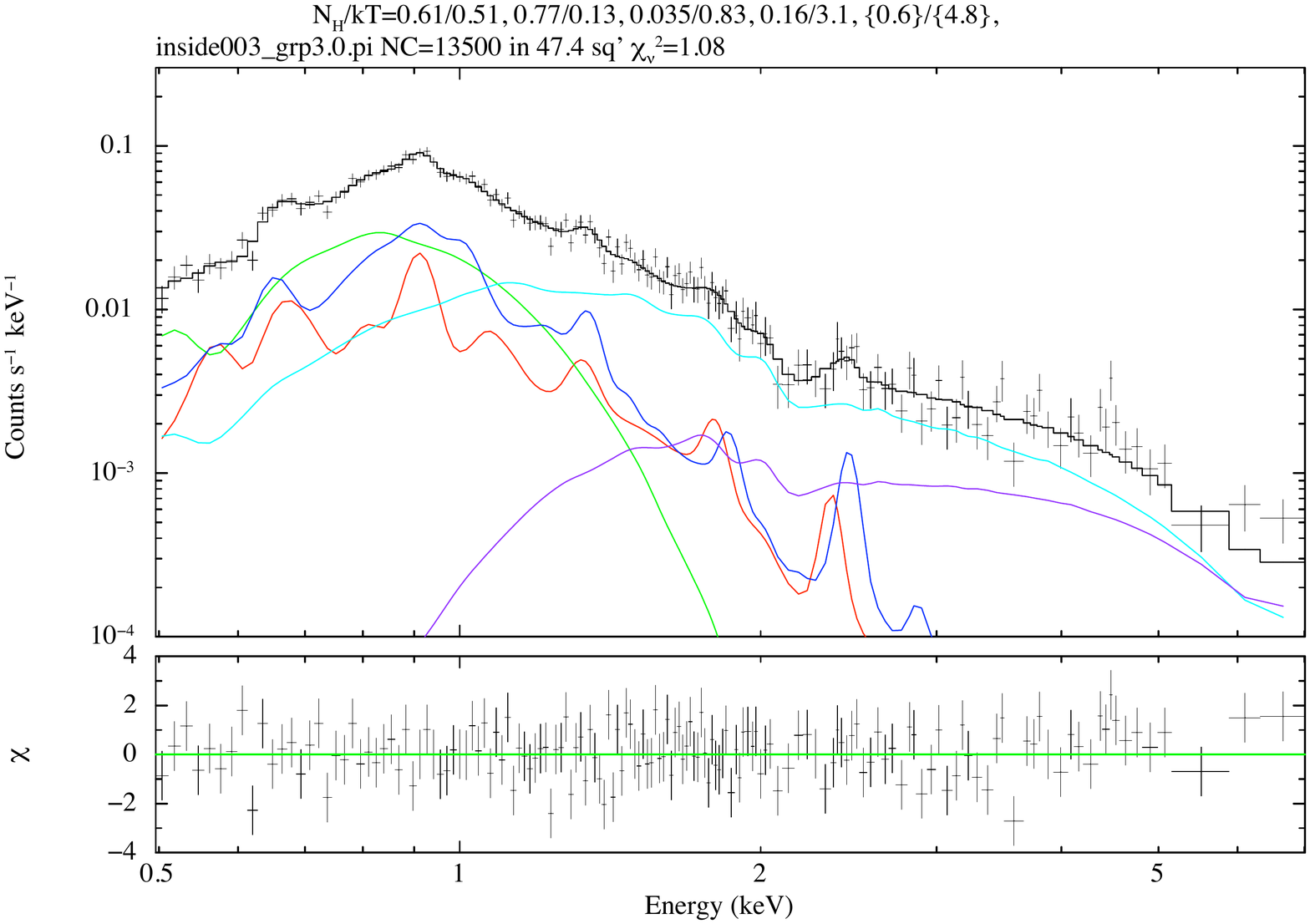}
\includegraphics[width=0.32\textwidth]{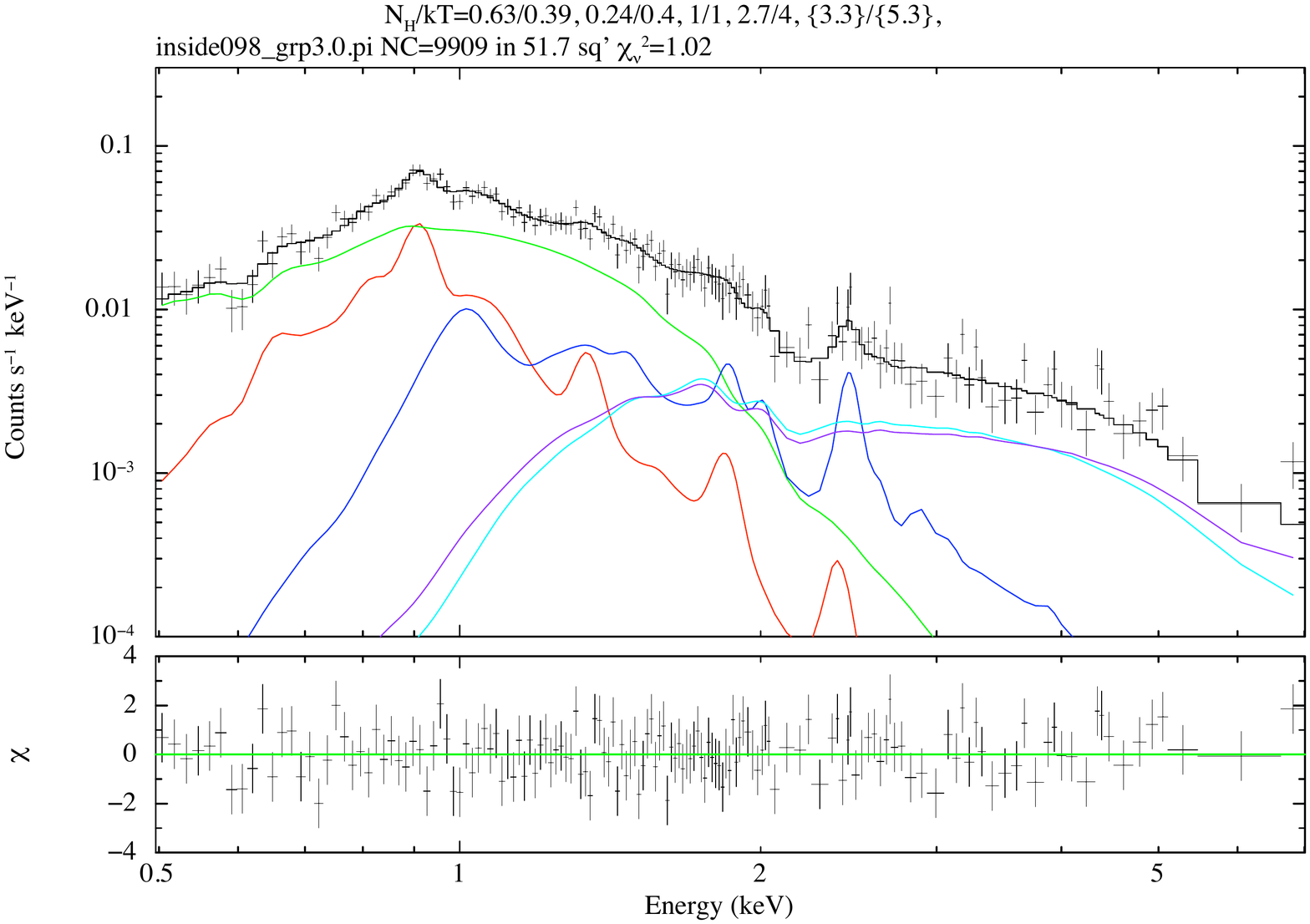}
\includegraphics[width=0.32\textwidth]{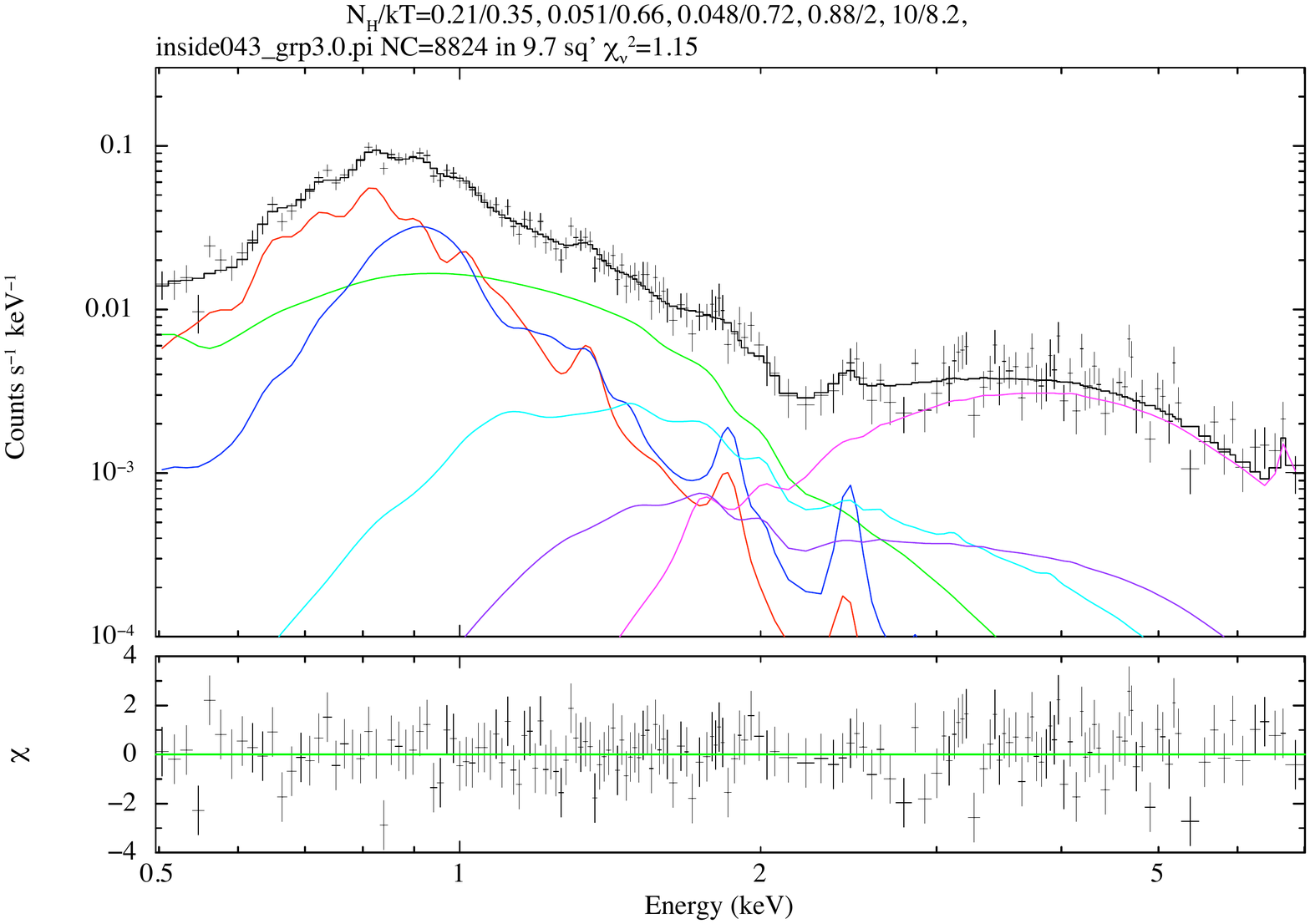}
\includegraphics[width=0.32\textwidth]{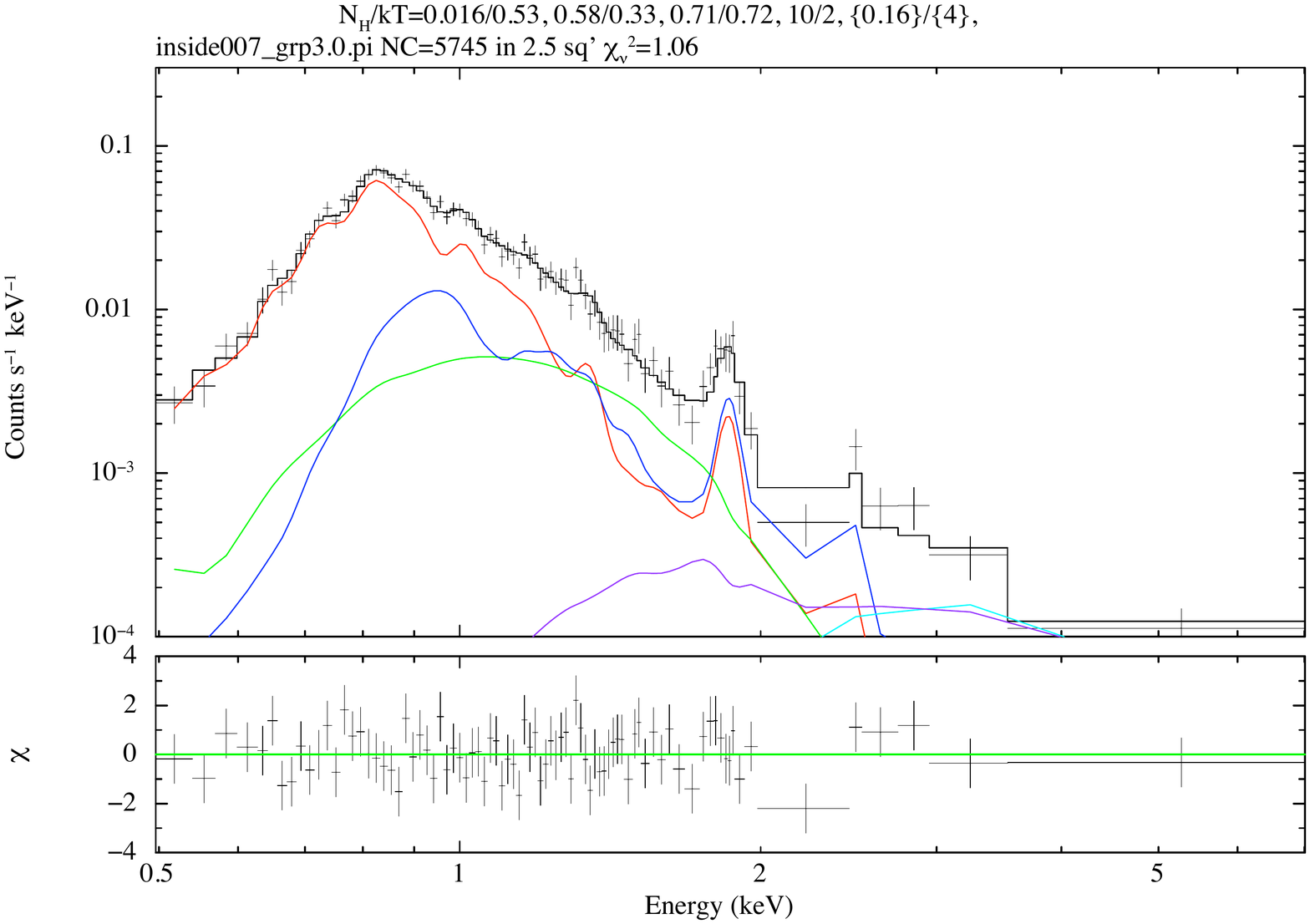}
\includegraphics[width=0.32\textwidth]{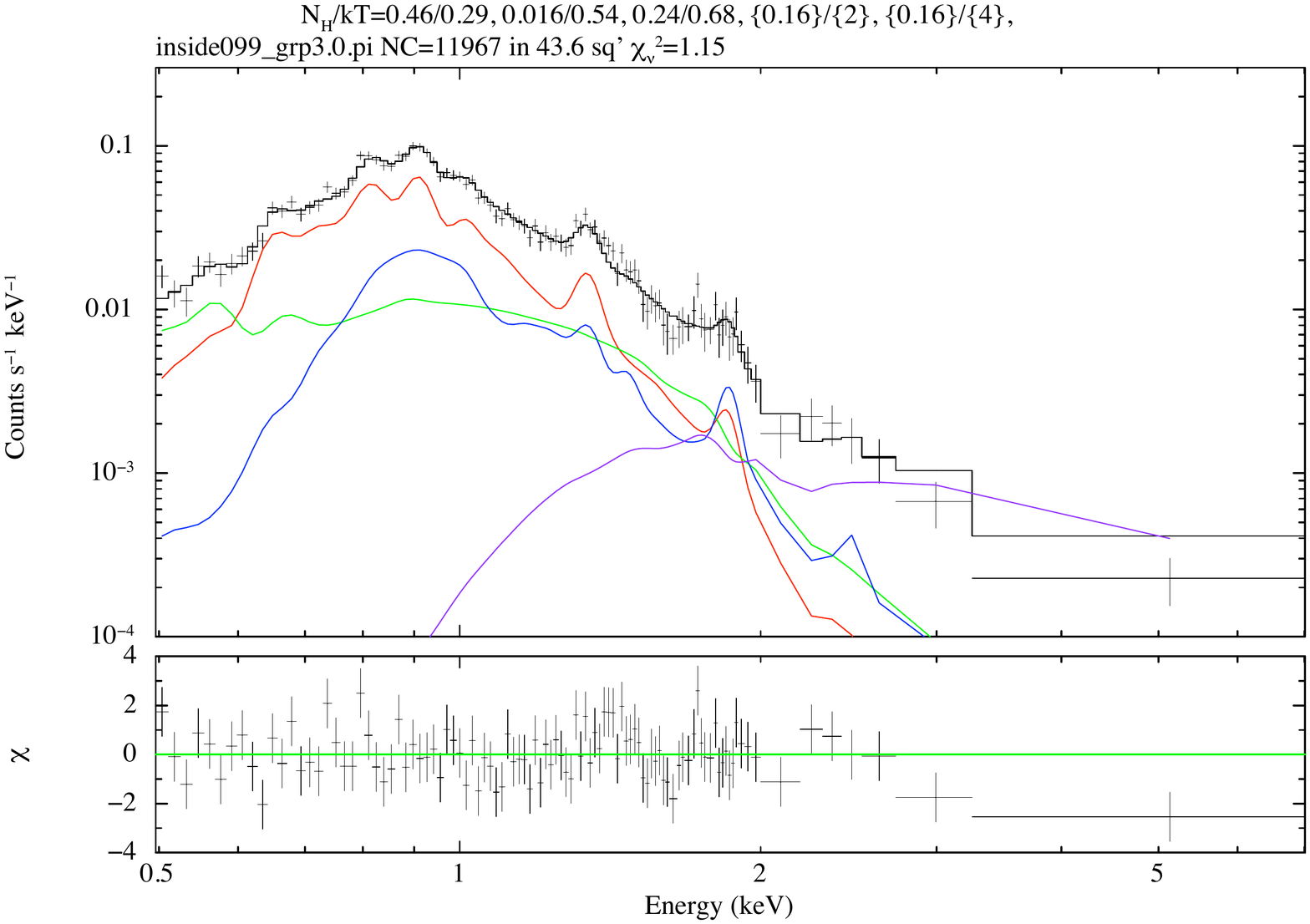}
\includegraphics[width=0.32\textwidth]{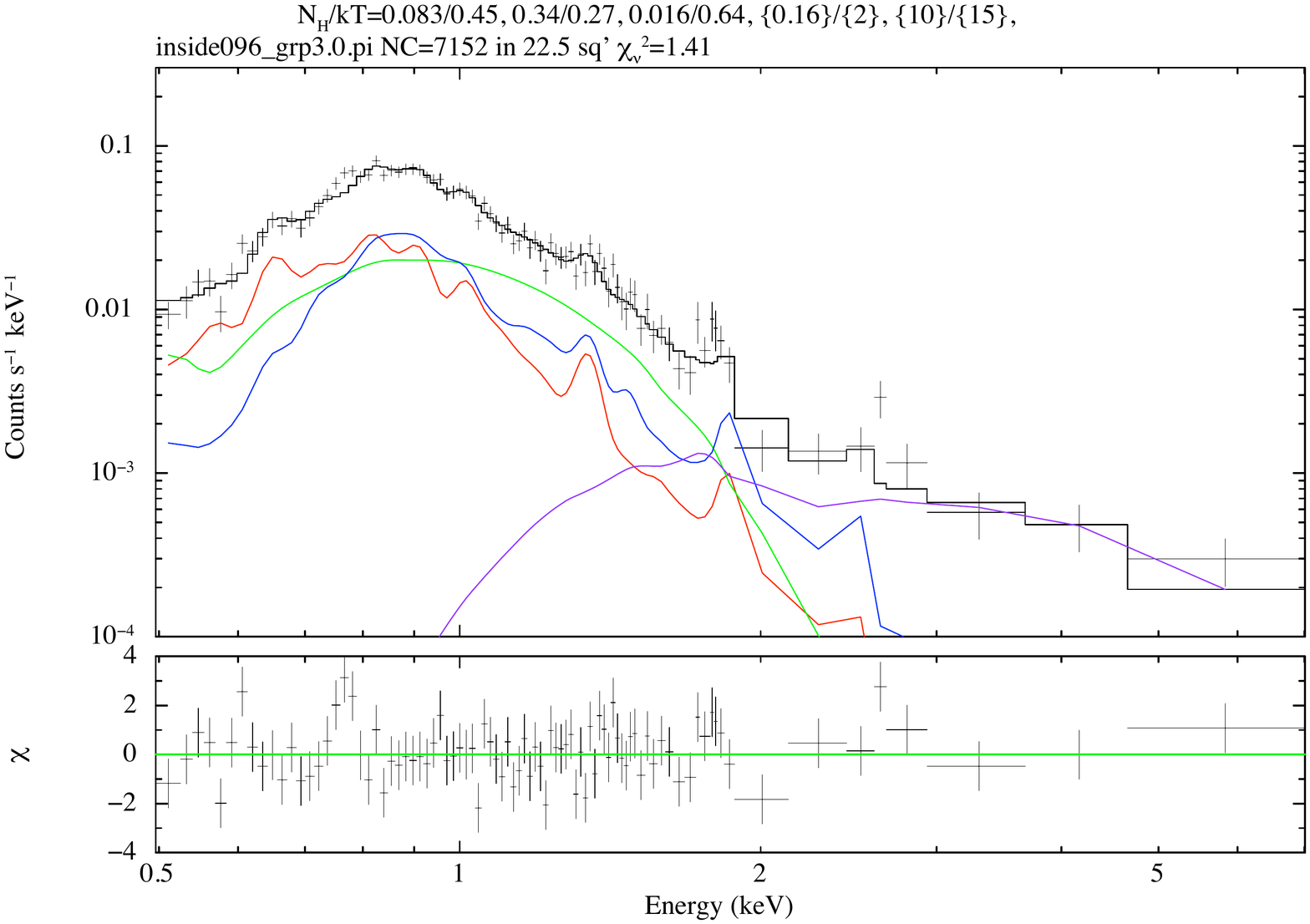}
\includegraphics[width=0.32\textwidth]{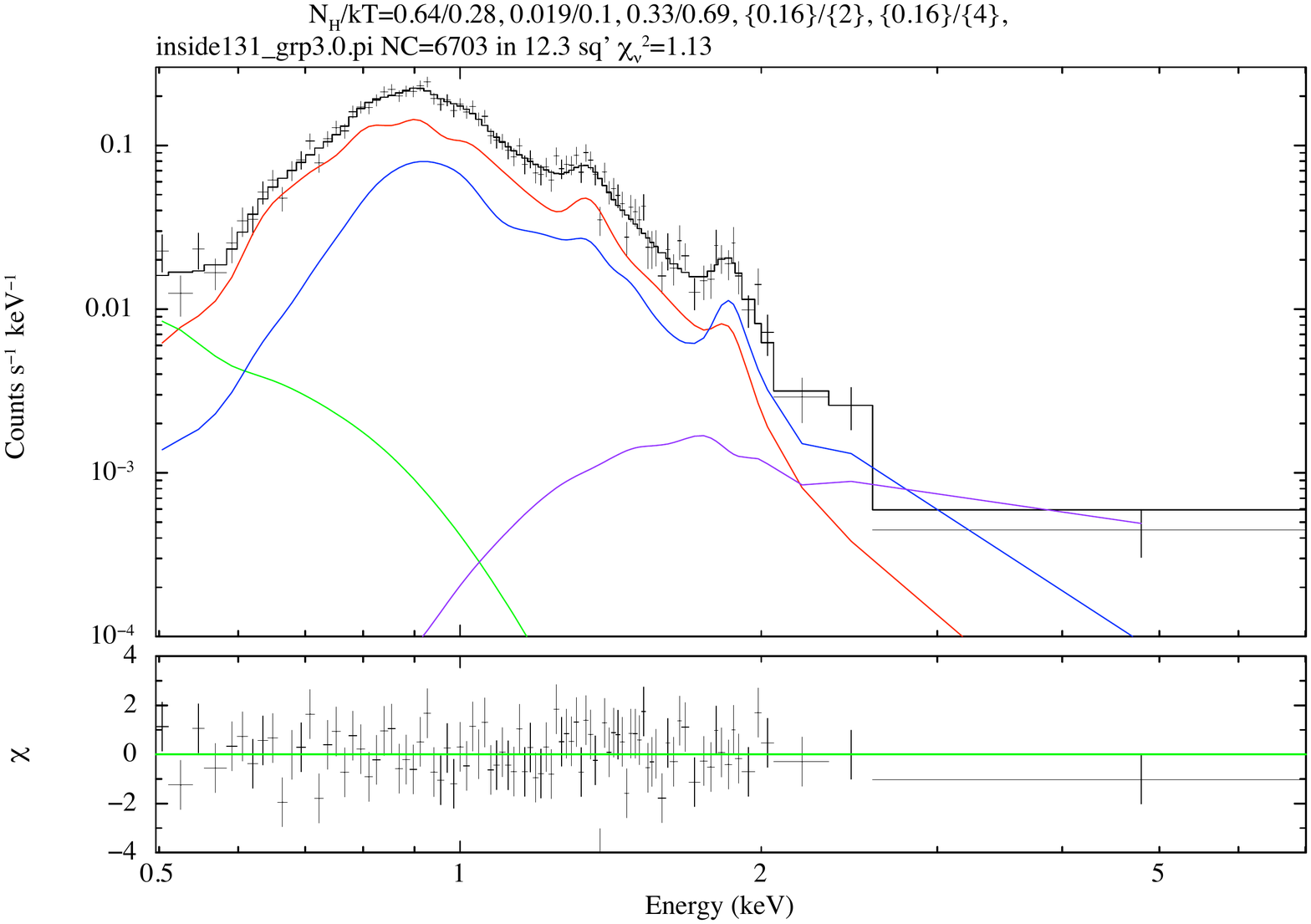}
\includegraphics[width=0.32\textwidth]{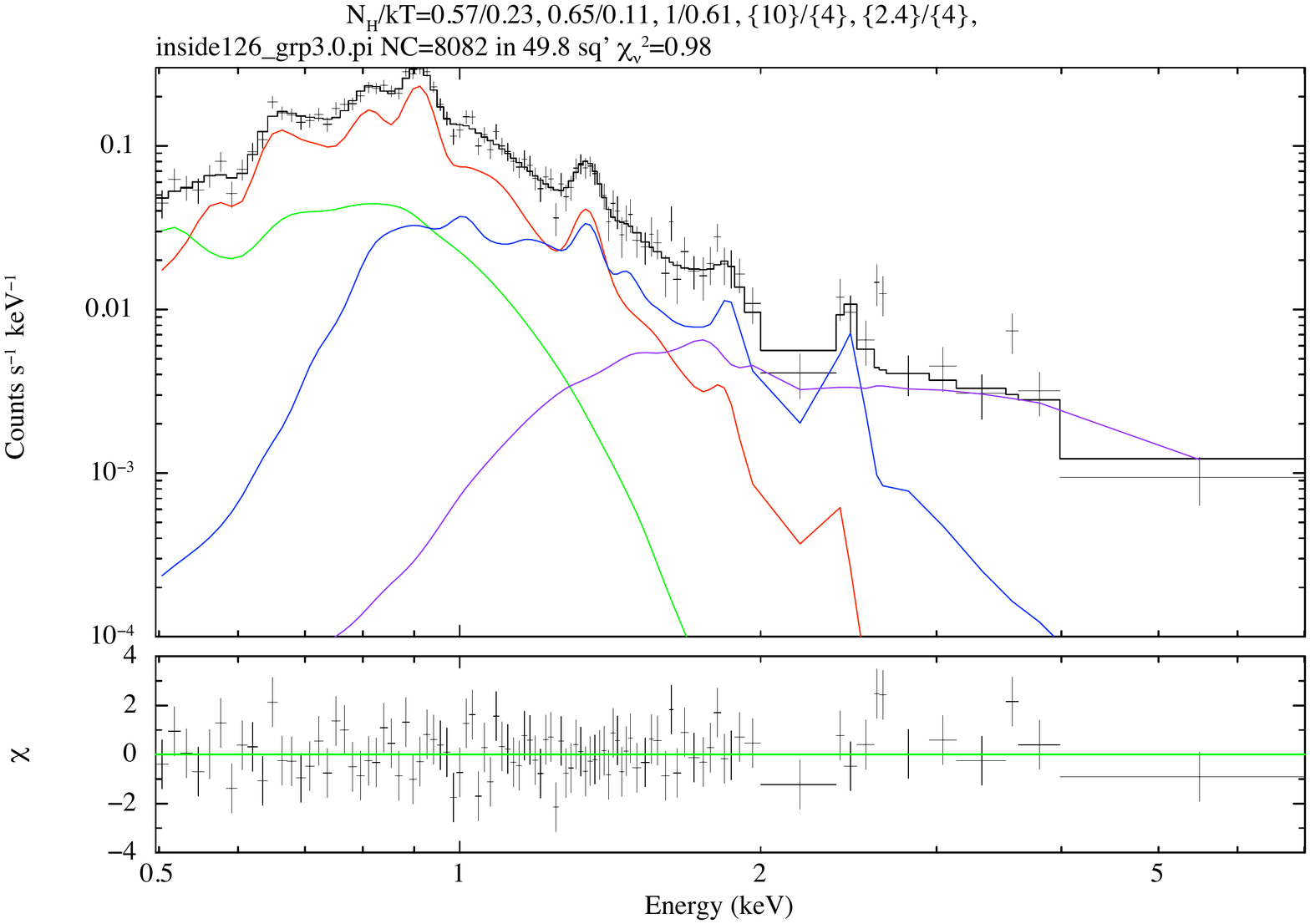}
\includegraphics[width=0.32\textwidth]{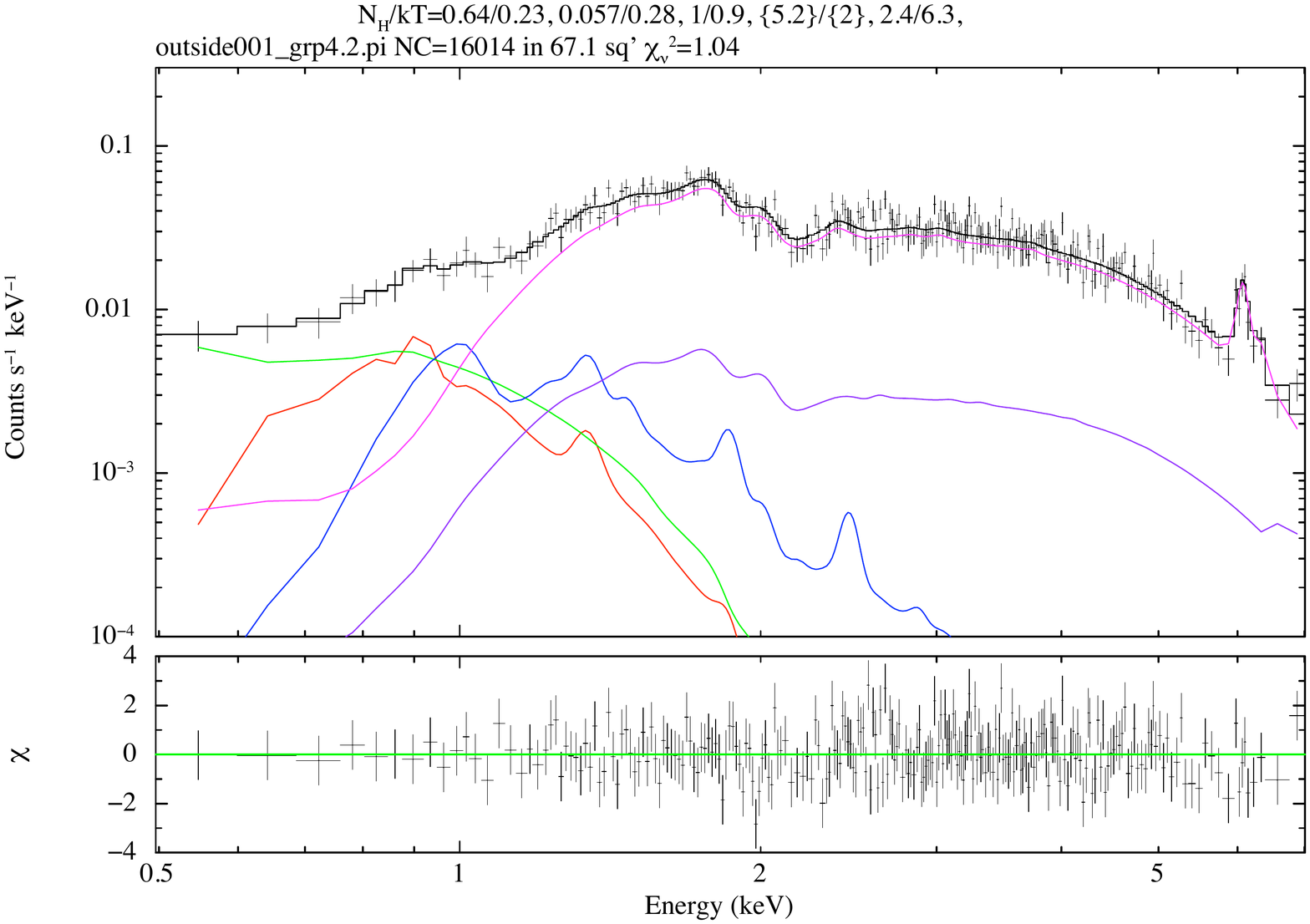}
\caption{Sample tessellate spectra; the electronic version of this figure shows the spectra for all tessellates.  It may be necessary to zoom this figure to read the details.  Each model component is shown, with the following color coding:  kT1=red, kT2=green, kT3=blue, kT4=cyan, kT5=magenta, kT6=purple.  Parameters in curly brackets were frozen in the fit; they were generally not needed for a good fit so their normalizations were set to zero.  Component 6 was fully frozen (including normalization) in all fits so it is not included in the heading of each spectrum.
(a)  Inside003 (containing Tr14).
(b)  Inside098 (containing Tr15 and the region between Tr14 and Tr15).
(c)  Inside043 (containing part of Tr16 and adjacent to the mask blocking $\eta$~Car).
(d)  Inside007, illustrating a spectrum with a strong Si line.
(e)  Inside099, illustrating a spectrum with a strong Fe line.
(f)  Inside096, illustrating a spectrum with a strong unmodeled line at $\sim$0.76~keV; this line is seen in several spectra.  
(g)  Inside131, an intrinsically bright tessellate in the eastern ``hook'' region.
(h)  Inside126, an intrinsically bright tessellate at the western edge of the field.
(i)  Outside001, the cluster of galaxies behind the South Pillars.
} 
\label{fig:spectra}
\end{center}
\end{figure}

A sampler showing some of the variety present in the tessellate spectra is presented in Figure~\ref{fig:spectra} (all tessellate spectra are included in an online-only version of this figure).  These examples are taken from the complete compendium of tessellate spectra, which is provided as an electronic-only figure. Each plot gives a brief title encoding some of the tessellate properties and fit parameters (more easily read by zooming the figure).  Absorbing columns (in units of $10^{22}$~cm$^{-2}$) and plasma temperatures (in keV) for kT1 -- kT5 are shown on the first title line.  Parameters in curly brackets were frozen in the fit; kT6 is not listed because all of its parameters were frozen to the same values for all tessellates (see Table~\ref{tbl:components}).  The second line gives the tessellate name and spectrum signal-to-noise grouping encoded into the spectrum's file name (the ``.pi'' file), the net counts in the spectrum, the area in square arcminutes, and the reduced $\chi^{2}$ of the fit.


Please note that the spectral fits were not correlated between tessellates, i.e., a given tessellate's fit was independent from those of surrounding tessellates.  We remind the reader that the ordinate axis in all {\it XSPEC} plots is shown in uncalibrated units (the observed event rate within the extraction region). ÊThus, even for point sources, inferring relative flux by comparing {\it XSPEC} plots from two sources is not reliable, because the PSF fraction and effective area can vary between the extractions. ÊFor diffuse extractions, count rate values in these plots provide no information about astrophysical surface brightness because the geometric area of tesselates varies significantly.  Also, the scheme that {\it AE} uses to combine and calibrate the multiple observations \citep[][Section 6.1]{Broos10} that contribute to individual tessellates results in large tessellate-to-tessellate variations in the ``exposure time'' (FITS keyword EXPOSURE), which {\it XSPEC} uses to normalizeÊthe ordinate axis in plots.

To get around these technical difficulties, Table~\ref{tbl:tess_properties} Column~6 reports a simple  photometric quantity, ``photon surface flux'' \citep[][Section~7.4]{Broos10}, in the soft energy band (0.5--2~keV).  This quantity appropriately accounts for all calibration details, including instrumental background subtraction, so it can be used to compare the average apparent surface brightness between tessellates (thus a tessellate that appears dark in Figure~\ref{fig:tessellates}a would have a large value of photon surface flux in Table~\ref{tbl:tess_properties}).

The first row of Figure~\ref{fig:spectra} (panels a--c) shows tessellates that contain populous, unobscured young stellar clusters.  While we have removed all detected X-ray point sources associated with these clusters before performing this diffuse spectral analysis, these tessellates must contain unresolved point sources.  To some degree, Component 4 (shown in cyan) accounts for the hard spectral component of these sources and/or for obscured young stars associated with these clusters.  Tessellate inside043 (panel c) also includes residual emission from \etacar and its X-ray nebula; this is seen primarily as excess hard emission taken up by Component 5 (in magenta).  

The second row of Figure~\ref{fig:spectra} (panels d--f) shows examples of prominent spectral lines featured in the tessellate spectra.  Tessellate inside007 (panel d) exhibits a strong Si-K$\alpha$ line at 1.86~keV.  Tessellate inside099 (panel e) shows a prominent Fe-L line at $\sim$0.8~keV.  In panel f (inside096) we see a prominent feature at $\sim$0.76~keV that is poorly modeled; it dominates the fit residuals and leads to a poor reduced $\chi^{2}$ for this tessellate.  This unmodeled feature is clearly seen in at least 22 tessellates.  Details are given in Section~\ref{sec:badfits}.

The third row of Figure~\ref{fig:spectra} (panels g--i) features regions of bright diffuse emission.  Tessellate inside131 (panel g) is located in the eastern ``hook'' region, while inside126 (panel h) is the westernmost tessellate in the survey.  While inside131 is bright in apparent surface brightness maps as well as intrinsic ones, inside126 only becomes bright in intrinsic maps, after correction for absorption.  This is explained by the spectral fit parameters:  the relatively high absorbing columns for all three NEI components in inside126 result in a large absorption correction.  Again we caution readers to regard such large corrections with skepticism; if our spectral model is incorrect, the intrinsic surface brightness of diffuse emission in this part of the field may in fact be fainter than we have found.

The spectral fit parameter values for all tessellates are compiled in Table~\ref{tbl:diffuse_spectroscopy}.  Many fit parameters are not well-constrained; errors are shown in the table when {\it XSPEC} successfully calculated them.  No errors are given for NEI density-weighted timescales because they were almost never successfully determined by {\it XSPEC}; thus these quantities should be treated as order-of-magnitude estimates only.  To facilitate comparison between tessellates (which all have different areas), we define ``surface'' emission measures, the X-ray emission measure per square parsec.  While abundances of O, Ne, Mg, and S were sometimes supersolar in the fits, their values are not reported in Table~\ref{tbl:diffuse_spectroscopy}, again because the exact numbers were poorly constrained.  Si and Fe abundances were better established due to prominent line features in the spectra, so they are reported, with errors when possible.

\setlength{\tabcolsep}{0.03in}
\renewcommand{\arraystretch}{2.0}
 \begin{deluxetable}{@{}rr*{3}{l}*{3}{l}*{3}{l}*{3}{l}*{2}{l}r*{3}{c}@{}}   

\centering \rotate \tabletypesize{\tiny} \tablewidth{0pt}

\tablecaption{Spectral Fits for Diffuse Tessellates
\label{tbl:diffuse_spectroscopy}}

\tablehead{
\multicolumn{2}{c}{Diffuse Region\tablenotemark{a}} &
\multicolumn{15}{c}{Spectral Fit Parameters for NEI Components\tablenotemark{b}} &
\multicolumn{3}{c}{Inferred Diffuse Emission\tablenotemark{c}} \\ 
\multicolumn{2}{c}{\hrulefill} &
\multicolumn{15}{c}{\hrulefill} &
\multicolumn{3}{c}{\hrulefill}\\
\multicolumn{2}{c}{Tessellate} &
\multicolumn{3}{c}{Absorptions} &     
\multicolumn{3}{c}{Temperatures} &     
\multicolumn{3}{c}{Timescales} &     
\multicolumn{3}{c}{Surface Emission Measures} &     
\multicolumn{2}{c}{Abundances} & 
\colhead{Quality} &
\multicolumn{3}{c}{Surface Brightnesses}  \\
\multicolumn{2}{c}{\hrulefill} &
\multicolumn{3}{c}{\hrulefill} &
\multicolumn{3}{c}{\hrulefill} &
\multicolumn{3}{c}{\hrulefill} &
\multicolumn{3}{c}{\hrulefill} &
\multicolumn{2}{c}{\hrulefill} &
\multicolumn{1}{c}{\hrulefill} &
\multicolumn{3}{c}{\hrulefill}   \\
\colhead{Name} & \colhead{area} & 
\colhead{$\log N_{H1}$} & \colhead{$\log N_{H2}$} & \colhead{$\log N_{H3}$} & 
\colhead{$kT1$}         & \colhead{$kT2$}         & \colhead{$kT3$}         & 
\colhead{$\log {\tau}1$}& \colhead{$\log {\tau}2$}& \colhead{$\log {\tau}3$}&  
\colhead{$\log SEM1$}    & \colhead{$\log SEM2$}    & \colhead{$\log SEM3$}    &  
\colhead{Si} &  \colhead{Fe} &     
\colhead{$\chi^2/DOF$} &
\colhead{$\log$ SB1$_{tc}$} & \colhead{$\log$ SB2$_{tc}$} & \colhead{$\log$ SB3$_{tc}$} 
\\
\colhead{} & \colhead{(pc$^2$)} & 
\multicolumn{3}{c}{(cm$^{-2}$)} & \multicolumn{3}{c}{(keV)} & \multicolumn{3}{c}{(cm$^{-3}$ s)} & \multicolumn{3}{c}{(cm$^{-3}$ pc$^{-2}$)} & 
\multicolumn{2}{c}{(solar units)} &  
&
\multicolumn{3}{c}{(erg s$^{-1}$ pc$^{-2}$)} 
\\
\numberthecolumn & \numberthecolumn &
\numberthecolumn & \numberthecolumn & \numberthecolumn & 
\numberthecolumn & \numberthecolumn & \numberthecolumn & 
\numberthecolumn & \numberthecolumn & \numberthecolumn & 
\numberthecolumn & \numberthecolumn & \numberthecolumn & \numberthecolumn & \numberthecolumn &
\numberthecolumn &
\numberthecolumn & \numberthecolumn & \numberthecolumn  
\setcounter{column_number}{1}
}



\startdata
inside001  &     6.6 &$20.2\phd$                  & $21.7\phd$                  & $22.0\phd$                      & $0.65\phd_{-0.04}^{\cdots}$ & $0.26\phd$                  & $0.83\phd_{-0.2}^{\cdots}$       &$12.4\phd$                       & $8.00\phd$                 & $13.7\phd$                 & $ 54.4\phd_{-0.08}^{+0.19}$ & $ 56.1\phd_{\cdots}^{+0.6}$ & $ 54.1\phd_{\cdots}^{+0.2}$    & $4.0\phd_{-1.1}^{\cdots}$ &                           &  102/ 75 &  31.60 &   31.83 &   31.35  \\
inside002  &     2.6 &$21.2\phd$                  & $21.7\phd$                  & $21.4\phd$                      & $0.55\phd$                  & $0.48\phd$                  & $1.1\phd$                        &$13.0\phd$                       & $8.00\phd$                 & $10.9\phd$                 & $ 54.9\phd_{\cdots}^{+0.2}$ & $ 55.2\phd_{-0.9}^{+1.4}$   & $ 54.2\phd_{\cdots}^{+0.9}$    & $1.5\phd_{\cdots}^{+0.8}$ &                           &   64/ 69 &  32.13 &   31.46 &   31.90  \\
inside003  &    21.2 &$21.8\phd_{-0.2}^{\cdots}$  & $21.9\phd$                  & $20.5\phd_{\cdots}^{+0.9}$      & $0.51\phd_{\cdots}^{+0.4}$  & $0.13\phd_{-0.01}^{+0.03}$  & $0.83\phd_{-0.1}^{+0.9}$         &${\phn}9.98\phd$                 & $8.00\phd$                 & $11.5\phd$                 & $ 54.5\phd_{\cdots}^{+0.7}$ & $ 57.6\phd$                 & $ 53.7\phd_{-0.3}^{+0.2}$      &                           &                           &  149/138 &  32.09 &   32.52 &   31.19  \\
inside004  &     2.2 &$20.2\phd$                  & $21.2\phd$                  & $20.2\phd$                      & $0.43\phd_{-0.3}^{+0.5}$    & $0.29\phd_{\cdots}^{+0.2}$  & $0.59\phd_{\cdots}^{+0.7}$       &$12.2\phd$                       & $8.00\phd$                 & $13.7\phd$                 & $ 54.3\phd_{\cdots}^{+0.3}$ & $ 56.0\phd_{\cdots}^{+0.7}$ & $ 54.5\phd_{\cdots}^{+0.2}$    & $4.1\phd_{-1.2}^{\cdots}$ & $2.7\phd_{-0.7}^{+1.1}$   &  100/ 74 &  31.77 &   31.86 &   32.04  \\
inside005  &     2.7 &$21.3\phd_{\cdots}^{+0.4}$  & $20.2\phd$                  & $20.2\phd$                      & $0.27\phd_{-0.05}^{+0.06}$  & $0.44\phd_{-0.1}^{\cdots}$  & $0.61\phd_{-0.03}^{+0.03}$       &$12.7\phd$                       & $8.32\phd$                 & $13.7\phd$                 & $ 55.0\phd_{\cdots}^{+0.3}$ & $ 55.3\phd_{\cdots}^{+0.5}$ & $ 54.6\phd_{-0.08}^{+0.18}$    & $2.9\phd_{-0.9}^{+0.9}$   & $2.6\phd_{-0.7}^{+1.3}$   &   91/ 70 &  32.12 &   31.54 &   32.06  \\
\enddata

\tablecomments{
Table~\ref{tbl:diffuse_spectroscopy} is available in its entirety in the electronic edition of the Journal.  The first few lines are shown here for guidance regarding its form and content. 
}
               
\tablenotetext{a}{
Properties of the {\it WVT Binning} tessellates:
\\Col.\ (1): Diffuse region label.
\\Col.\ (2): Geometric area of the tessellate in square parsecs, irrespective of point source masking.
}

\tablenotetext{b}{
All fits used the {\em source} model ``TBabs*vpshock + TBabs*vpshock + TBabs*vpshock + TBabs*apec + TBabs*apec + TBabs*apec'' in {\it XSPEC}.
\\Col.\ (3)-(5): The best-fit value for the extinction column density ({\it TBabs} components).
\\Col.\ (6)-(8): The best-fit value for the plasma temperature ({\it vpshock} components).
\\Col.\ (9)-(11): The electron-density-weighted ionization timescale for the plasma ({\it vpshock} components).
\\Col.\ (12)-(14): The ``surface emission measure'' (X-ray emission measure per unit area) for the plasma ({\it vpshock} components), assuming a distance of 2.3~kpc.
\vspace{\baselineskip}\noindent
\\Abundances in the plasma models were tied together.  Abundances of He, C, N, Al, Ar, and Ca were frozen at 1.0 $Z_\odot$.  Abundances of Si and Fe are reported in Cols.\ (15)-(16); when omitted the abundance was frozen at 1.0 $Z_\odot$.  The Ni abundance was tied to Fe.  Abundances of O, Ne, Mg, and S  have values greater than solar in some fits but those values are not well-constrained so they are not reported here.  
\\
\vspace{\baselineskip}\noindent
Uncertainties represent 90\% confidence intervals.
More significant digits are used for uncertainties $<$0.1 in order to avoid large rounding errors; for consistency, the same number of significant digits is used for both lower and upper uncertainties.
Uncertainties are missing when {\it XSPEC} was unable to compute them or when their values were so large that the parameter is effectively unconstrained.  
}

\tablenotetext{c}{
All inferred intrinsic diffuse plasma properties assume a distance of 2.3~kpc.
\\Cols.\ (18)-(20): Absorption-corrected surface brightness of the diffuse emission in Carina ({\it vpshock} components) in the total band (0.5--7~keV).
}

\end{deluxetable}  

\subsection{Spectral Fit Parameter Maps \label{sec:maps}}
  
We can visualize these spectral fits by making parameter maps showing various spectral fit parameter values for each tessellate.  These are shown in several figures in this section; in all examples, black indicates high values and white shows low values, with images scaled the same to facilitate comparison.   A histogram of the mapped parameter value is included in the lower right corner of each map, with the number of tessellates on the ordinate axis and the parameter value on the abscissa.  All coordinates are celestial J2000.  Most parameter maps show the logarithm of the parameter values; exceptions are the plasma temperature maps (Figure~\ref{fig:ktmaps}), elemental abundance maps (Figure~\ref{fig:abundmaps}), and the goodness-of-fit map (Figure~\ref{fig:other}c).

\subsubsection{NEI Plasma Components \label{sec:diffusecomps}}  

\begin{figure}[htbp] 
\begin{center}  
\includegraphics[width=0.32\textwidth]{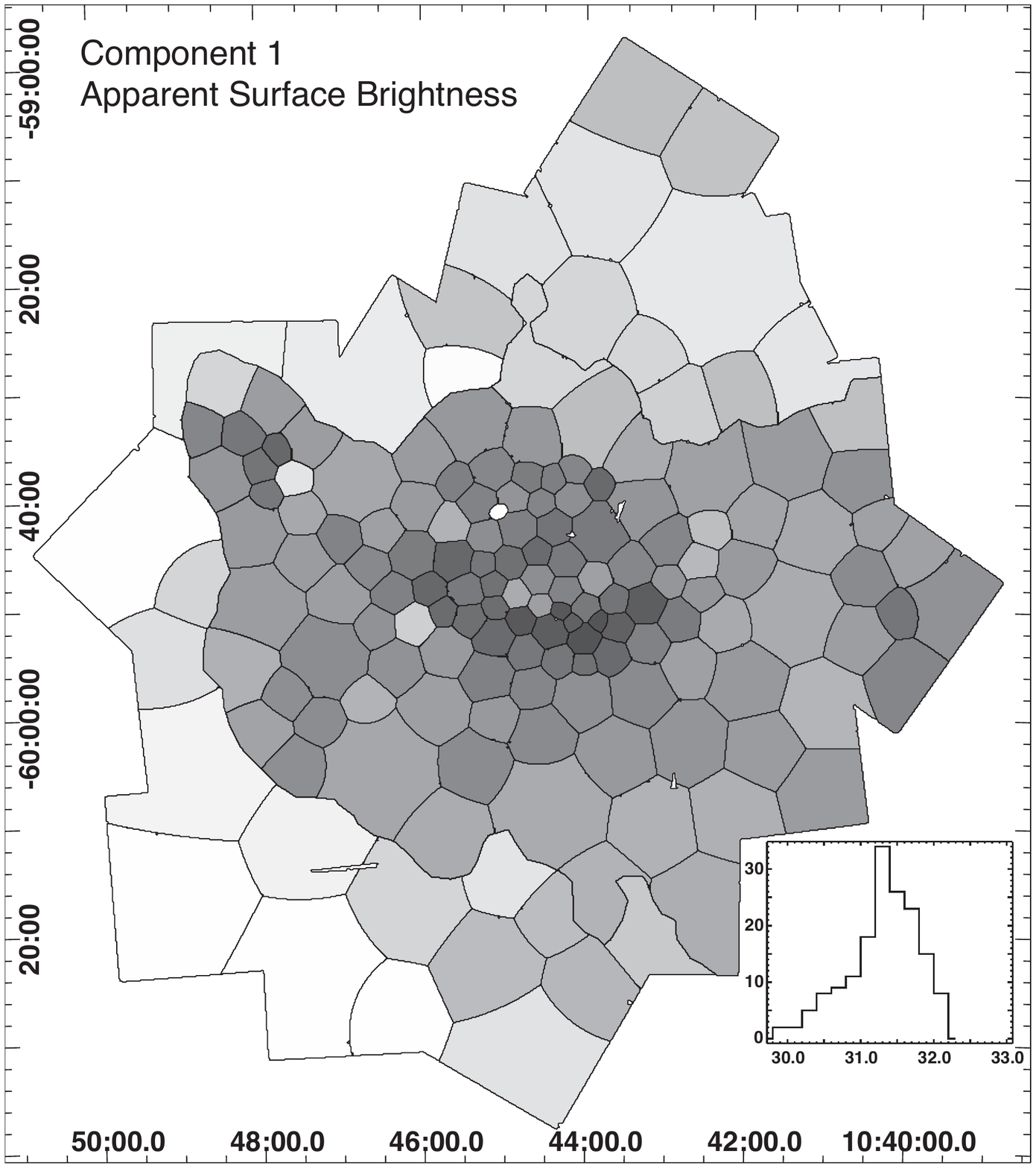}
\includegraphics[width=0.32\textwidth]{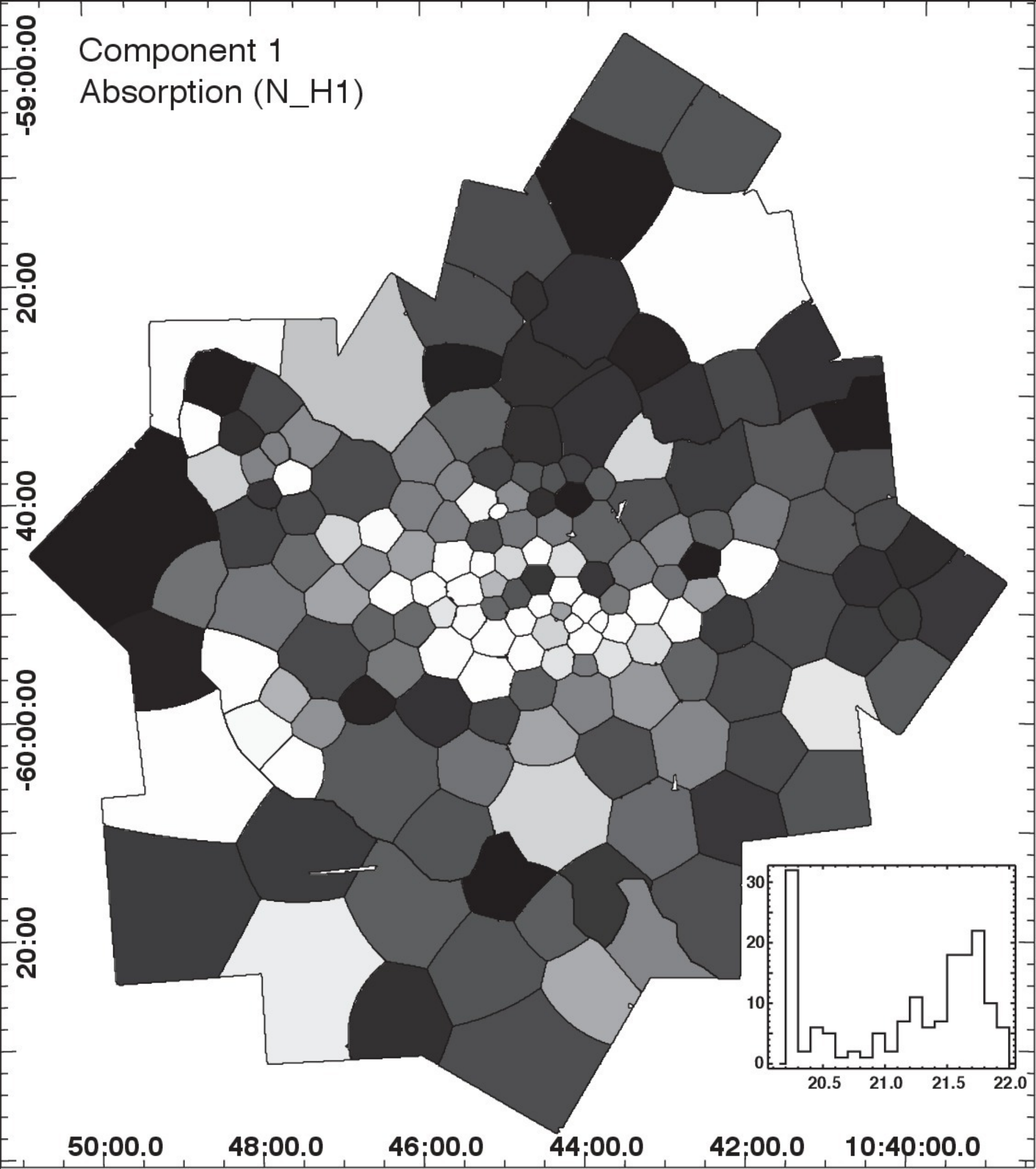}
\includegraphics[width=0.32\textwidth]{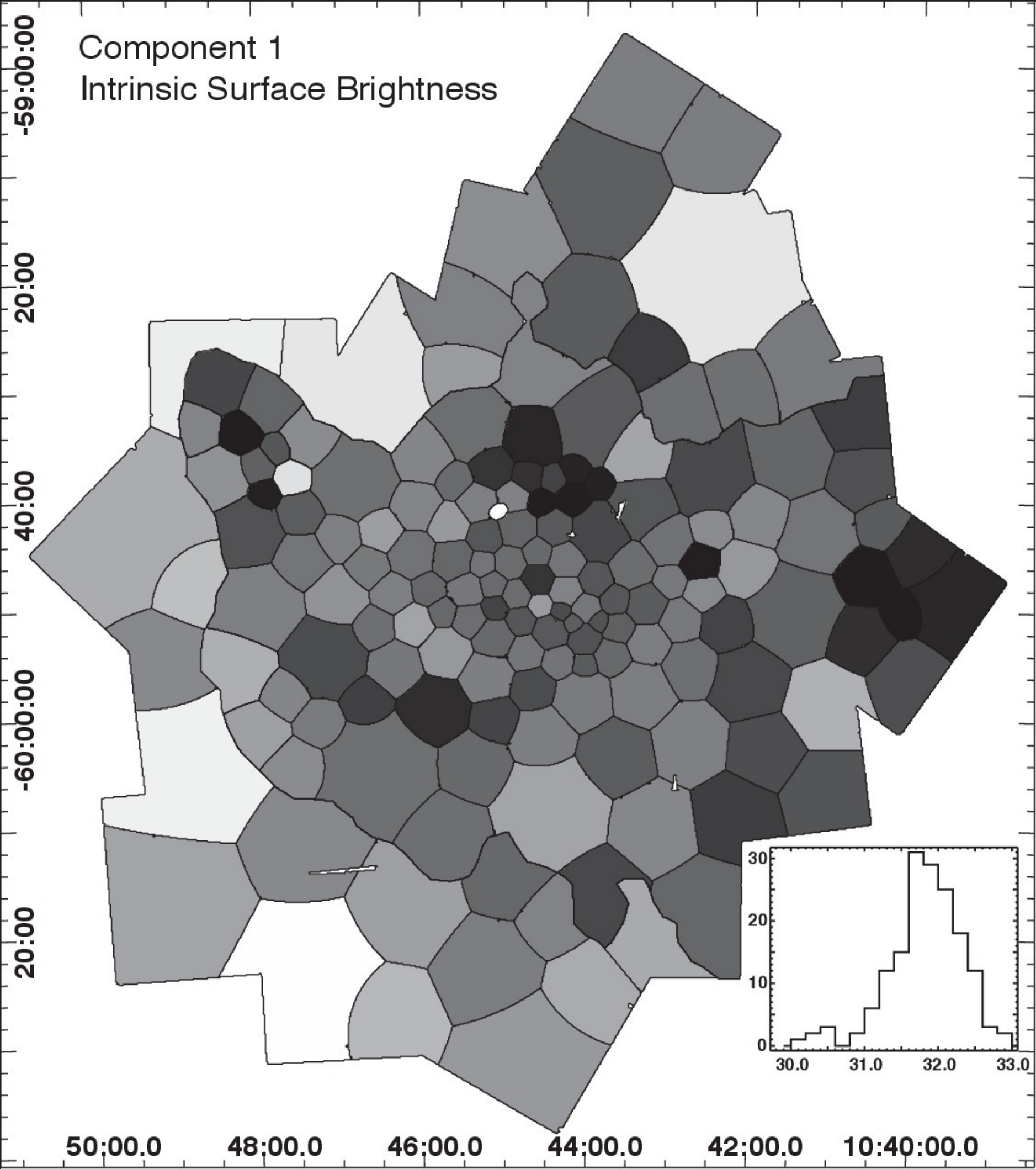}
\includegraphics[width=0.32\textwidth]{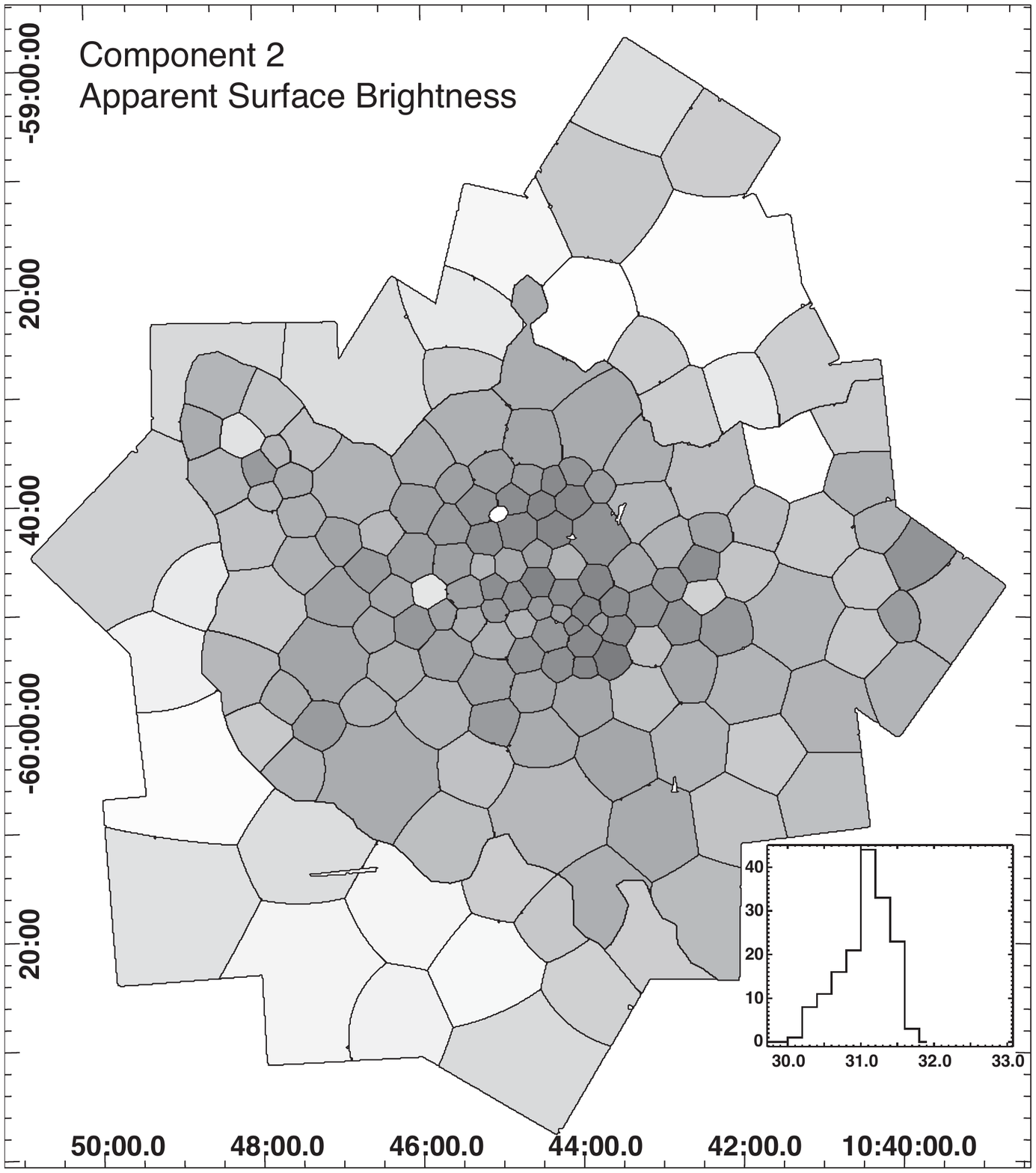}
\includegraphics[width=0.32\textwidth]{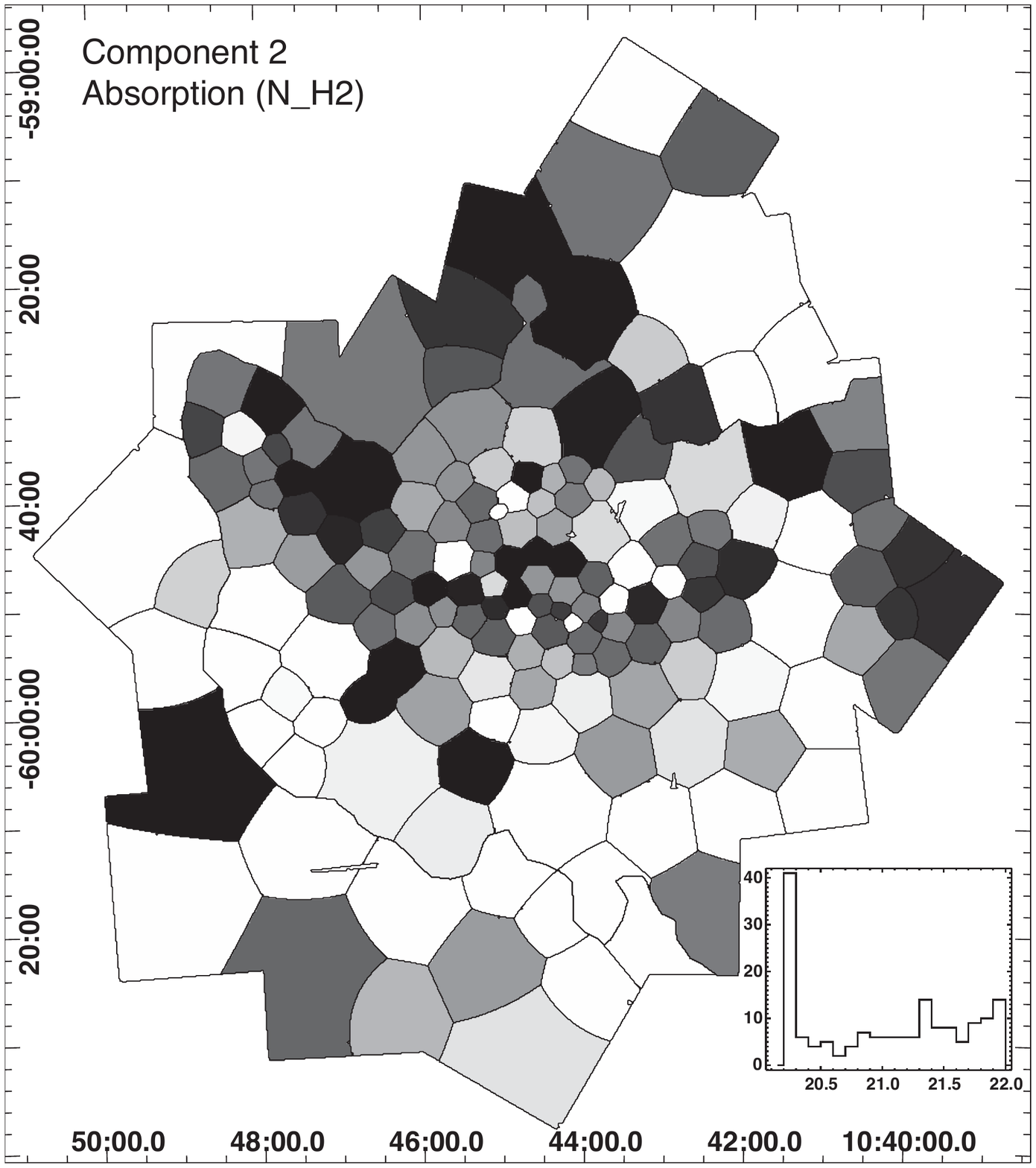}
\includegraphics[width=0.32\textwidth]{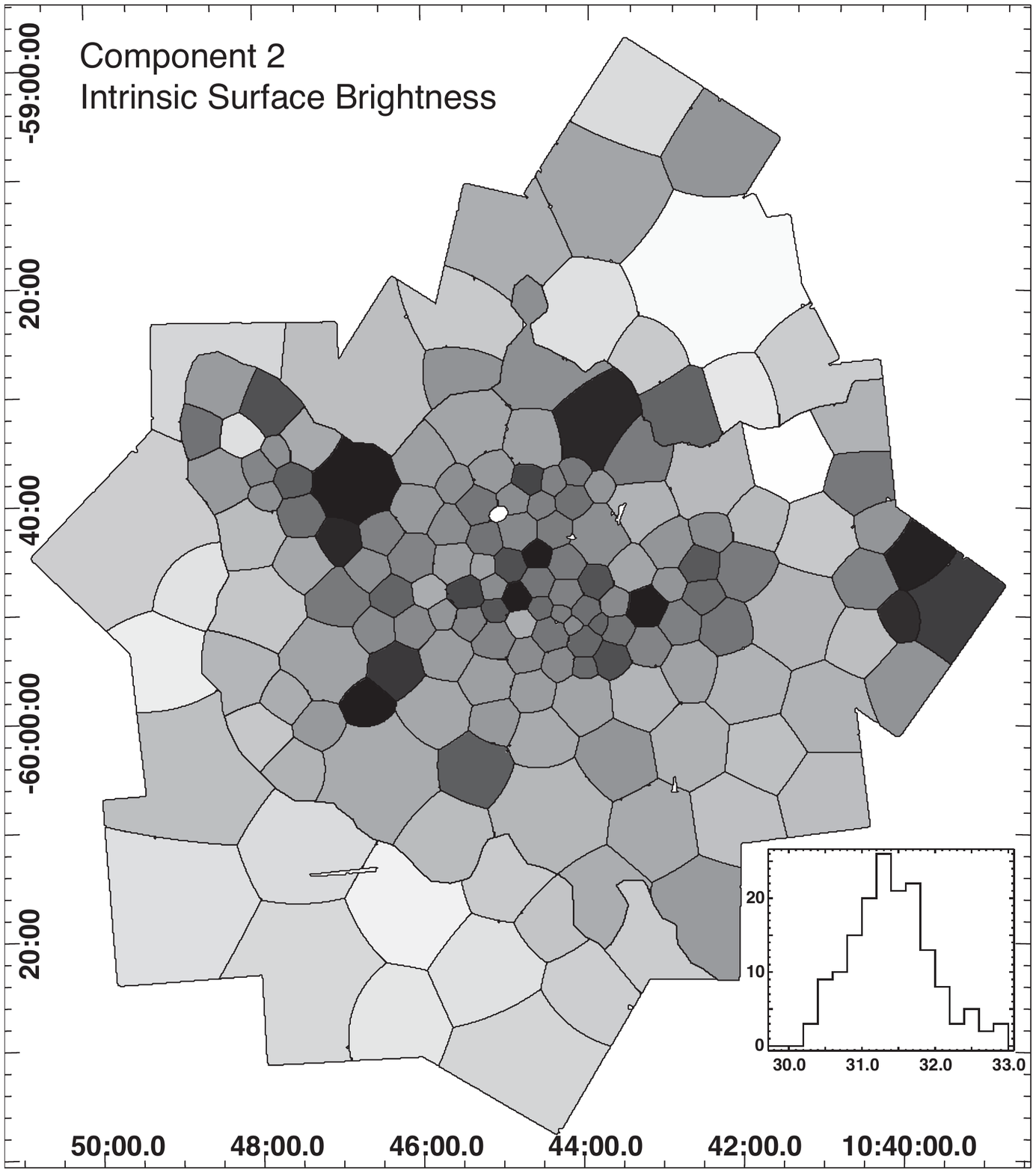}
\includegraphics[width=0.32\textwidth]{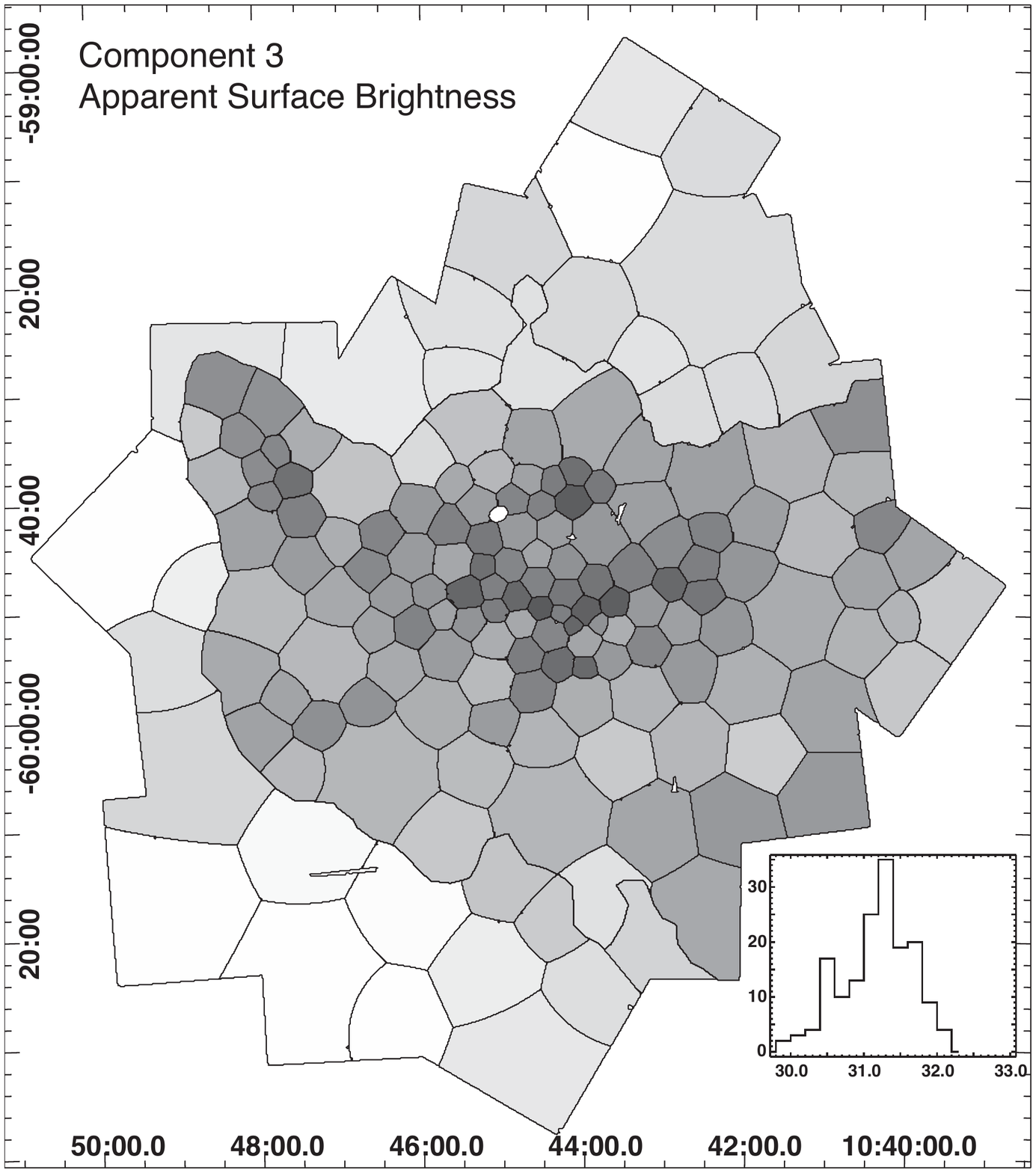}
\includegraphics[width=0.32\textwidth]{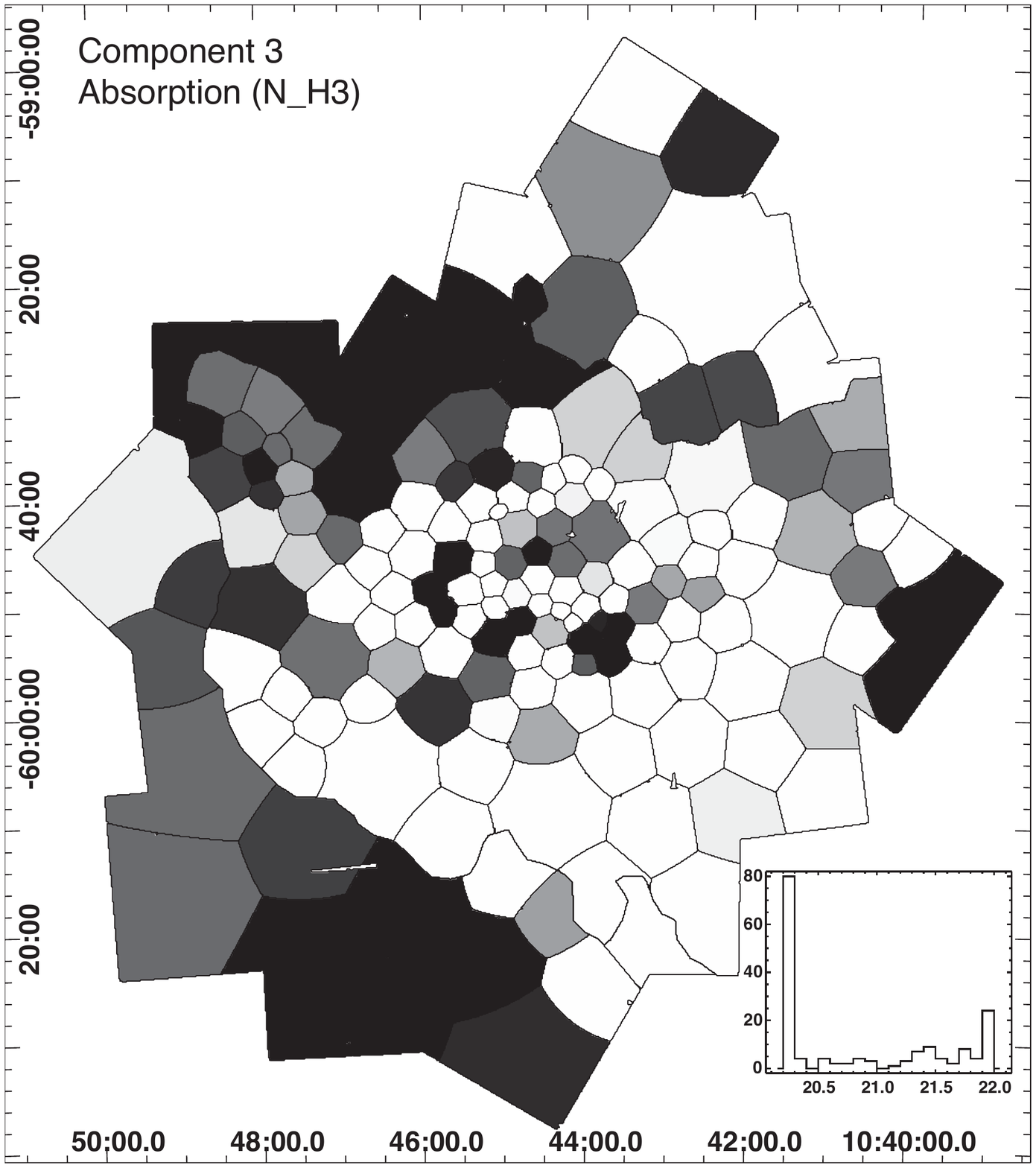}
\includegraphics[width=0.32\textwidth]{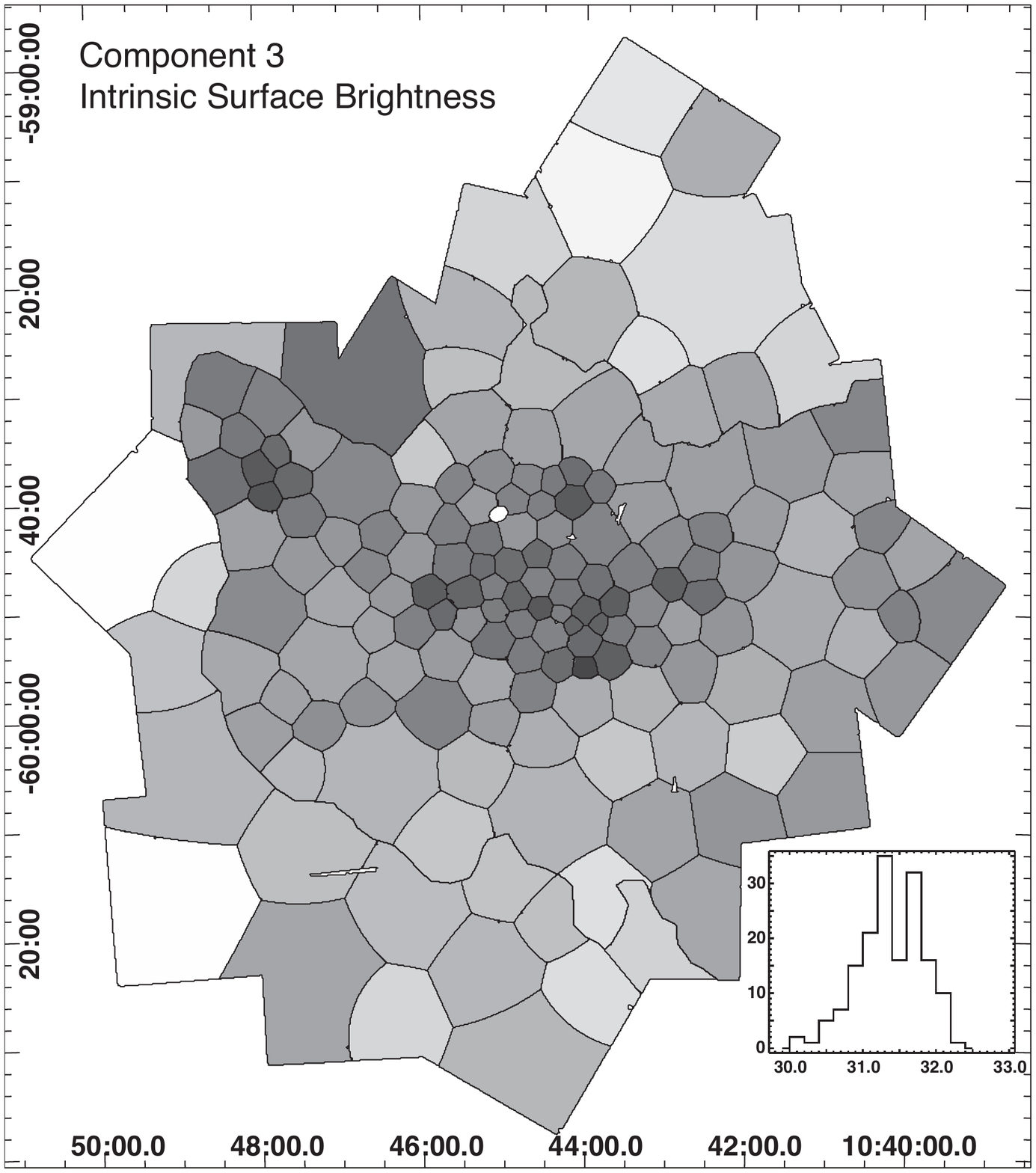}
\caption{Diffuse emission broken down by NEI component.
(a)  The first NEI component's apparent surface brightness.
(b)  The first NEI component's absorbing column, $N_{H1}$.
(c)  The first NEI component's intrinsic surface brightness.
(d)--(f)  Same as (a)--(c) but for the second NEI component.
(g)--(i)  Same as (a)--(c) but for the third NEI component.
All apparent and intrinsic surface brightness maps have the same scaling, with $30.2 < \log SB < 32.6$, where $SB$ stands for surface brightness in units of erg~s$^{-1}$~pc$^{-2}$, assuming a distance to Carina of 2.3~kpc \citep{SmithN06a}.  The absorption maps are all scaled with $20.2 < \log N_{H}$~(cm$^{-2}$) $< 21.9$.  Here and in all component maps, darker tessellates represent higher values.
} 
\label{fig:neimaps}
\end{center}
\end{figure}

\begin{figure}[htb] 
\begin{center}  
\includegraphics[width=0.32\textwidth]{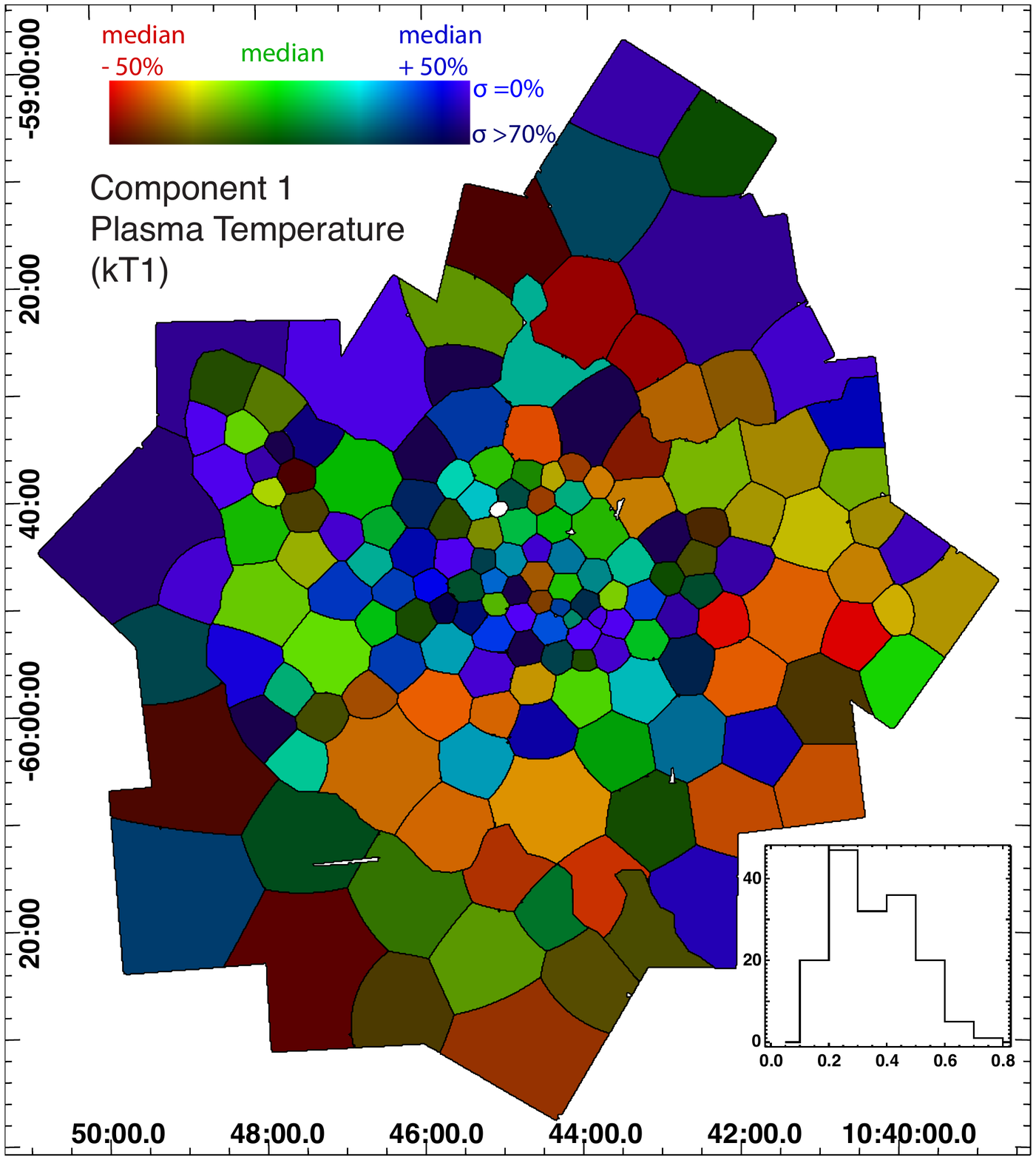}
\includegraphics[width=0.32\textwidth]{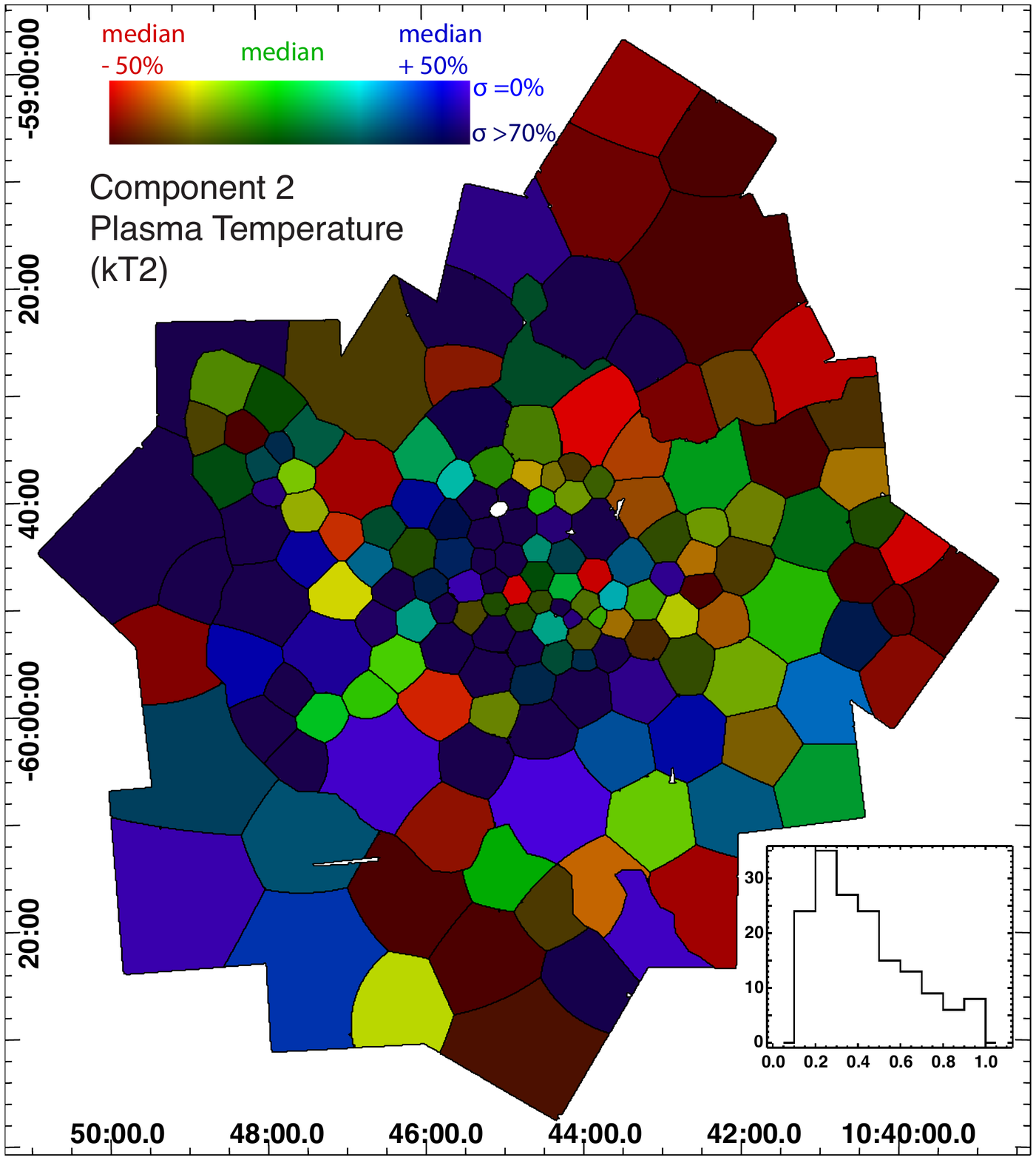}
\includegraphics[width=0.32\textwidth]{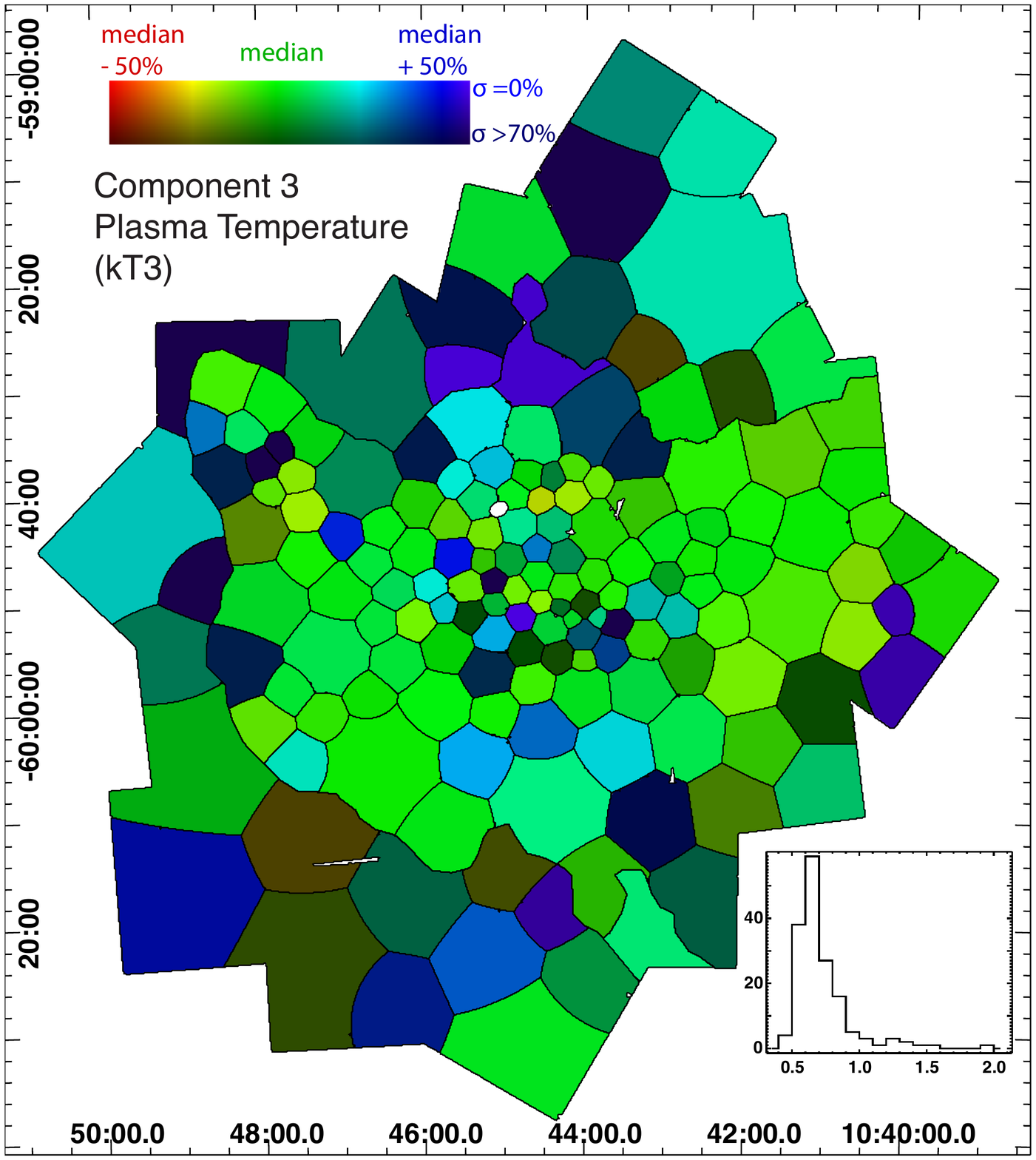}
\caption{Value/confidence maps of diffuse emission plasma temperatures.
(a)  The first NEI component, kT1 (median = 0.33~keV).
(b)  The second NEI component, kT2 (median = 0.37~keV).
(c)  The third NEI component, kT3 (median = 0.65~keV).
} 
\label{fig:ktmaps}
\end{center}
\end{figure}

First we concentrate on the NEI model components (kT1 -- kT3) because we assume that they trace Carina's diffuse X-ray emission.  Figure~\ref{fig:neimaps} shows the apparent surface brightness, the absorption, and the intrinsic surface brightness for each NEI component separately.  Component 1 (Figure~\ref{fig:neimaps} top row) is a soft plasma (median kT1 = 0.33~keV; see Figure~\ref{fig:ktmaps}a below); it is the strongest contributor to the diffuse X-ray surface brightness in most tessellates.  Its absorbing column is relatively high across the whole field, except in a broad arc across the field center; this arc of low Component 1 absorption contributes substantially to the apparent surface brightness of Carina's diffuse emission.  Removing that obscuring screen, we see that Component 1 is actually brightest in the region between Tr14 and Tr16 and at the western edge of the survey.  It has substantial emission measure across the whole field, except in a few outside tessellates.
  
Component 2 is fainter than Component 1 both in apparent and intrinsic surface brightness, for most tessellates.  It is also quite soft in many tessellates (median kT2 = 0.37~keV) but can range to higher temperatures (see Figure~\ref{fig:ktmaps}b).  Its absorption is typically very low (pegged at its minimum allowed value) across the southern half of the survey but is higher across the central arc, to the northeast, and at the western edge of the field.  Its intrinsic surface brightness is generally higher across the central arc and exceeds Component 1 in a few tessellates.

Component 3 is a harder plasma (median kT3 = 0.65~keV; see Figure~\ref{fig:ktmaps}c).  It exhibits about the same apparent surface brightness as the other NEI components.  Its absorbing column is roughly similar to that of Component 2, with minimal absorption across much of the southern half of the field, but it shows high columns towards the South Pillars, on the southeast side of the field.  Absorption across the central arc is patchy for this component, with some very low and some very high values; again the northeast region and the western edge of the survey show high absorption for this component.  Its intrinsic emission is relatively flat across the field, with some enhancement across the central arc.  Its emission measure dominates the other NEI components in a few tessellates, but generally it is fainter.

Not surprisingly, a general impression from Figure~\ref{fig:neimaps} is that absorption strongly shapes the apparent surface brightness of Carina's diffuse emission.  The differences between the absorption maps for each NEI component is more surprising; these three emission components do not appear to be co-located behind the same obscuring screen.  The central arc is minimally obscured in Component 1, quite obscured in Component 2, and patchy in Component 3.  The eastern ``hook'' region is quite obscured for all three components; this explains why it looks bluer in Figure~\ref{fig:patsmooth}.  

Figure~\ref{fig:ktmaps} provides more detail on the plasma temperature distributions for the three NEI components.  Rather than showing simple value maps (as we did for the surface brightnesses and absorbing columns), here we have computed value/confidence maps, which combine fit parameter values with the errors on those values by mapping the values to a range of colors (hues) and the uncertainties to a range of brightnesses \citep{Broos10}.  The convention is that parameter values near the median value are close to green, those larger than the median value are bluish, and those smaller than the median value are reddish.  A highly certain low parameter value would be bright red; a highly uncertain low value would be maroon.  Under this color model, adding uncertainty can be thought of as adding black paint to a bright color.  
In the legends, the shorthand $\sigma$ stands for the average of the upper and lower legs of the 90\% confidence interval expressed as a percentage of the best-fit parameter value.

The plasma temperature for Component 1, shown in Figure~\ref{fig:ktmaps}a, is relatively high across the central arc and eastern hook regions, and relatively low to the south, west, and north.  Outside tessellates are typically more uncertain than inside tessellates because they have lower surface brightness and are larger, thus they are potentially averaging over a wider range of physical conditions in the plasma.  

Component 2 was allowed a wider range of plasma temperature in the spectral fits (Figure~\ref{fig:ktmaps}b); its distribution covers its full range but is peaked toward softer values.  Its value/confidence map shows little temperature correlation between tessellates, except that the northwestern edge of the survey appears softer and the southeastern edge harder (although these kT2 values are often quite uncertain).  

Component 3 (Figure~\ref{fig:ktmaps}c) was allowed yet a wider range of plasma temperatures but the fit values are strongly peaked around kT3 $\sim$ 0.6~keV.  Its value/confidence map is predominantly green because the median value of kT3 is close to its lower limit (0.4~keV).  The values of this component tend to have low uncertainties, at least for the inside tessellates.

\begin{figure}[htb] 
\begin{center}  
\includegraphics[width=0.32\textwidth]{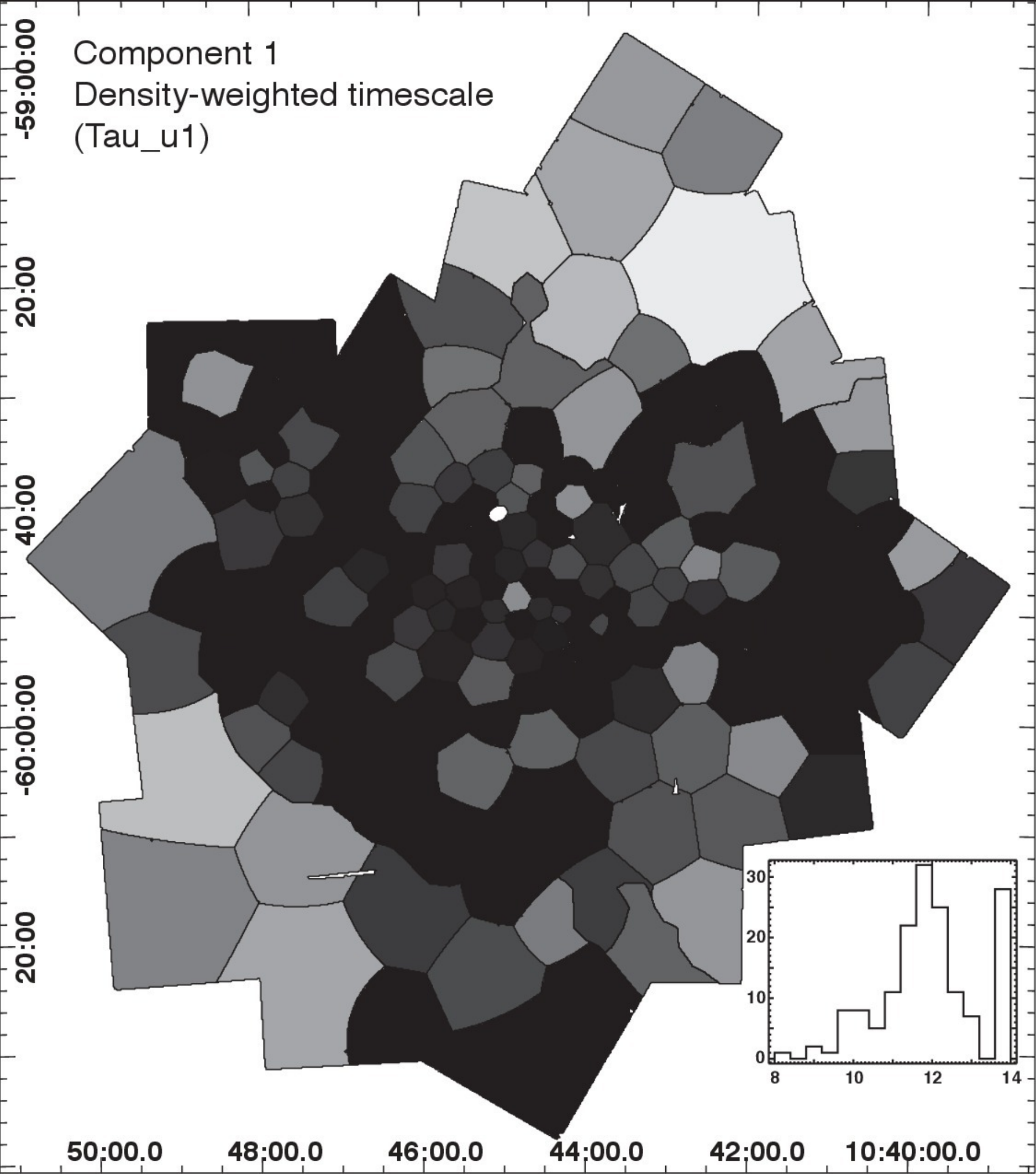}
\includegraphics[width=0.32\textwidth]{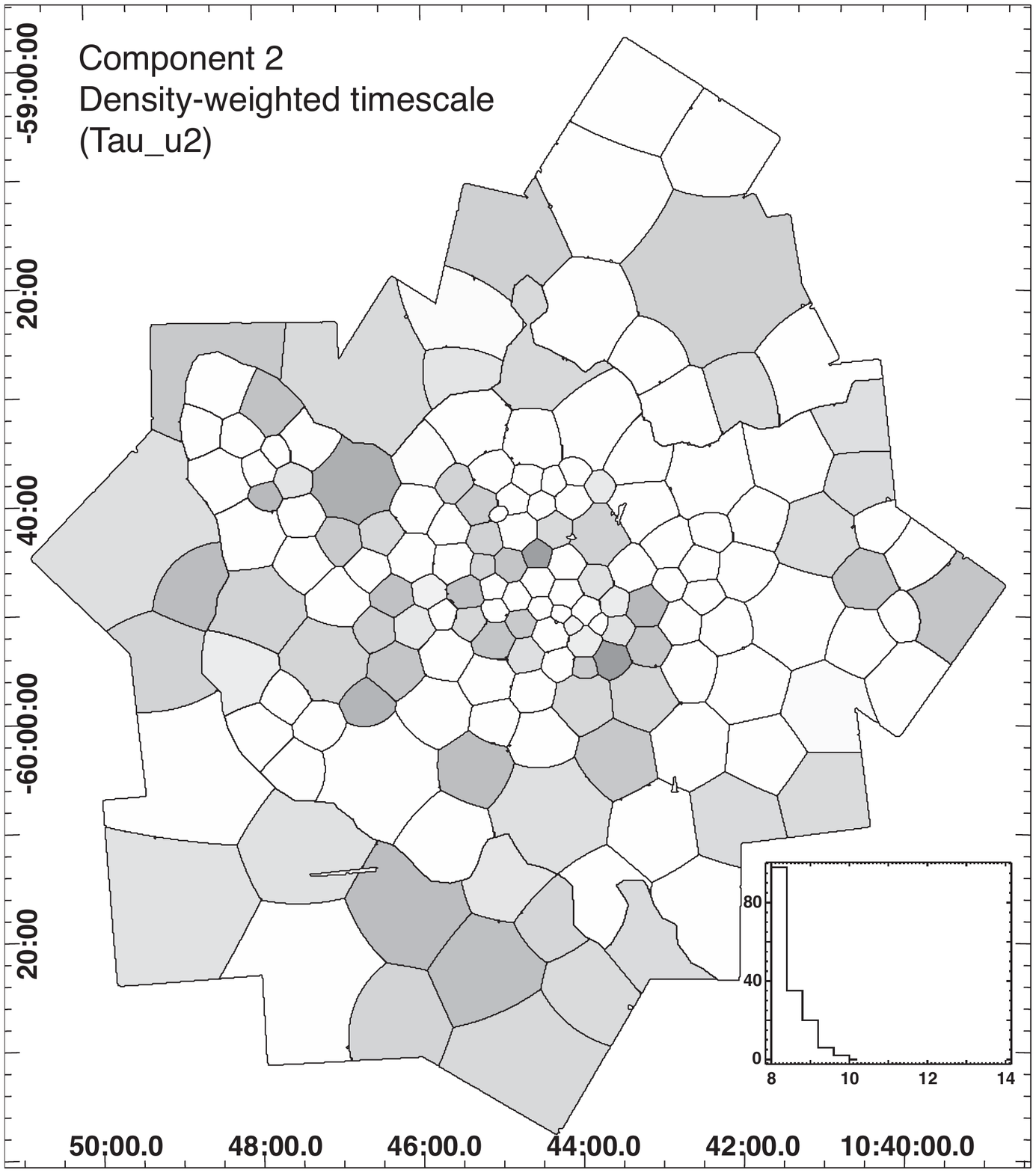}
\includegraphics[width=0.32\textwidth]{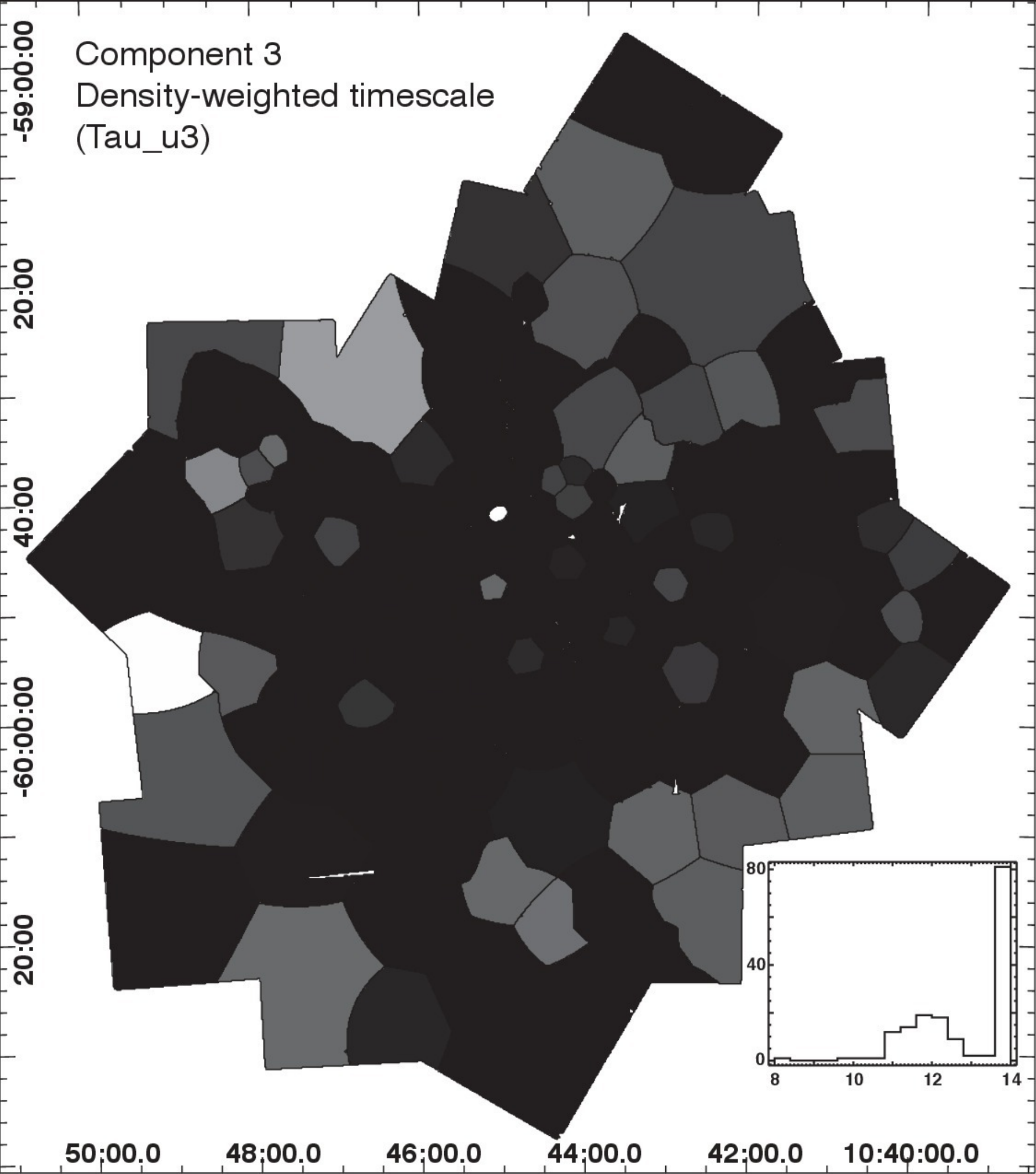}
\caption{Density-weighted timescales for the diffuse emission plasma components.
(a)  The first NEI component.
(b)  The second NEI component.
(c)  The third NEI component.
All maps are scaled with $8 < \log \tau_{u} < 12$ in units of cm$^{-3}$s.
} 
\label{fig:taumaps}
\end{center}
\end{figure}

An important characteristic of the {\it vpshock} model (and all NEI plasmas) is the time $t$ for the plasma to return to equilibrium \citep{SmithR10}.  This time is always scaled by the electron density $n_{e}$; in the {\it vpshock} model \citep{Borkowski01} it is represented by the quantity $\tau_{u} = n_{e}t$ where the $u$ refers to the upper limit of the ionization timescale. Large values of $\tau_{u}$ imply long timescales and high densities and suggest that the plasma is in CIE.  Value maps of this ionization timescale for our three NEI components are shown in Figure~\ref{fig:taumaps}.  

The scaling in Figure~\ref{fig:taumaps} is deliberate; \citet{SmithR10} show that most ions have reached collisional ionization equilibrium (CIE) by $\log \tau_{u} = 12$ for the temperatures exhibited by our NEI plasma components, so tessellates that appear black in Figure~\ref{fig:taumaps} are unlikely to have suffered recent shocks, for the plasma component in question.  Thus it appears that both Component 1 and Component 3 are largely in equilibrium, while Component 2 is strongly NEI, generated by a very low-density plasma in a recent shock or outburst.  We will consider the ramifications of these spectral fit results later.

In our exploration of spectral models (Section~\ref{sec:model}), we found that allowing supersolar abundances for some elements significantly improved the goodness-of-fit, although the actual abundance values were often not well-constrained.  Visual examination of the spectra (Section~\ref{sec:spectra}) showed that the Si-K$\alpha$ line at 1.86~keV was prominent for some tessellates, much less so for others.  We also found that the Fe-L line complex at $\sim$0.8~keV was featured in some spectra.  Thus our automated fits were performed with Si and Fe abundances thawed but restricted to lie in the range 1.0--5.0 times solar values.  All other abundances were frozen at solar values; the abundances of all three {\it vpshock} plasma components were linked.  After automated fitting was completed, each spectrum was re-fit by hand, thawing the abundances of O, Ne, Mg, and/or S as needed to improve the fits; again abundances were linked among the three {\it vpshock} components and ranges were limited to 1.0--5.0 times solar values.  

\begin{figure}[htb] 
\begin{center}  
\includegraphics[width=0.32\textwidth]{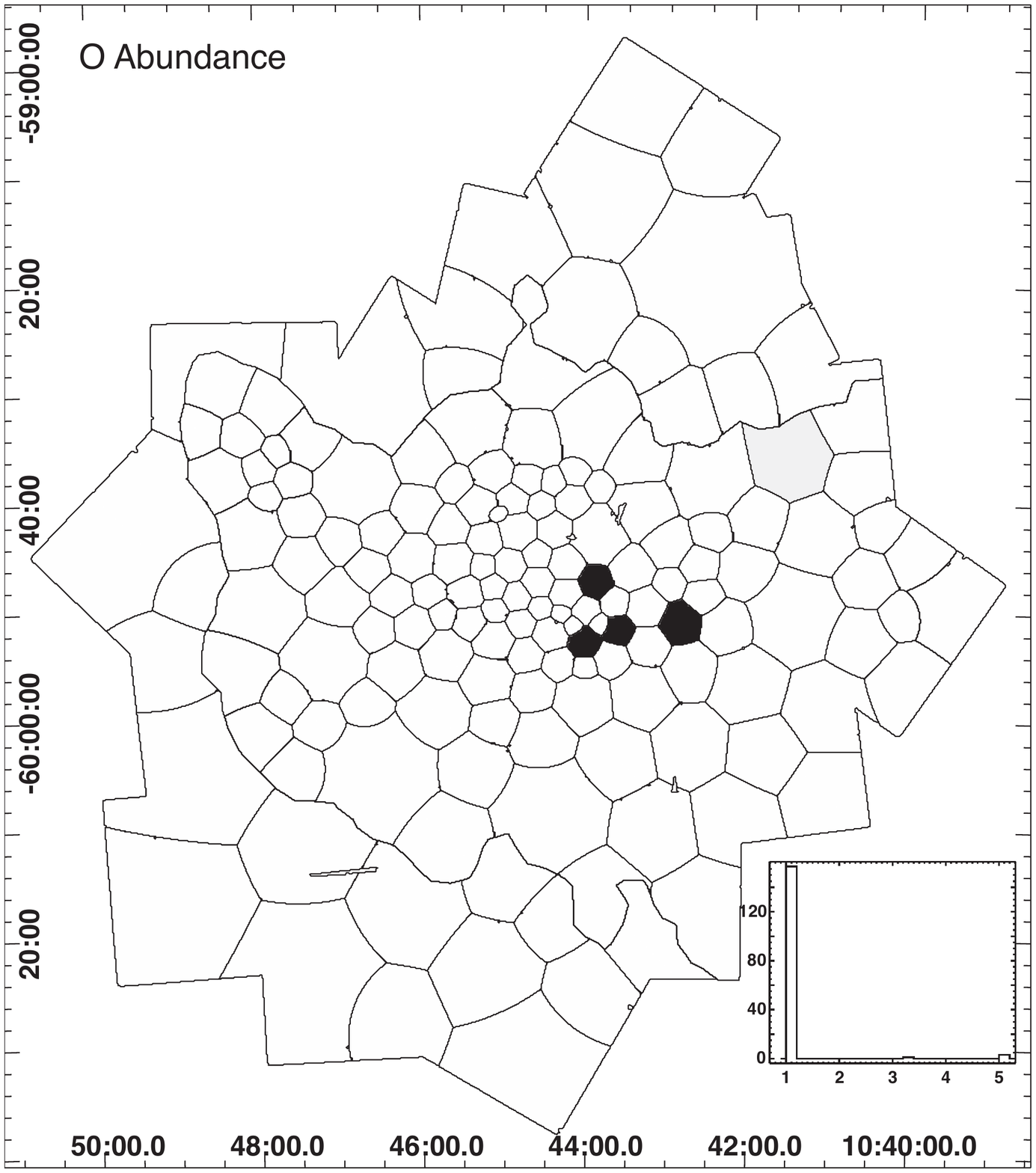}
\includegraphics[width=0.32\textwidth]{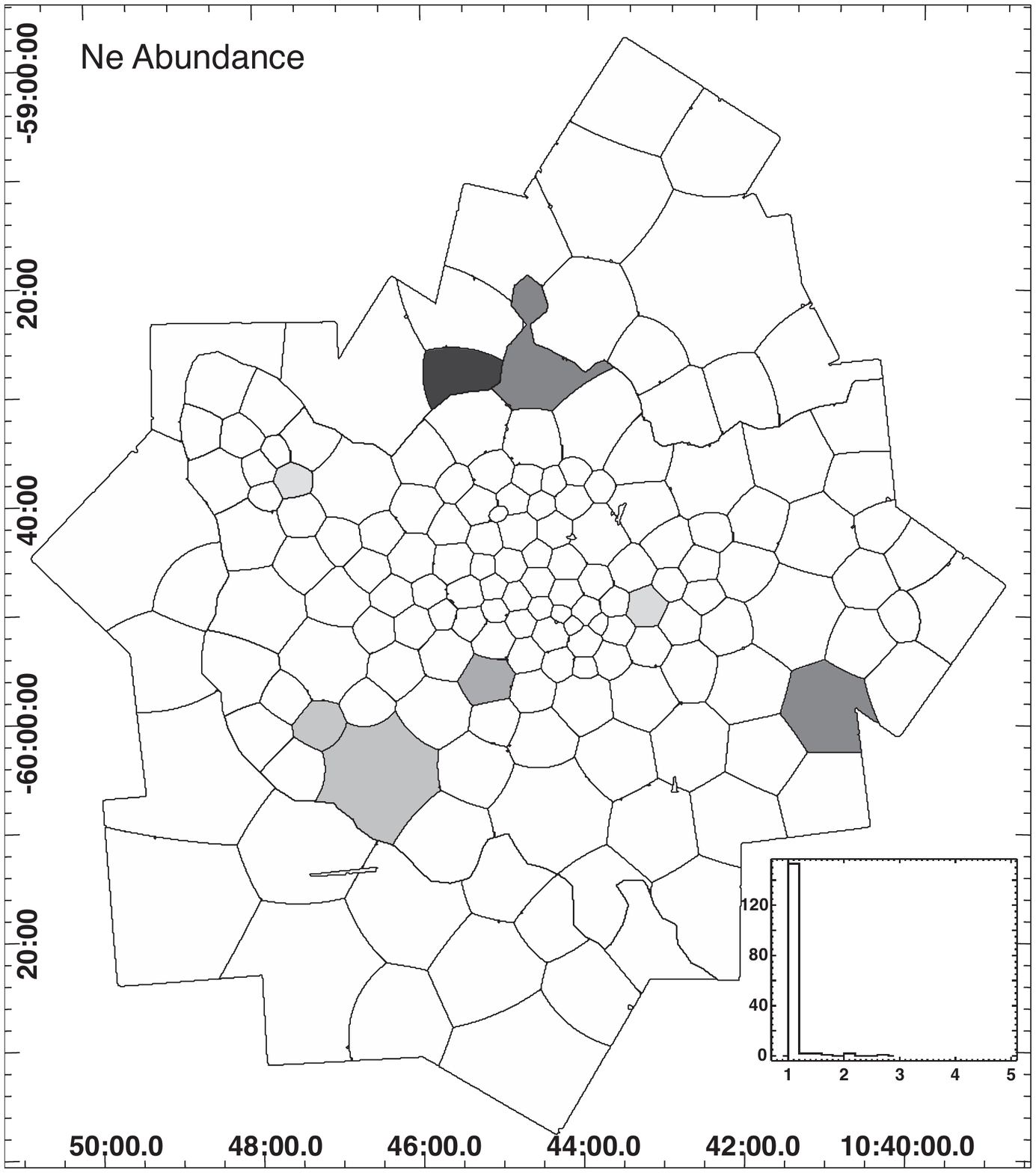}
\includegraphics[width=0.32\textwidth]{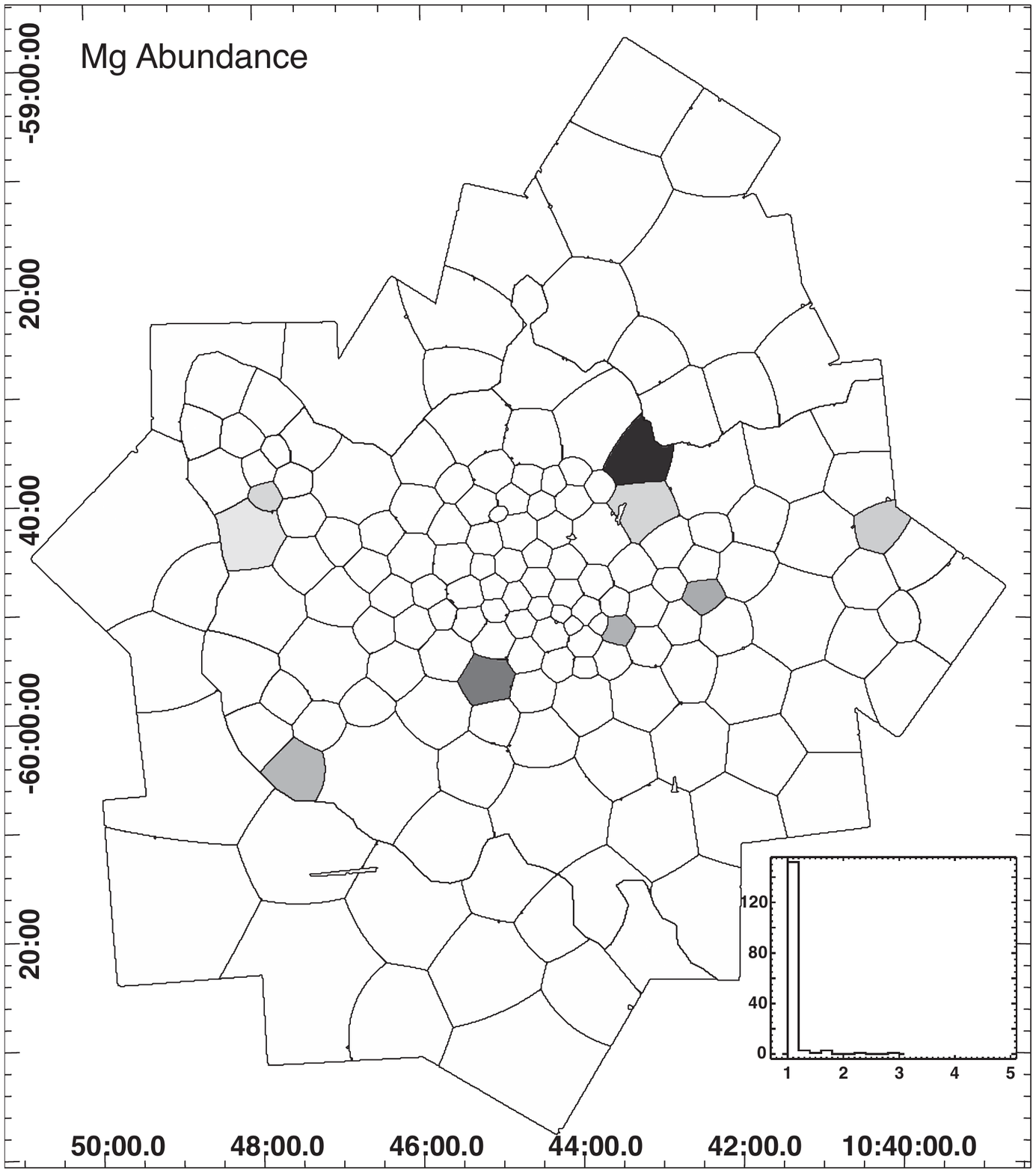}
\includegraphics[width=0.32\textwidth]{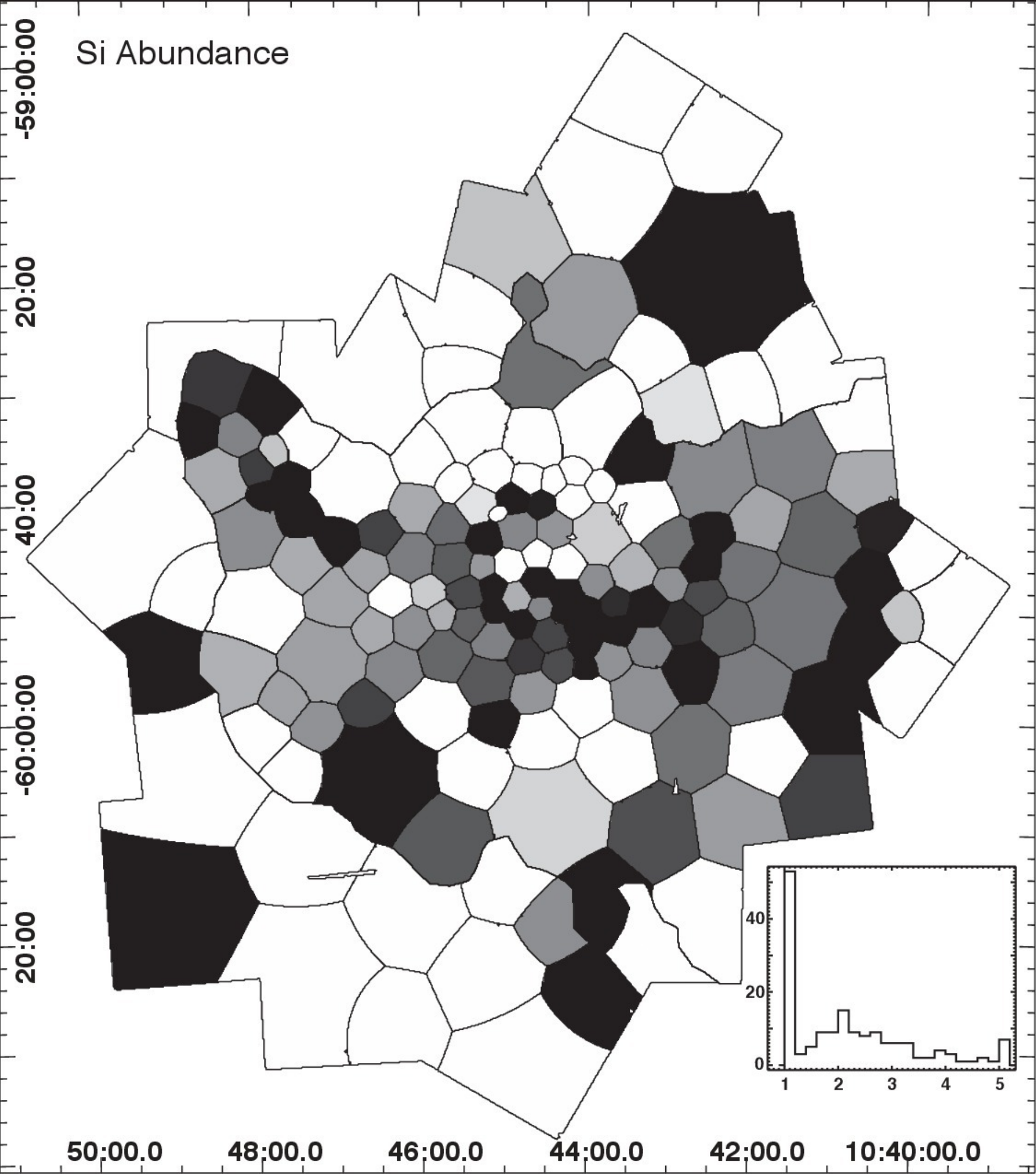}
\includegraphics[width=0.32\textwidth]{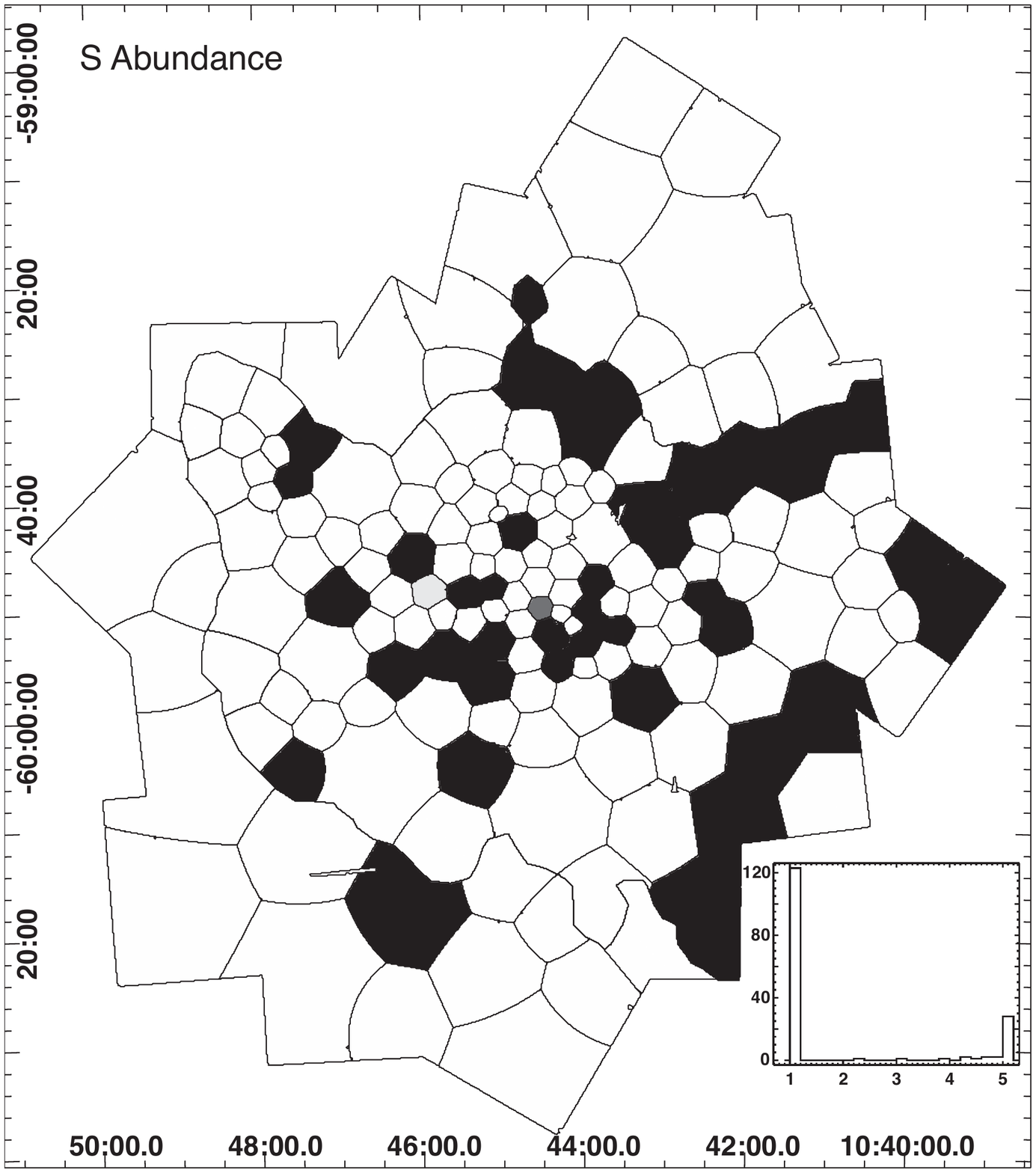}
\includegraphics[width=0.32\textwidth]{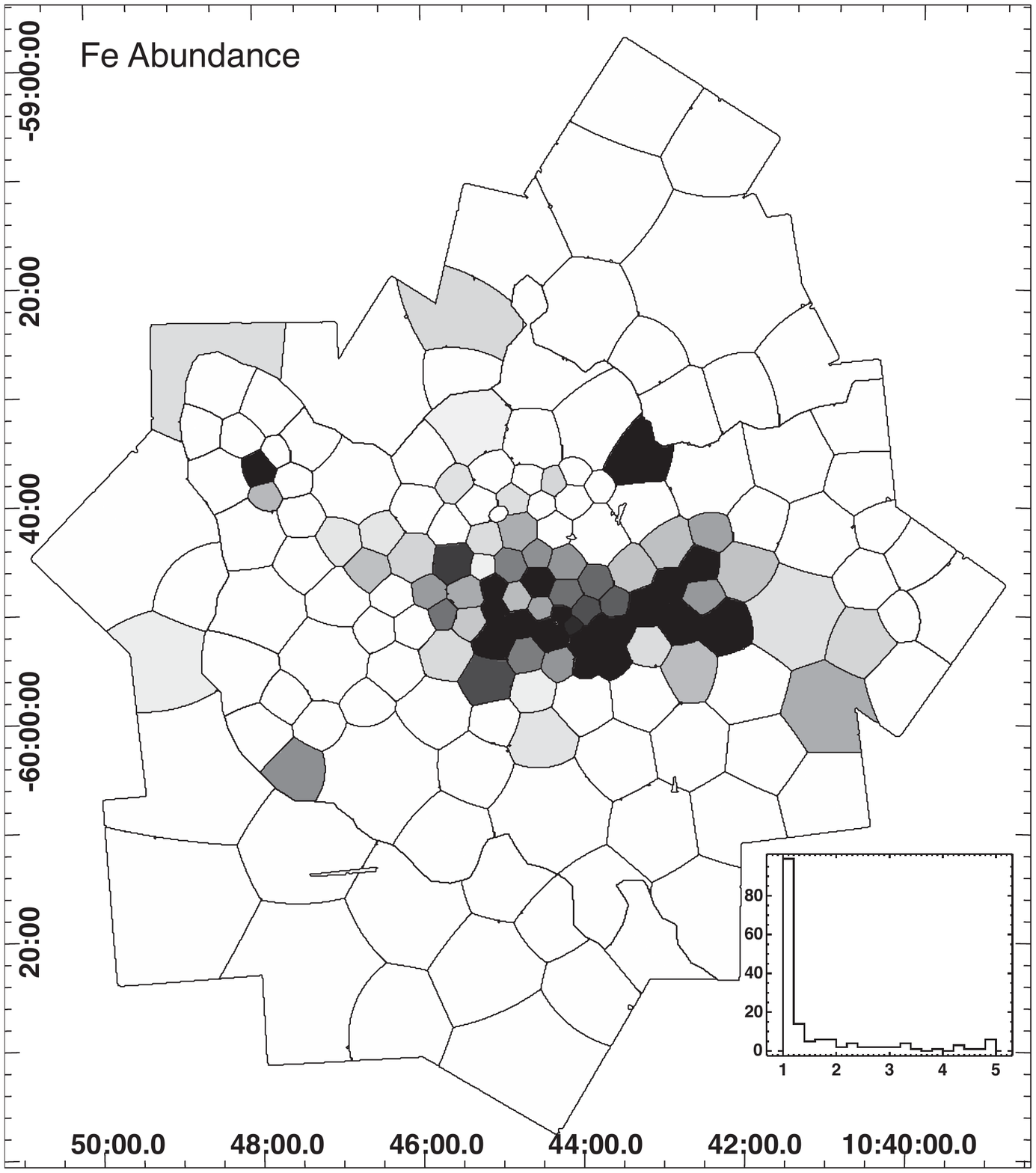}
\caption{Abundance maps for:
(a) oxygen
(b) neon
(c) magnesium
(d) silicon
(e) sulfur
(f) iron.
Most abundances were frozen at solar values and appear white in these maps.  All maps are scaled between 1 and 3 times solar values.  
} 
\label{fig:abundmaps}
\end{center}
\end{figure}

The resulting abundance maps are shown in Figure~\ref{fig:abundmaps}.  These maps show that enhanced abundances of O, Ne, and Mg were needed for very few tessellates, without any prominent spatial pattern.  Enhanced Si was required quite often, as we predicted from the prominent Si~K-$\alpha$ line seen in many spectra; there are some spatial concentrations in the regions of bright diffuse emission in the central arc and in the eastern arm.  The S abundance is not well-constrained because there are few counts at such high energies ($>$2~keV) in our spectra; many regions that suggest S enhancement also show Si enhancement, but the overall appearance of S enhancement is quite patchy.  The Fe abundance, however, is striking in its spatial correlation with the brightest diffuse emission in the central arc.  Unlike Si, the eastern X-ray arm is not particularly enhanced in Fe.  Note once again that there was no correlation between tessellates in the spectral fitting.

We hesitate to overinterpret the actual abundance values; as mentioned above, they are not particularly well-constrained by the data in many cases.  Abundances were not allowed to go below solar values because it is unlikely that massive star winds and/or cavity supernovae could generate plasmas with sub-solar abundances.  There is a long history in X-ray astronomy where unexpectedly low abundances in fits to low-resolution spectra often indicate an oversimplified model.  \citet{Buote98} showed that single-temperature fits to {\em ASCA} data typically returned sub-solar abundances, while in two-temperature fits the abundances were much closer to solar.  In a different approach, \citet{Strickland98} and \citet{Pittard10a} showed that fits to synthetic data generated from complex theoretical models typically return lower abundances than put into the model.  Both approaches demonstrate that fitting relatively simple models to highly complex emission can return inaccurate parameters.

\subsubsection{Hard Emission Components \label{sec:ptcomps}}  

\begin{figure}[htb] 
\begin{center}  
\includegraphics[width=0.32\textwidth]{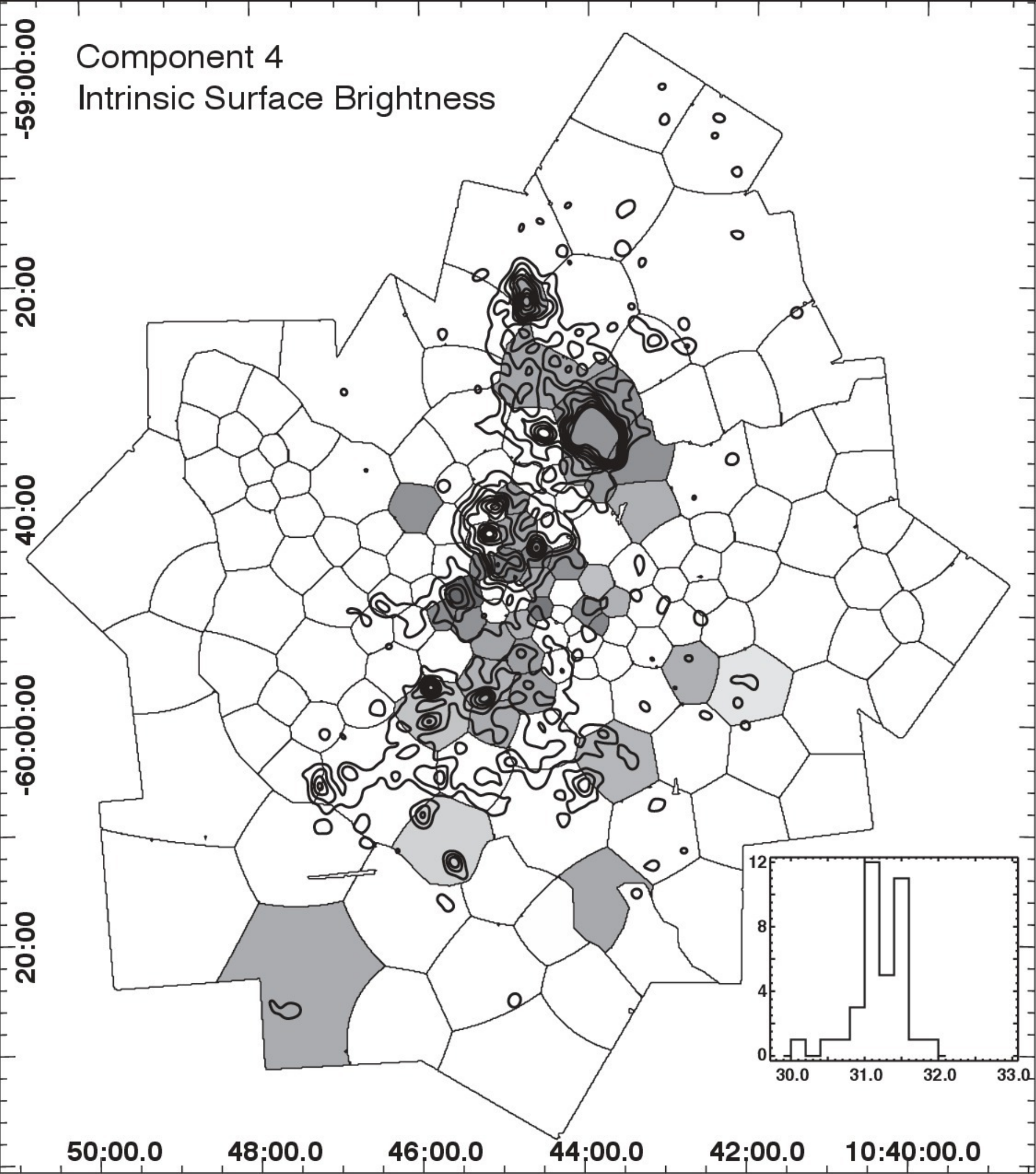}
\includegraphics[width=0.32\textwidth]{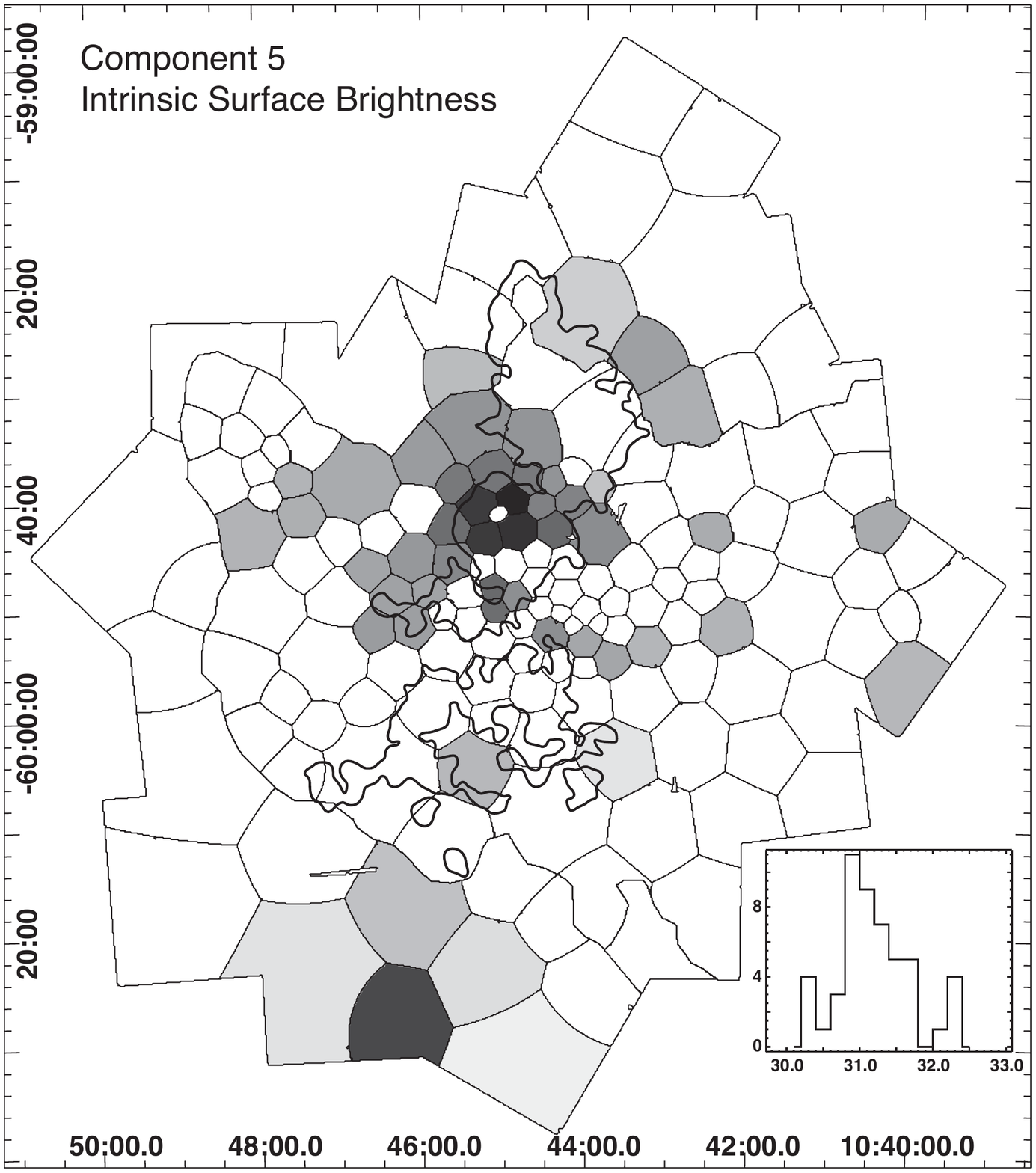}
\includegraphics[width=0.32\textwidth]{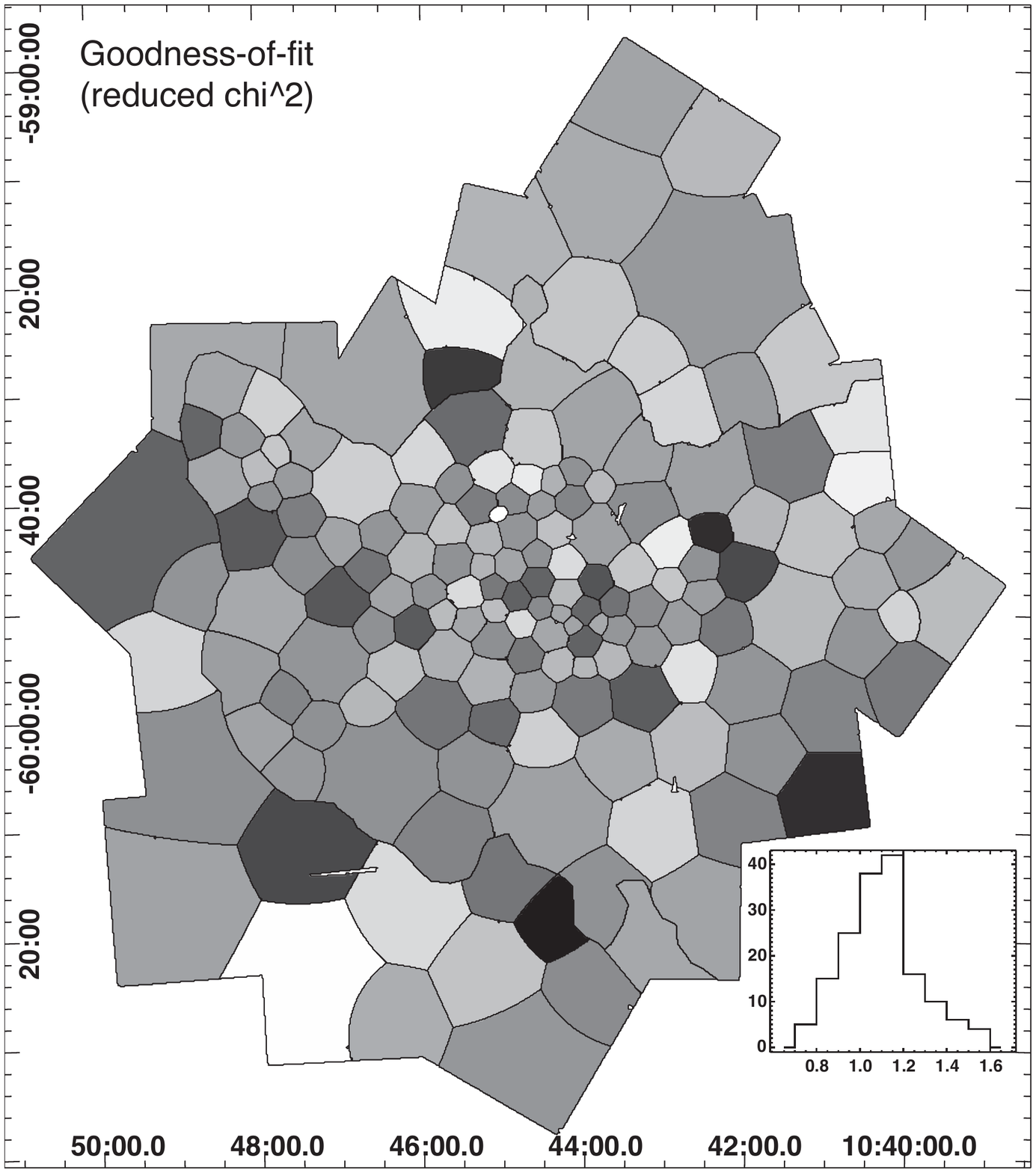}
\caption{Other spectral fit parameters. 
(a) A map of the intrinsic surface brightness of {\it apec} model component kT4.  Contours of point source surface density from \citet{Feigelson11} outline Carina's stellar clusters.  The good correlation between this map and the stellar contours confirms that this {\it apec} component is tracing the hard emission from unresolved pre-MS stars in the CCCP.
(b) A map of the intrinsic surface brightness of {\it apec} model component kT5.  Simplified contours outlining areas of high point source surface density from \citet{Feigelson11} are shown.  
Both surface brightness maps have the same scaling, with $30.2 < \log SB < 32.6$, to match the scaling in Figure~\ref{fig:neimaps}.  Note that the value for tessellate outside001 is incorrect, since all surface brightness calculations assume the Carina distance but the kT5 emission in outside001 is dominated by the cluster of galaxies, which is of course at a much larger distance.
(c) A goodness-of-fit map, showing the reduced $\chi^{2}$ for the spectral fit of each tessellate.  The map is scaled over its full range, $0.71 < $ reduced $\chi^{2} < 1.59$.    
} 
\label{fig:other}
\end{center}
\end{figure}

In addition to the three {\it vpshock} components designed to characterize Carina's diffuse emission, our spectral model (Section~\ref{sec:model}) includes three harder {\it apec} thermal plasma components that are supposed to account for the more obscured, unresolved pre-MS stars (kT4; median $\sim$2.5~keV), other hard sources such as \etacar and the cluster of galaxies (kT5; median $\sim$6~keV), and a constant component representing the hard X-ray background (kT6, frozen at 10~keV).  In Figure~\ref{fig:other}a we show the intrinsic surface brightness map of Component 4, overlaid with the contours of CCCP X-ray point source surface density from \citet{Feigelson11}.  These contours trace out several well-known clusters (Tr14, Tr15, Tr16, Bochum~11, Collinder~228, and the Treasure Chest) and several smaller concentrations of sources, some discovered through X-ray analysis.  Since Component 4 was designed to represent the contribution to Carina's ``diffuse'' emission from the hot thermal plasma component of unresolved pre-MS stars, Figure~\ref{fig:other}a serves as a test of its success in this regard.  

The correlation between the Component 4 intrinsic surface brightness map and the contours of X-ray stellar surface density is generally good -- it appears that the kT4 model component is usually tracking the distribution of the large number of unresolved young stars in the field.  We can compute the total X-ray luminosity contributed to the CCCP by Component 4; this is determined by multiplying the intrinsic surface brightness in each tessellate by the tessellate area, then summing over all tessellates.  The result is that the total-band absorption-corrected luminosity of Component 4, $L4_{tc} = 5.0 \times 10^{33}$~erg~s$^{-1}$.  This should be a rough estimate of the hot component of the composite unresolved pre-MS star population across the entire CCCP mosaic.  

Figure~\ref{fig:other}b shows the intrinsic surface brightness map of Component 5, the hard {\it apec} plasma designed to account for unusual sources of emission such as the cluster of galaxies and $\eta$~Car.  While it does serve these purposes (it is bright in the tessellates most affected by these objects), it also shows an interesting anticorrelation with Component 4, often having substantial emission measure in tessellates adjacent to those that were bright for Component 4.  This perhaps indicates that Component 5 is also tracing an obscured unresolved stellar component in Carina, at least in some tessellates.  In many cases, kT4 has run to its upper limit (4 keV) while kT5 has run to its lower limit (also 4 keV), so both of these model components may be representing the same population of obscured young stars.


\subsection{Soft Stellar Components \label{sec:softptcomps}}  

While obscured pre-MS stars are well-represented by the hard thermal plasma component kT4, we expect there to be tens of thousands of unobscured young stars in Carina that also show a softer plasma component with $kT \sim 0.86$~keV ($T \sim 10$~MK) \citep{Preibisch05}, lying just below our detection limit in this shallow survey.  They should enhance the surface brightness of what we are assuming to be truly diffuse emission.  In our model parameterization, we might expect Components 2 and/or 3, the harder {\it vpshock} plasmas, to show the imprint of this stellar contribution.  This effect may be subtly apparent in the parameter maps shown above:  in Figure~\ref{fig:neimaps}f, the Component 2 intrinsic surface brightness of tessellate inside003 (covering Tr14) is bright; in Figure~\ref{fig:ktmaps}c, kT3 is high for tessellates inside003 and inside098 (covering Tr15).  So while some contamination from unresolved stars certainly must affect our diffuse emission maps, it appears that these effects are subtle and do not dominate the results.  This is perhaps because our tessellates cover large areas that extend, in many cases, well beyond the regions of enhanced stellar surface density.  

\begin{figure}[htb] 
\begin{center}  
\includegraphics[width=0.48\textwidth]{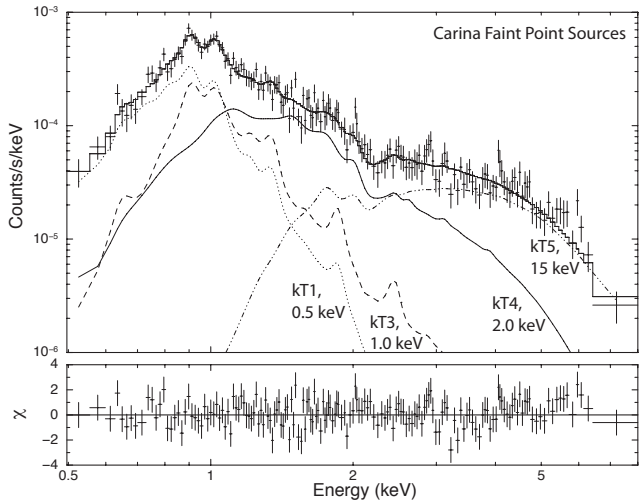}
\caption{The composite spectrum of 200 faint detected point sources in Carina, containing $\sim$6000 net counts.  Absorptions (in units of $1 \times 10^{22}$~cm$^{-2}$) are $N_{H1} = 0.0$, $N_{H3} = 0.8$, $N_{H4} = 0.4$, $N_{H5} = 4$.
} 
\label{fig:ptsrcs}
\end{center}
\end{figure}

As mentioned above, we also expect a large number of foreground stars in our field, contributing a small amount of soft, unobscured flux to Carina's true diffuse emission.  We can address both of these soft contaminants by examining a composite spectrum of the 200 faintest point sources that we do detect; the brightest part of the undetected foreground and pre-MS stellar population that is contaminating our maps of Carina's diffuse emission should have a very similar spectral shape.  In Figure~\ref{fig:ptsrcs} we show this composite, or ``stacked'' spectrum, another standard data product from {\it AE}.  

This spectrum includes foreground stars and AGN as well as pre-MS stars in Carina.  The fit incorporates the same components as our diffuse emission model, but Component 2 (the short-timescale NEI plasma) and Component 6 (the frozen hard thermal plasma representing the cosmic X-ray background and Galactic Ridge emission in the diffuse fits) were not needed for this fit, so they are absent in Figure~\ref{fig:ptsrcs}.
  
A reasonable interpretation of this fit is that the unabsorbed soft component kT1 traces mainly the unresolved foreground stars, the absorbed, harder kT3 traces primarily the soft component of the pre-MS stars, kT4 traces the hard component of the pre-MS stars as usual, and kT5 traces mostly AGN.  If this is true, then we can estimate the contribution to Carina's diffuse emission by foreground stars and the soft component of Carina's pre-MS stars by using a simple scaling argument.  The intrinsic luminosities of kT1, kT3, and kT4 in the composite point source spectrum (Figure~\ref{fig:ptsrcs}) are, respectively, 
$L1_{tc,pt} = 0.7 \times 10^{30}$~erg~s$^{-1}$, $L3_{tc,pt} = 2.5 \times 10^{30}$~erg~s$^{-1}$, and $L4_{tc,pt} = 1.5 \times 10^{30}$~erg~s$^{-1}$.  Thus we can compute the component luminosity ratios 
$L1_{tc,pt}$/$L4_{tc,pt}$ = 0.5 and $L3_{tc,pt}$/$L4_{tc,pt}$ = 1.7.

In Section~\ref{sec:ptcomps} we estimated the total intrinsic luminosity of unresolved pre-MS stars using the Component 4 intrinsic surface brightness map and got $L4_{tc} = 5.0 \times 10^{33}$~erg~s$^{-1}$.  From the ratios computed above, then, we estimate that the total intrinsic luminosity of all unresolved foreground stars in the CCCP is $2.5 \times 10^{33}$~erg~s$^{-1}$ and the total intrinsic luminosity of the soft component of all unresolved pre-MS stars is $8.3 \times 10^{33}$~erg~s$^{-1}$.

\subsection{Unmodeled Spectral Lines \label{sec:badfits}}

Lastly, the goodness-of-fit statistic in our spectral fitting, represented by the reduced $\chi^2$ in Figure~\ref{fig:other}c, is generally acceptable for most tessellates.  As can be seen by examining the fit residuals in Figure~\ref{fig:spectra} (preferably the online-only version that shows fits to all tessellate spectra), poor fits are certainly indicating that different physics is sometimes at work, because they are generally caused by line features in the data that are not present in the models, or vice-versa.  Adjusting abundances or adding more NEI or CIE components cannot remedy these problems, because the lines in the data often appear at the wrong energies compared to what these models predict.   

\begin{figure}[htb] 
\begin{center}  
\includegraphics[width=0.49\textwidth]{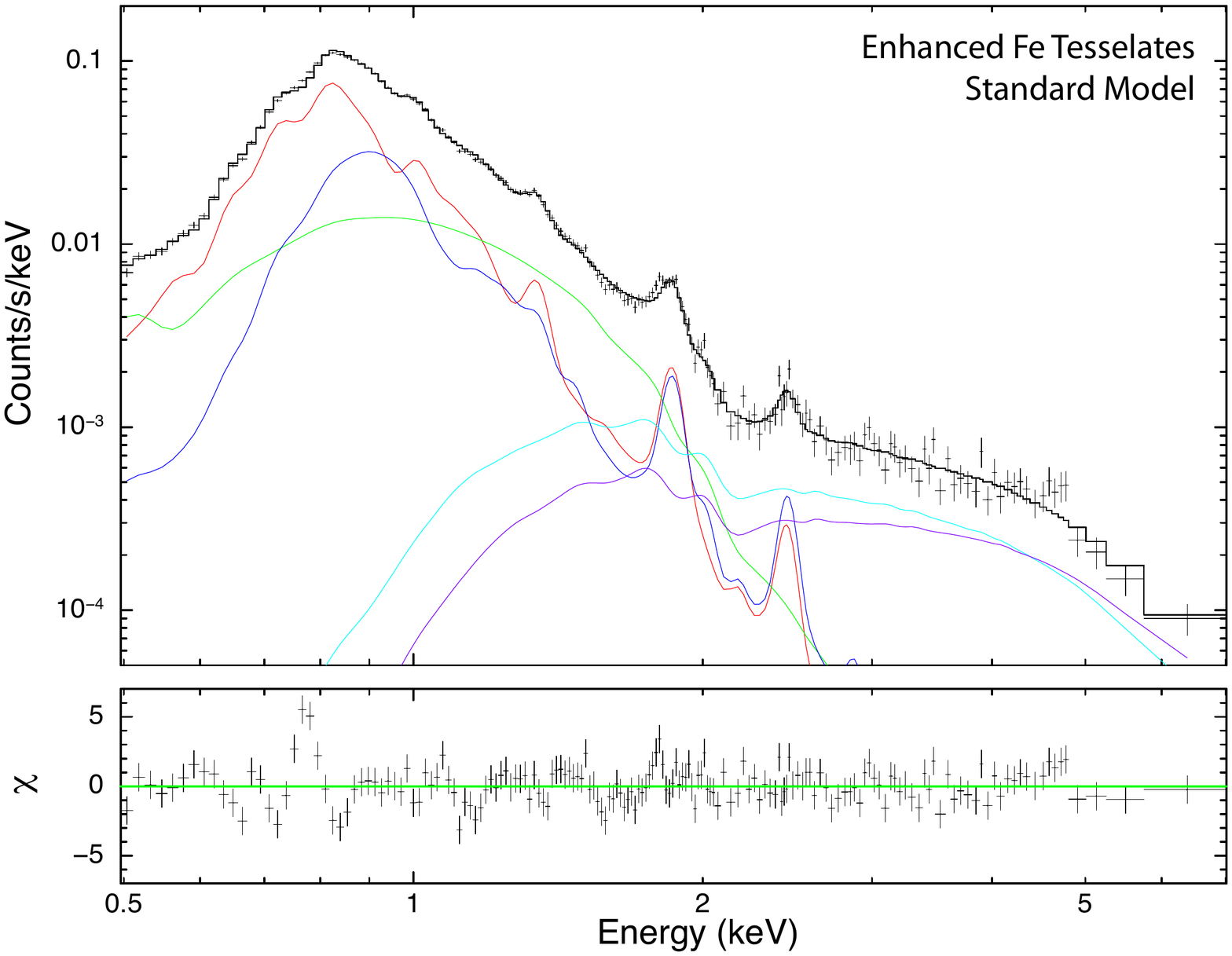}
\includegraphics[width=0.49\textwidth]{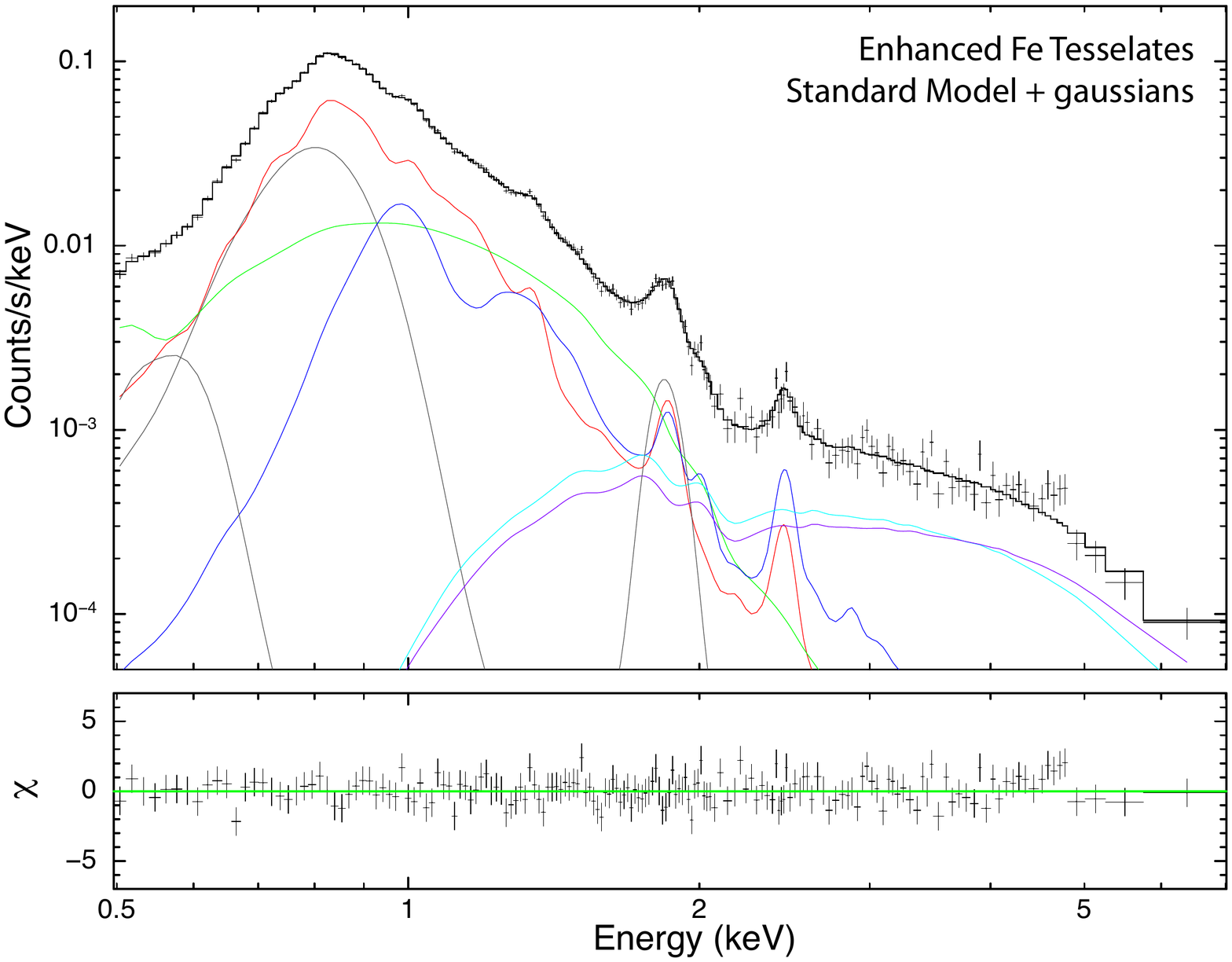}
\includegraphics[width=0.5\textwidth]{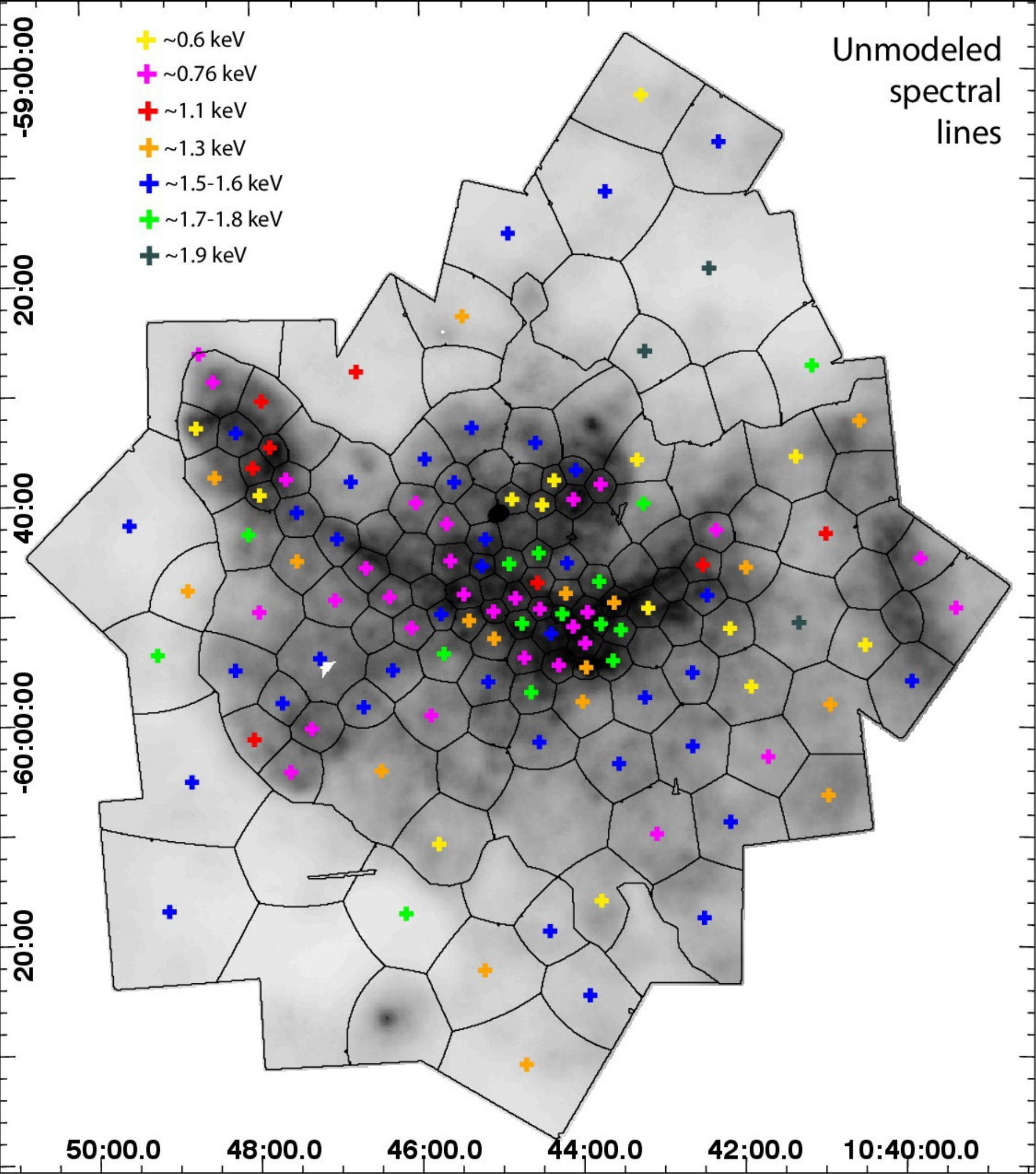}
\caption{Unmodeled spectral lines.
(a--b) Stacked spectra of the tessellates with enhanced Fe emission (from Figure~\ref{fig:abundmaps}f).  This composite spectrum has $\sim$200,000 counts.
(a)  Fit with our usual model gives these NEI component parameters:  ($N_{H}$, kT) = (0.0e22, 0.5), (0.2e22, 0.35), (0.0e22, 0.7) in units of (cm$^{-2}$, keV).  Reduced $\chi^{2}$ is 1.92.
(b)  Fit with our usual model, plus unobscured gaussian lines at 0.56, 0.76, and 1.85~keV (shown in dark grey), gives these NEI component parameters:  ($N_{H}$, kT) = (0.2e22, 0.6), (0.0e22, 0.34), (0.4e22, 0.9) in units of (cm$^{-2}$, keV).  Reduced $\chi^{2}$ is 1.00. 
(c)  A depiction of the most prominent unmodeled spectral line in each tessellate, overlaid on a soft-band smoothed image of the apparent surface brightness.  Note that most tessellates show more than one unmodeled line, a fact not depicted in this map.  Unmarked tessellates show no prominent unmodeled lines below 2~keV.
} 
\label{fig:stacking}
\end{center}
\end{figure}

To try to understand these features and infer what physics we have left out of our spectral modeling, we stacked the spectra from the tessellates that appear to show enhanced Fe emission (Figure~\ref{fig:abundmaps}f) and fit the resulting spectrum with our usual model, then examined the residuals.  This fit is shown in Figure~\ref{fig:stacking}a.  It is statistically poor, with clearly correlated residuals.  In Figure~\ref{fig:stacking}b, we have repeated the fit, adding 3 gaussians to represent spectral lines.  It is difficult to know what absorbing column to use for these gaussian components, since it is not clear where the emission is coming from.  We decided on a conservative approach and chose the lowest absorbing column in our usual model (Figure~\ref{fig:stacking}a); this happens to be close to zero for two of the three NEI components, so we added gaussians with no absorption.  The gaussian line energies, widths, and normalizations were all free to vary; final line energies are 0.56, 0.76, and 1.85~keV, with widths $\sigma$ = 0.03, 0.09, and 0.04~keV respectively.

The intrinsic surface brightness estimated from these two fits is the same to within 10\%.  Including the surface brightness in the gaussian lines increases the total surface brightness estimate from the second model by 20\%; this emission is neglected in our usual model, which leaves the lines unfitted so their flux is not included.  We note that 20\% should be considered a lower limit to the actual surface brightness contained in the gaussian components, as we conservatively estimated that they have zero absorbing column. 

Clearly adding gaussians to model these possible emission lines can improve our spectral fits.  Virtually every tessellate shows correlated residuals that probably indicate unmodeled emission lines (see Figure~\ref{fig:spectra} and the corresponding online-only figure).  In addition to the lines shown in Figure~\ref{fig:stacking}, correlated residuals are often seen at (approximate) soft energies of 1.1, 1.3, 1.5, and/or 1.9~keV (not all lines appear in all tessellates).  

Figure~\ref{fig:stacking}c gives a map of the most prominent unmodeled line in each tessellate, for tessellates that show such features (note again that most tessellates show multiple unmodeled lines; this map shows only the most prominent one).  Occasionally harder correlated residuals appear in the tessellate spectra, at roughly 2.1, 2.5, 3.2, 4.5, and/or $>$6~keV.  By adding gaussians at appropriate energies, we can bring essentially all fits into the acceptable regime where the reduced $\chi^{2} < 1.1$.  We will offer an explanation for at least some of these line features in Section~\ref{sec:charge} below.

    
\section{DISCUSSION \label{sec:discussion}}

\subsection{Mass-loading Processes}

Circumstellar, interstellar, and intergalactic environments usually show broad distributions of density and temperature due to a combination of supersonic turbulence and temperature-dependent cooling.  Thermodynamically stable regions of the temperature-density plane tend to contain more material, giving rise to the concept of thermal phases.  The interaction between these phases is a key phenomenon, as the evolution and morphology of large-scale flows can ultimately be regulated by mass-loss from objects of much smaller size.  Mass-loss from clouds (clumps, globules, and knots) may occur through hydrodynamic ablation and through thermal or photoionized evaporation \citep[see e.g.,][and references therein]{Pittard07}.  The response of such multiphase environments to the impact of winds and shocks is central to the investigation of feedback in star and galaxy formation.

Recent work has made extensive use of numerical hydrodynamics to investigate the interaction of a flow with a single cloud, focusing on the effects of radiative cooling \citep[e.g.,][]{Yirak10}, thermal
conduction \citep[e.g.,][]{Orlando08}, ordered magnetic fields \citep[e.g.,][]{Shin08}, and turbulence \citep{Pittard09,Pittard10b}.  Because of the extra computational cost, there have been far fewer
numerical simulations of a flow interacting with a system of clouds \citep[e.g.,][]{Poludnenko02,Pittard05}.  

However, the global effects of mass-loss from a large number of clouds (a process termed ``mass-loading'') can be studied using either similarity solutions or direct numerical modeling.  A key feature of mass-loading is that it tends to drive the ambient flow to Mach number unity, by the slowing and pressurization of supersonic flows and through the acceleration of subsonic flows \citep{Hartquist86}.  Studies of mass-loading in \hii regions, wind-blown bubbles, supernova remnants, superbubbles, and starburst superwinds are summarized in \citet{Pittard07}.  Mass-loading has been suggested to explain the broad H$\alpha$ emission seen in starburst regions \citep[e.g.,][]{Westmoquette09}.  Emission from boundary layers is discussed by \citet{Hartquist93} and \citet{Binette09}.  The turbulent flows expected from the interaction of multiple stellar winds and supernova remnants drive turbulent diffusion, resulting in the transfer of material from dense to rarefied gas \citep{Avillez02,Avillez04,Avillez05}, although the relevance of traditional mass-loading studies to this process remains to be demonstrated.

\subsection{Carina's Total Intrinsic Diffuse X-ray Emission \label{sec:total}}

\begin{figure}[htb] 
\begin{center}  
\includegraphics[width=0.48\textwidth]{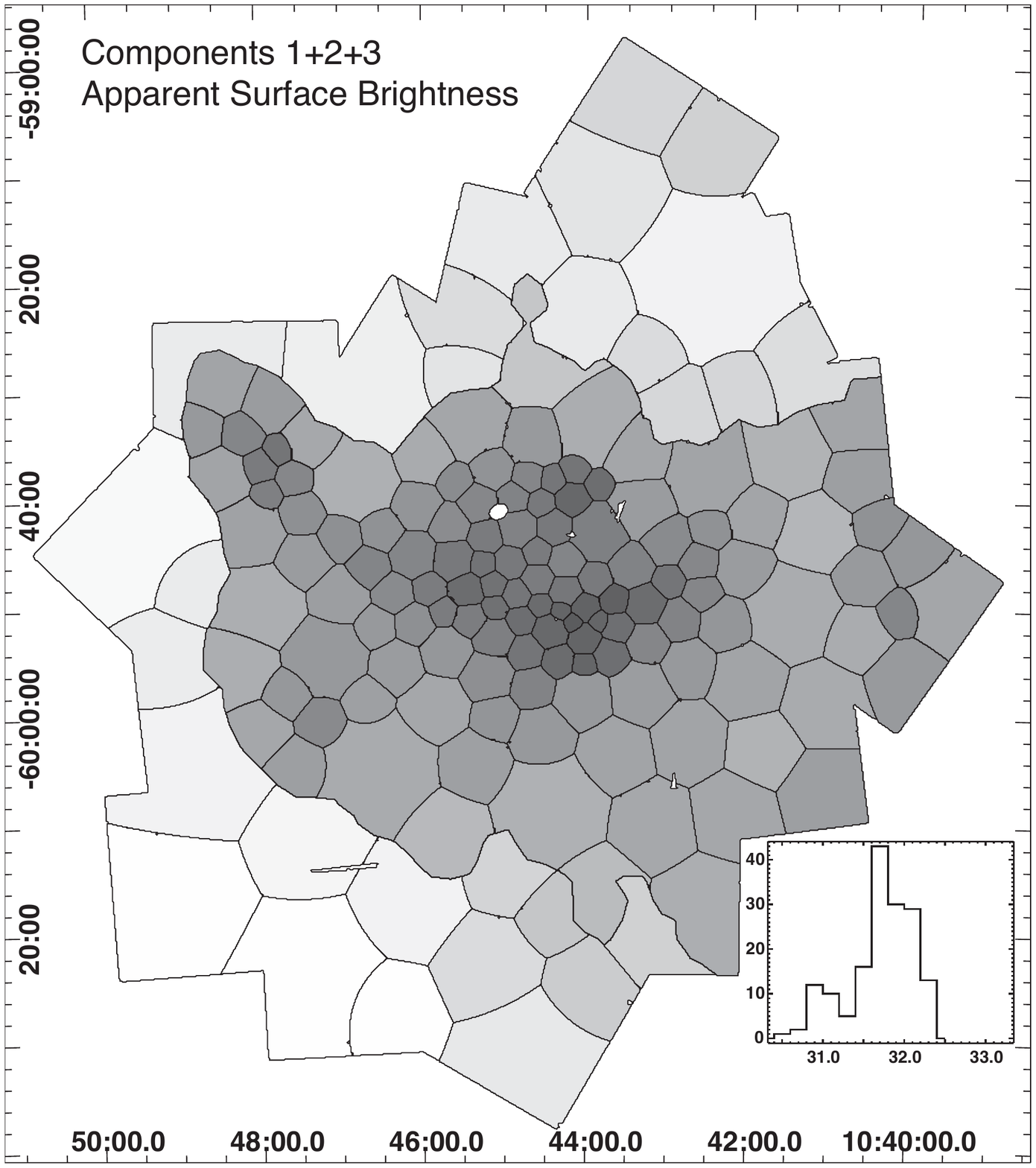}
\includegraphics[width=0.48\textwidth]{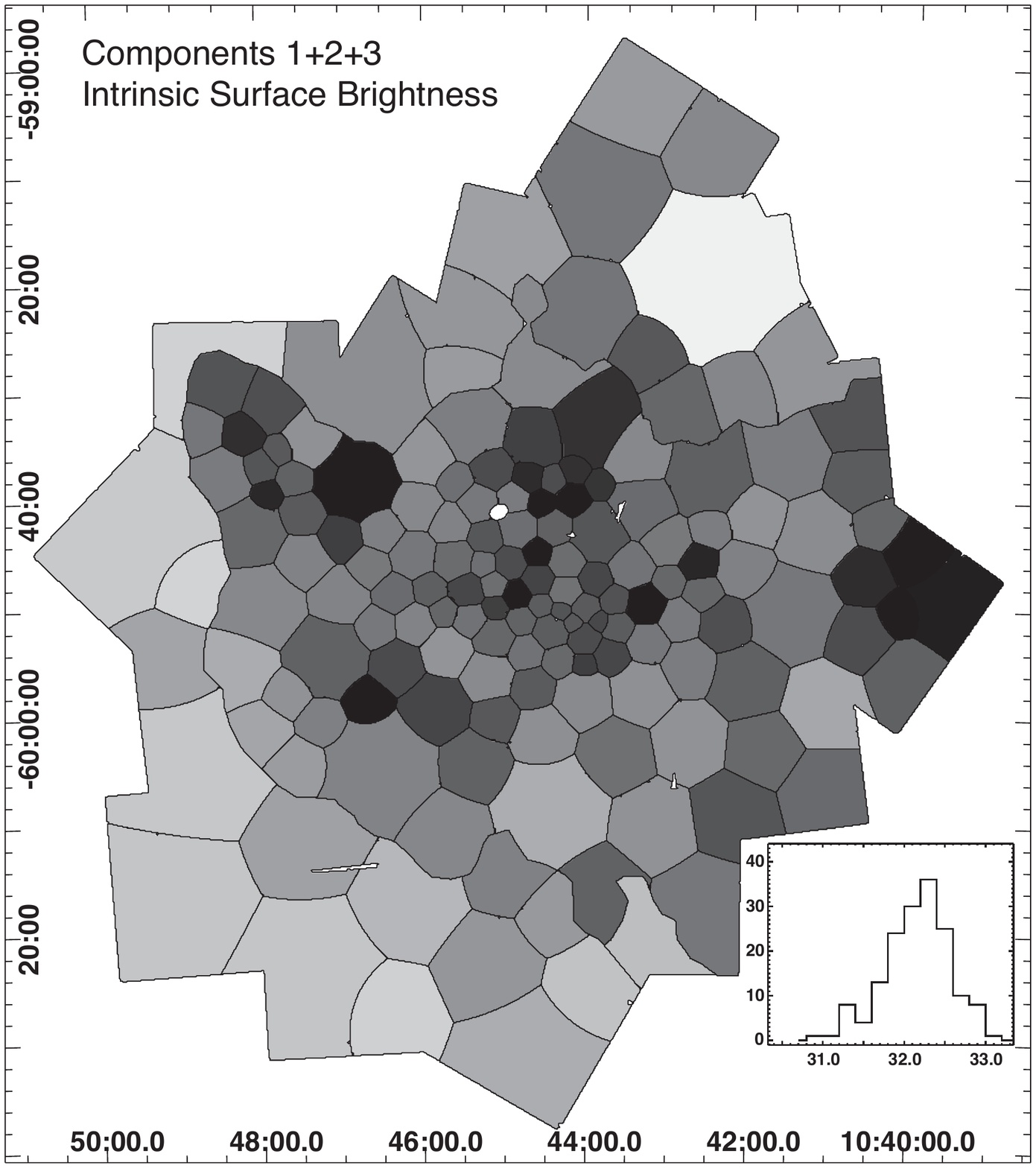}
\caption{Diffuse emission surface brightness.
(a)  Apparent surface brightness of the 3 NEI model components combined.
(b)  Intrinsic surface brightness of the 3 NEI model components combined.
Both images are scaled the same, with $30.8 < \log SB < 32.8$, where again $SB$ stands for surface brightness in units of erg~s$^{-1}$~pc$^{-2}$.  
} 
\label{fig:srfbrt}
\end{center}
\end{figure}

To summarize our tessellation and spectral fitting results, Figure~\ref{fig:srfbrt} shows the apparent and intrinsic surface brightness of Carina's diffuse X-ray emission, isolated from most point source and background components by summing the surface brightness maps from the three NEI components in Figure~\ref{fig:neimaps}.  The similarity between the original apparent surface brightness map (Figure~\ref{fig:tessellates}b) and the apparent surface brightness as calculated from the emission measures of our three NEI model components  (Figure~\ref{fig:srfbrt}a) shows that our model has done a reasonable job of characterizing the diffuse plasma emission in Carina.  The intrinsic surface brightness map in Figure~\ref{fig:srfbrt}b is more speculative, relying on the fidelity of our model parameterization and our absorption estimates.  As described in Section~\ref{sec:badfits} above, it does not include any emission from prominent unmodeled spectral lines that we see in our fit residuals.
  
How much could it be corrupted by unrelated emission?  The total-band, absorption-corrected luminosity of the emission shown in Figure~\ref{fig:srfbrt}b (obtained by multiplying the total surface brightness by the area in each tessellate then summing the resulting luminosities over all tessellates) is $L123_{tc} = 3.2 \times 10^{35}$~erg~s$^{-1}$.  In Section~\ref{sec:softptcomps} above we calculated that the relevant contribution to this total from the soft component of pre-MS stars was $\sim 8.3 \times 10^{33}$~erg~s$^{-1}$, or 2.6\% of this emission.  Unresolved foreground stars contribute $\sim 2.5 \times 10^{33}$~erg~s$^{-1}$, or 0.8\% of this emission.  \citet{Getman11} also showed that the contribution to the diffuse emission from unresolved foreground stars is $\sim$1\%.  Thus unresolved point sources are contributing only $\sim$3\% of the diffuse luminosity, confirming that most of Carina's apparently diffuse emission is truly diffuse.  

We established in Section~\ref{sec:frgd} that foreground soft diffuse emission (Local Hot Bubble, SWCX) are also minimal contributors to Carina's soft diffuse emission.  We speculated that there might be emission from an X-ray halo around a dark cloud at the same distance as the Coalsack, which shows such emission \citep{Andersson04}, but that Carina's diffuse X-ray emission appears anticoincident with other features of its ISM (e.g., dense ionized gas), implying that most of the emission comes from the Carina complex, not a foreground diffuse component.  The parameter maps from our spectral fitting support this (see Figure~\ref{fig:neimaps}); all three NEI model components show substantial absorption correlated between tessellates, implying that the emission they model is not always due to a minimally-absorbed foreground screen.  Although the NEI components may represent some foreground emission, especially for tessellates where they suffer minimal obscuration, they more likely model emission originating in the Carina complex for most tessellates.  

Given these considerations, we might conservatively say that Carina generates a total-band X-ray luminosity of $\sim 3 \times 10^{35}$~erg~s$^{-1}$ from hot plasmas, from the part of the complex captured by the CCCP survey.  If both of Carina's superbubbles are filled with hot plasma emitting X-rays (thus the X-ray emission extends well beyond the CCCP survey area), and if we added in the luminosity from the unmodeled emission lines, this total luminosity number might double.

\subsection{The Multiwavelength Context \label{sec:multi}}

\begin{figure}[htb] 
\begin{center}  
\includegraphics[height=0.35\textheight]{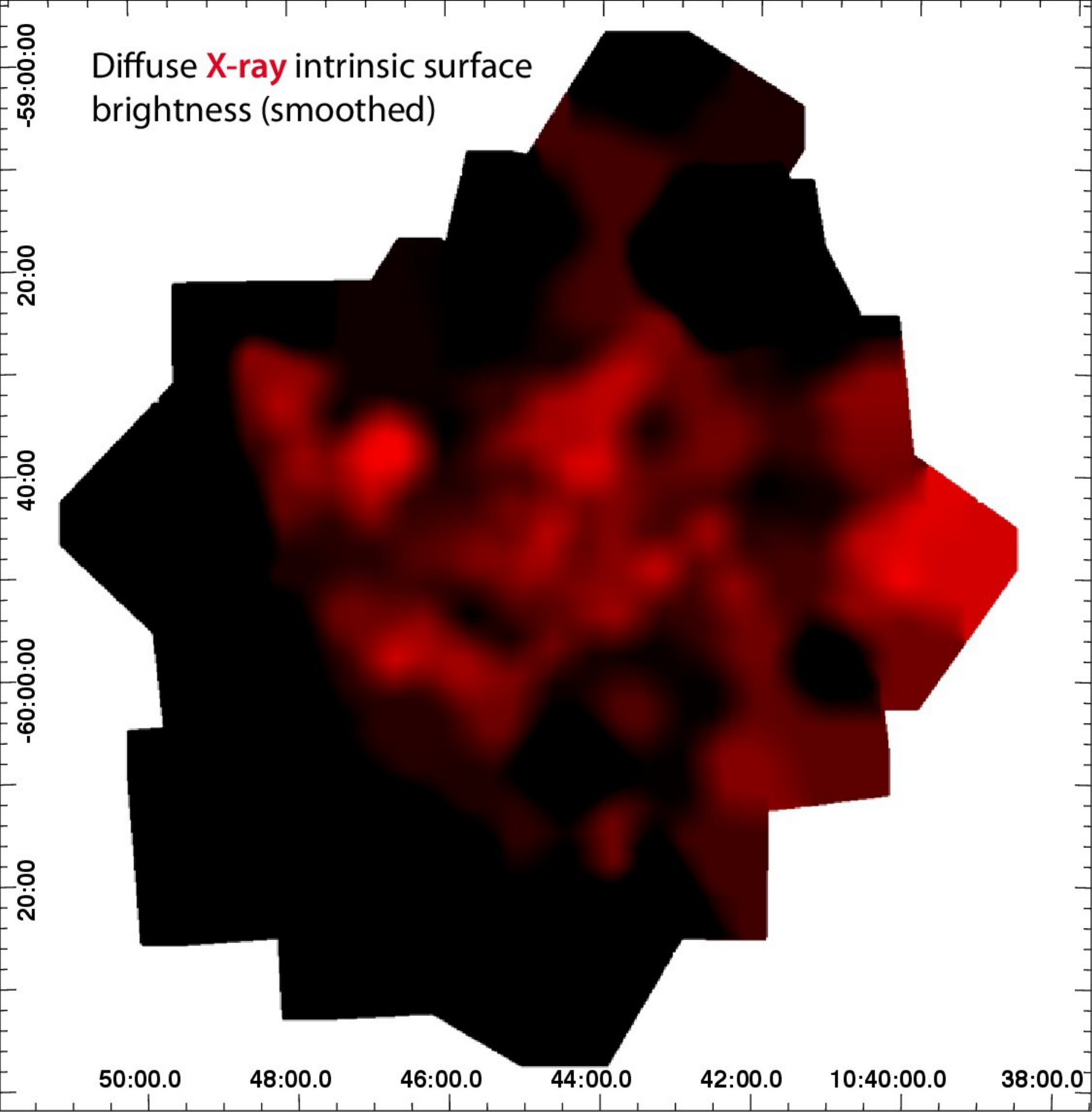}
\includegraphics[height=0.35\textheight]{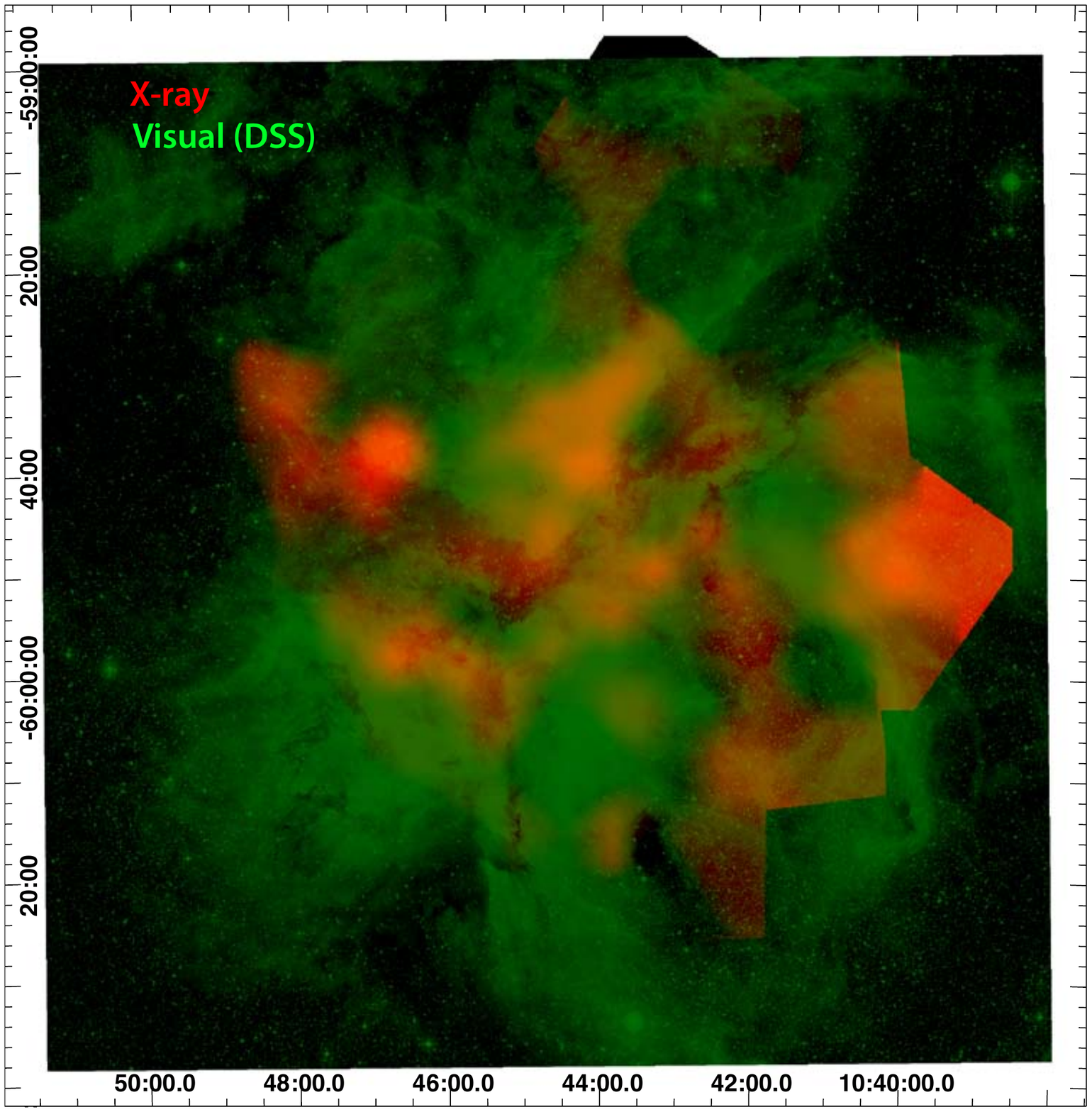}
\includegraphics[height=0.35\textheight]{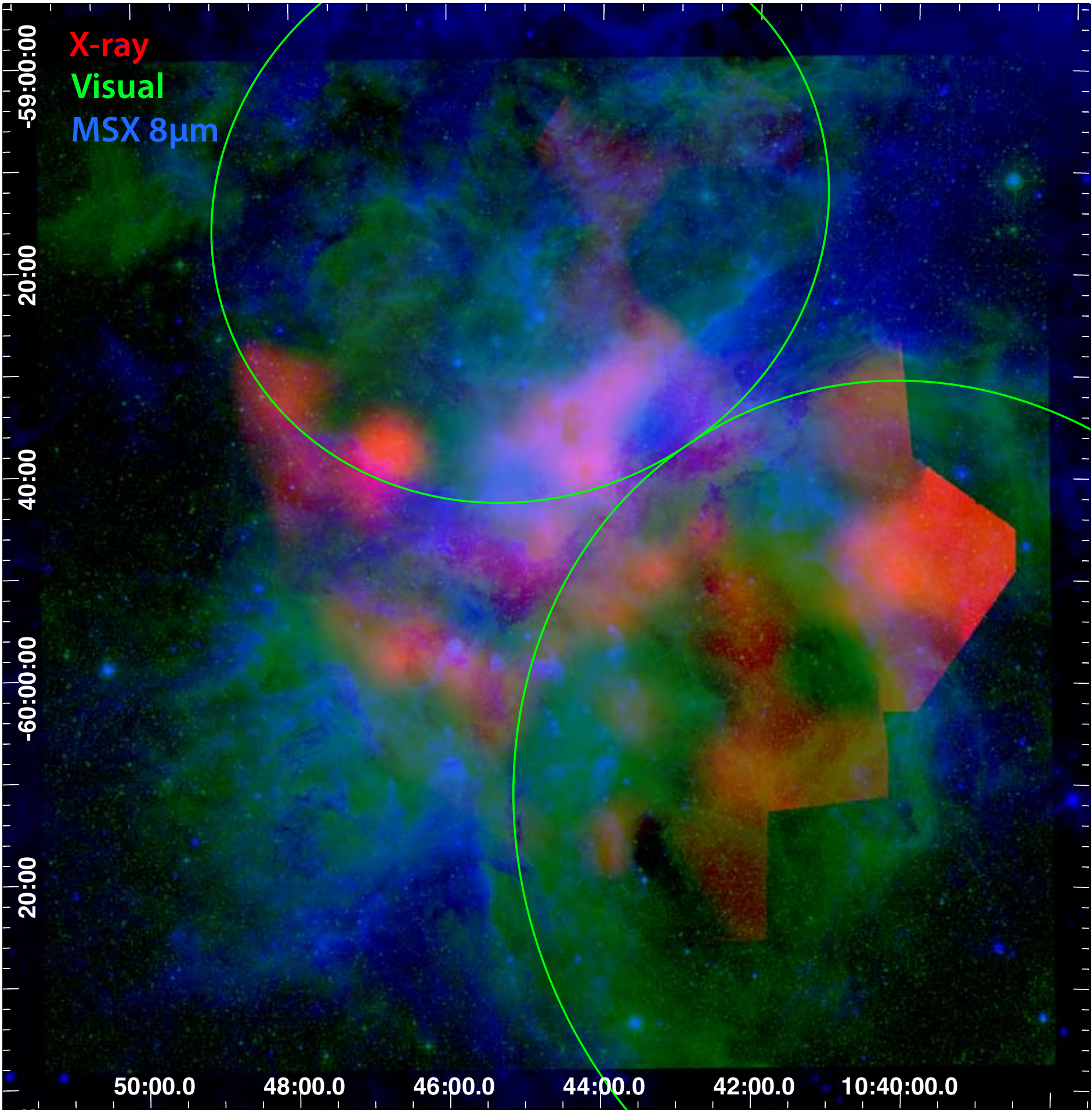}
\includegraphics[height=0.35\textheight]{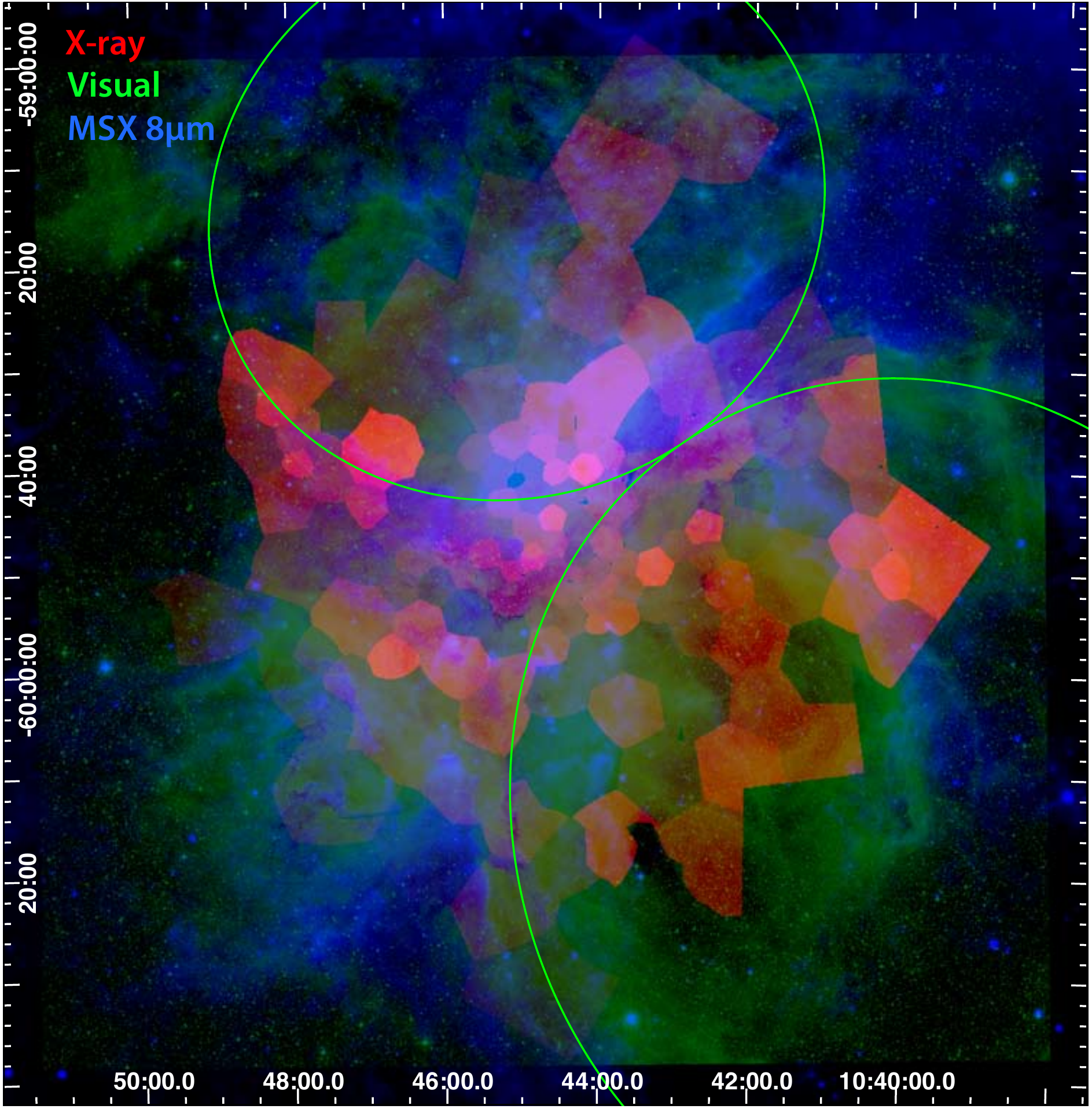}
\caption{A multiwavelength depiction of Carina to place the tessellate maps in context.
(a)  Carina's intrinsic diffuse X-ray emission from Figure~\ref{fig:srfbrt}b, now smoothed to suppress the ``cartoon'' effect of the tessellation.
(b)  The same scene as (a), now with the visual DSS image superposed to show Carina's dense ionized gas.  (c)  The same scene as (b), now including the 8~$\mu$m {\it MSX} image tracing PAH emission and heated dust.  The green ovals roughly outline the superbubble lobes.
(d)  The same multiwavelength images as in (c), but now using the unsmoothed tessellate map of the intrinsic diffuse X-ray emission (from Figure~\ref{fig:srfbrt}b) for comparison (and truth in advertising).
} 
\label{fig:multi}
\end{center}
\end{figure}

To test the fidelity of our intrinsic diffuse emission map (Figure~\ref{fig:srfbrt}b) and to assess its implications, Figure~\ref{fig:multi} attempts to place our estimate of Carina's intrinsic diffuse emission in a multiwavelength context.  In panel (a), the diffuse X-ray emission intrinsic surface brightness from Figure~\ref{fig:srfbrt}b has been smoothed to reduce the stark appearance of its tessellate edges then displayed in red; panel (b) adds in DSS data tracing dense ionized gas in green and panel (c) adds in 8~$\mu$m {\it MSX} data tracing PAH and heated dust emission in blue.  In panel (d), the actual tesselate map (Figure~\ref{fig:srfbrt}b) is shown, rather than the smoothed version presented in panels (a)--(c).  In panels (c) and (d), the green ovals coarsely outline Carina's bipolar superbubble lobes, as seen in the {\it MSX} data.  

Many other multiwavelength representations of the Carina Nebula are shown in the CCCP {\em Special Issue} \citep[e.g.,][]{Townsley11a}.  In particular, CO contours from the study by \citet{Yonekura05} are shown in \citet{Getman11} and \citet{Povich11}.  Recent sub-millimeter work \citep{Preibisch11a} shows many clumps of cold material that could be contributing to the mass-loading processes described above.

Although interpreting these multiwavelength images is not easy, there are a few general observations that we can make.  Carina's diffuse X-ray emission remains quite clumpy even when the intervening absorption affecting the appearance of apparent surface brightness maps (e.g., Figure~\ref{fig:patsmooth}) is removed.  The most obvious large-scale feature is a general anticorrelation between X-ray emission and dense ionized gas; in Figure~\ref{fig:multi}b, the X-ray emission is often bright where the DSS emission is faint, and vice versa.  There appears to be little X-ray-emitting hot gas pervading the South Pillars.  While some hot gas seems to thread through the complex ISM of the northern superbubble, it is not predominant there, except in the arc along its southeastern edge (the eastern arm of diffuse X-ray emission) and at its interface with the southern superbubble.  

While this plasma appears to pervade the southern superbubble's ISM more thoroughly than what we saw in the northern superbubble, it still seems to lie mainly inside regions that are outlined by DSS emission from ionized gas.  This volume-filling (as opposed to edge-brightened) geometry of the diffuse X-ray emission, especially in the southern superbubble, indicates that mass-loading is probably an important process in Carina's ISM.  The fact that the {\em intrinsic} diffuse X-ray emission map (Figure~\ref{fig:multi}) remains highly structured implies that Carina's superbubbles are not simple bubbles with uniformly-evacuated interiors and uniformly-dense edges; if they were, the diffuse X-ray emission should smoothly trace the superbubble interiors.  Instead, it seems that all of Carina's ISM is a complex network of smaller cavities; the superbubbles seem to be more like honeycomb structures with interior surfaces separating narrow cavities filled with X-ray-emitting plasma.  

To add even more complexity, the diffuse X-ray emission is not confined to the superbubbles, though; it is quite prominent across the central V-shaped dust lanes and in a long ``crevice'' at the northern edge of the South Pillars that is dark in the DSS image.  There is no simple correlation between the brightness of the diffuse X-ray emission and the locations of massive stars in Carina.  Thus O-star winds and/or supernova blast waves may have to travel far in Carina's cavities to encounter enough cold surfaces (cloud walls, mass-loading clumps) to contribute to the diffuse X-ray emission that we see.

\subsection{Charge Exchange:  The Missing Physics? \label{sec:charge}}

A recent {\em XMM-Newton} study of the starburst galaxy M82 \citep{Ranalli08} showed prominent lines in the fit residuals around 0.78 and 1.23~keV.  Ranalli et al.\ propose that these unmodeled lines are due to charge exchange (CE) between M82's hot galactic superwind (the result of merging multiple supernova remnants over many epochs) and cold neutral clouds in the galaxy's ISM.  The process is described by \citet{Lallement04}:  as hydrogenic or helium-like ions from the hot plasma impinge on the cold neutral clouds, electrons freed from the neutrals are captured into high-excitation states in the ions; subsequent line emission (with no continuum) from both the ions and neutrals (which presumably have recaptured an electron into a high-excitation state) results.  Lallement notes that CE emission is more likely to come from lighter elements (thus producing softer X-ray emission lines); it scales as $n_{n}Un_{e}^{-2}$, where $n_{n}$ is the neutral gas density, U is the relative velocity between the hot gas and the neutral gas, and $n_{e}$ is the density of the hot gas.  Thus it is most important for the lowest-density hot plasmas hitting high-density cold clouds; Lallement predicts that it is important for galactic superwinds hitting cold halo clouds and, conversely, high-velocity clouds hitting galactic halos, but not for young supernova remnants, confirming the work of \citet{Wise89}.  Interestingly, Wise and Sarazin note that Fe-L lines could also be affected by CE, because they are generated by species such as Fe{\scriptsize XVII} or Fe{\scriptsize XIX} that have similar ionization conditions as lighter species such as O{\scriptsize VII} or O{\scriptsize VIII}.  

\citet{Ranalli08} also note the results of laboratory work by \citet{Djuric05}, where CE emission from a neutral Mg line at 1.254~keV (close to one of the line energies seen in M82) was produced by firing highly-charged ions at olivine, augite (a pyroxine), and quartz, materials that might make up comet surfaces.  A strong neutral Si line at 1.739~keV was also seen in these experiments, along with weaker lines of several other elements, including O and Fe.  Ranalli et al.\ note that their M82 emission lines could be caused by CE in highly-excited O{\scriptsize VIII} ions (0.78~keV) and neutral Mg (1.254~keV) and that this might indicate the presence of, e.g., olivine grains in M82's cold clouds.  Clearly the CE lines that are present in an X-ray spectrum will depend not only upon the densities and relative velocities of the neutral cloud and the hot plasma, but also upon the elemental make-up of the cloud, including its dust grains, and the plasma abundance.  Cold ISM components will produce emission lines from neutral or near-neutral species while hot plasma components will produce emission lines from highly-ionized species; this presumably can result in a wide variety of emission lines below $\sim$2~keV in X-ray spectra.

Interestingly, we might expect a foreground X-ray halo around a dark cloud, such as that seen around the Southern Coalsack \citep{Andersson04}, also to exhibit CE emission.  Andersson et al.\ interpret the Coalsack halo as the result of the Upper Centaurus-Lupus Superbubble interacting with this dark cloud.  As described above, this interaction between a hot, rarefied plasma and a cold neutral cloud is exactly the kind of place where CE is likely to occur.  Further observations around the Coalsack to search for CE emission lines seem warranted.

So it seems plausible that the unmodeled emission lines that are so pervasive in Carina's diffuse X-ray spectra could be due to CE between its hot plasma components and its complex cold ISM.  The DSS and {\em MSX} images in Figure~\ref{fig:multi} illustrate the many cold surfaces available for the hot plasma to impinge upon; as noted by \citet{Lallement04}, the brightness of the CE emission depends strongly on the viewing angle of this thin interface layer, so tessellate spectra showing strong CE line emission might indicate primarily that the CE layer is viewed edge-on (tangentially) there.  In Figure~\ref{fig:stacking}c, we see ``threads'' of the strongest unmodeled lines weaving through the figure -- there are very few tessellates with a given prominent line that are not adjacent to another tessellate with the same prominent line.  This may suggest that we are tracing long ``ridges'' of contiguous hot/cold interfaces with specific molecules or dust grains, plasma abundances, and/or physical parameters that favor a given CE line.

In addition to the large-scale cold structures that we clearly see with DSS or {\em MSX} images, Carina very likely contains a plethora of unresolved cold clumps, perhaps left over from the original molecular cloud, that permeate the complex; above we described the mass-loading that can be caused by such structures and that can lead to volume-filling X-ray emission.  If such cold clumps pervade Carina---as we might expect, given the structures seen in the {\em HST} images of \citet{SmithN10} and in the sub-millimeter images of \citet{Preibisch11a}---CE may be occurring throughout the entire region, not just at the walls of larger structures.  This geometry might manifest itself as faint unmodeled emission lines in many of our tessellates.  Clearly we are limited by the CCCP sensitivity here; more photons would allow smaller tessellates and give us more ability to resolve Carina's complicated hot/cold interfaces.

The brightest unmodeled emission line has an energy of $\sim$0.76~keV.  This could be the same O{\scriptsize VIII} line that \citet{Ranalli08} find in M82, or it could be an Fe-L line.  Similarly the Fe{\scriptsize XVII} line at 0.81~keV that is included in the NEI and CIE thermal plasma models (and whose strength results in the prediction of supersolar Fe in these models) could really be due to other high-ionization states of O{\scriptsize VIII} that also generate lines at 0.81~keV.  The production of O, Fe, Mg, Si, and S X-ray emission lines from the destruction of grains in cold clouds seems plausible:  olivines can be Fe-rich, Mg-rich, or both; olivines, silicates, and iron sulfide grains are known to be plentiful in molecular clouds \citep{Keller02}.  Thus destruction of olivine, (Mg,Fe)$_{2}$SiO$_{4}$, quartz, SiO$_{2}$, and iron sulfide, FeS, at the conduction layer between the hot plasma and the cold clouds might explain the bulk of Carina's CE emission.  Perhaps even protoplanetary disks, also rich in olivines, iron sulfide, pyroxenes, and other silicates, could serve as a source for some of the CE emission that we see. 
Although these ideas are nothing but speculation at this stage, we advance them here as motivation for future work to prove or disprove.

\subsection{Inferred Physical Parameters in Carina's Superbubbles \label{sec:physics}}

We would like to approximate the physical conditions in the X-ray-emitting plasma.  The biggest difficulty with this exercise is assuming the geometry of the emitting volume, as Carina's diffuse emission morphology is extremely complex and much of it may come from surface interactions rather than from volume-filling hot plasmas.  As a first step, we attempt here to estimate the properties of the hot plasma in Carina's superbubbles, where estimating the volume is relatively straightforward and where we can assume that the emitting plasma is (at least partially) volume-filling.

Figure~\ref{fig:multi} showed ellipses that approximate the shape on the sky of Carina's superbubbles as determined from the 8~$\mu$m emission outlining them in {\em MSX} data.  If we assume that the 3-dimensional superbubble shape is a similar ellipsoid, we can approximate the volume sampled by tessellates contained within the superbubble outlines as the tessellate area multiplied by the depth through the ellipsoid calculated at the tessellate's center.  We assume that a tessellate contributes to the superbubble emission if its center is contained inside the relevant ellipse.  Rather than assuming a filling factor for each plasma component, it is kept separate as an explicit multiplier for the physical quantities.  Presumably it could be different for each NEI component.  Following the arguments and assumptions of \citet{Townsley03}, we have constructed Table~\ref{tbl:physics}, a summary of approximate physical parameters for each of the three NEI components that we used to model Carina's diffuse X-ray emission.  The only difference between these calculations and those in Townsley et al.\ is that here we make no correction for the passband of the X-ray luminosity, since we determined it using the total ACIS band (0.5--7~keV).  We warn readers that these quantities are rough estimates at best; we work with median thermal plasma temperatures and make no account for Component 2 being far from equilibrium.

\begin{deluxetable}{@{}ccccccccc@{}}
\centering  \tabletypesize{\footnotesize} \tablewidth{0pt}

\tablecolumns{9}
\tablecaption{Physical Properties of the Diffuse Plasma Components \label{tbl:physics}}

\tablehead{
\colhead{Parameter} & \colhead{~Scale factor~~} & 
\multicolumn{3}{c}{Northern Superbubble} & \colhead{~~~~~} & 
\multicolumn{3}{c}{Southern Superbubble}\\
\colhead{}{\hrulefill} & \colhead{}{\hrulefill} &
\multicolumn{3}{c}{\hrulefill} & \colhead{} &
\multicolumn{3}{c}{\hrulefill}\\
\colhead{} & \colhead{} &
\colhead{NEI 1} & \colhead{NEI 2} & \colhead{NEI 3} & 
\colhead{} &
\colhead{NEI 1} & \colhead{NEI 2} & \colhead{NEI 3} \\
\numberthecolumn & \numberthecolumn & 
\numberthecolumn & \numberthecolumn & \numberthecolumn &
\colhead{} &
\numberthecolumn & \numberthecolumn & \numberthecolumn  
\setcounter{column_number}{1}
}

\startdata
\multicolumn{9}{l}{\it Observed X-ray properties} \\
median $kT_x$ (keV)      & \nodata     &  0.32               &  0.33               & 0.72               &&  0.31              &  0.32              & 0.62               \\
$L_{tc}$ (erg~s$^{-1}$)  & \nodata     &$4.7 \times 10^{34}$ &$3.4 \times 10^{34}$ &$1.2 \times 10^{34}$&&$8.6 \times 10^{34}$&$2.7 \times 10^{34}$&$1.5 \times 10^{34}$\\
$V_x$ (cm$^3$)           & $\eta$      & $4.9 \times 10^{59}$& $4.9 \times 10^{59}$& $4.9 \times 10^{59}$                       && $7.1 \times 10^{59}$& $7.1 \times 10^{59}$& $7.1 \times 10^{59}$\\
                         &             &                     &                     &                    &&                    &                    & \\
\multicolumn{9}{l}{\it Derived X-ray plasma properties} \\
$T_x$ (K)                & \nodata     & $3.7 \times 10^6$   & $3.8 \times 10^6$   & $8.4 \times 10^6$  && $3.6 \times 10^6$  & $3.7 \times 10^6$  & $7.2 \times 10^6$  \\
$n_{e,x}$ (cm$^{-3}$)    &$\eta^{-1/2}$&  0.04               &  0.04               &  0.02              &&  0.05              &  0.03              &  0.02              \\
$P_x/k$ (K~cm$^{-3}$)  	&$\eta^{-1/2}$& $3 \times 10^5$     & $3 \times 10^5$     & $4 \times 10^5$    && $4 \times 10^5$    & $2 \times 10^5$    & $3 \times 10^5$    \\
$E_x$ (erg)              &$\eta^{1/2}$ & $3 \times 10^{49}$  & $3 \times 10^{49}$  & $4 \times 10^{49}$ && $6 \times 10^{49}$ & $3 \times 10^{49}$ & $4 \times 10^{49}$ \\
$\tau_{cool}$ (Myr)      & $\eta^{1/2}$& 20                  & 28                  & 106                && 22                 & 35                 & 85                 \\
$M_x$ (M$_\odot$)        & $\eta^{1/2}$&  10.2               &  10.2               &  5.1               &&  18.5              &  11.1              &  7.4               \\

\enddata

\tablecomments{Equations detailing how the derived properties were obtained from the observed properties are given in \citet{Townsley03}.  The quantity $\eta$ is a ``filling factor,'' $0 < \eta < 1$, accounting for partial filling of the superbubble volume with the X-ray-emitting plasmas.  The parameters in the table should be multiplied by the appropriate scale factor (Column 2) to account for this filling factor.  Derived plasma properties are proportional to $\eta^{1/2}$ and are thus only weakly sensitive to this correction.}

\end{deluxetable}

We can compare Carina's hot plasma properties to those found in M17, a massive star-forming complex at a comparable distance that also shows bright diffuse X-ray emission, but is so young that it is less likely to be fueling its diffuse emission by old, merged cavity supernovae; rather the powerful stellar winds from the young massive cluster NGC~6618 are likely responsible for its volume-filling hot plasma \citep{Townsley03}.  Although M17's X-ray-emitting plasma has temperatures similar to Carina's, M17's plasma is ten times denser and cools ten times faster than the plasma in Carina's superbubbles.  That said, Carina has many more O stars than M17, along with Wolf-Rayet stars and the powerful luminous blue variable $\eta$~Car, so we may well expect that some of its diffuse X-ray emission is wind-generated.  The fact that Carina's young massive stellar clusters are not peaks in its diffuse emission surface brightness must be telling us that the winds flow far from the stars that generate them before they interact with Carina's ISM and produce diffuse X-rays.

We can also compare the derived plasma properties in Table~\ref{tbl:physics} to those of M82's galactic superwind \citep{Ranalli08}; we will use the updated M82 values presented in \citet{Ranalli10}.  Of course the M82 study samples much larger volumes and measures much higher luminosities, thus the energy and mass in the X-ray-emitting gas are orders of magnitude larger than in Carina's superbubbles.  The densities, though, are comparable:  Carina's superbubbles have plasma densities similar to the central regions of M82.  The values are low compared to M17, consistent with conditions needed to make bright CE emission.

Cooling times for the hot (kT3) plasma components in Carina are similar to the center of M82, while those for the cooler plasma components are $\sim$5 times shorter.  They are still at least 20~Myr, though, implying that this rarefied gas remains hot for long time periods compared to the lifetimes of Carina's massive stars.  Just as M82's superwind is fueled primarily by the merged remains of old supernova remnants, much of the rarefied X-ray-emitting plasmas in Carina's superbubbles could come from the cavity supernovae that created them.  
 

    
\section{SUMMARY \label{sec:summary}}

The CCCP has shown that Carina's diffuse X-ray emission is real, morphologically complex, and likely generated by a mix of several physical processes, perhaps including stellar winds, cavity supernovae, and charge exchange with cold material on many spatial scales.  We summarize here a few of our findings and pose a few of the questions they raise.

The large-scale apparent morphology of the diffuse emission is due primarily to absorption.  NEI Component 1 has a ``window'' in its absorption map that defines the central bright arc.  The ``hook'' and eastern arm are blue in Figure~\ref{fig:patsmooth} because all three NEI components have high absorbing columns across that part of the field.  The emission measure in each NEI component also helps to define the appearance of Carina's diffuse emission.  Even though kT2 and kT3 have low absorbing columns across the south, the X-ray emission in this region looks red in Figure~\ref{fig:patsmooth} because kT1 is simply brighter there; if it didn't have such a high absorption column in the south, it would appear even brighter.
  
We cannot rule out the possibility that some diffuse X-ray emission in the CCCP is really foreground emission from halos around dark clouds at $\sim$200~pc.  Could a whole NEI component be due to such emission?  If so, we shouldn't need that component to characterize diffuse emission in other targets.  Such comparisons are made in a separate CCCP paper \citep{Townsley11b}.
  
In our spectral model, NEI Component 2 significantly improves the spectral fits for many tessellates, but its physical interpretation is enigmatic.  Is it truly a strongly-NEI plasma with a very short timescale and/or low density?  Is it instead a steeply-sloped synchrotron component, or some form of bremsstrahlung?  What other emission mechanism could produce such a soft spectrum with no strong lines?  If the emission represented by this model component is a thermal plasma far from equilibrium or synchrotron emission, it could be the last vestiges of one or more cavity supernovae.  These plasma parameters indicate that we have captured Carina's ISM in a state of change, undergoing a transition from NEI to CIE plasma conditions, as well as cooling, ionization, and mixing.
  
Using our assumed spectral model, we infer substantial abundance enhancements for silicon and iron and find a striking spatial concentration of enhanced iron emission just south of the western arm of the V-shaped dust lane that distinguishes visual images of the Carina Nebula.  Is this feature really due to enhanced iron, or does it simply reflect the NEI models' attempts to reproduce an emission feature that could in fact be due to O{\scriptsize VIII} CE?  Conversely, is the unmodeled $\sim$0.76~keV line really CE from O{\scriptsize VIII}, or could it be an Fe-L line, from CE or otherwise?  If the iron is real, where did it come from?  Could it be liberated by dust (e.g., olivine or iron sulfide) destruction when cold clouds or protoplanetary disks are impacted by hot plasma?  Another (less likely) explanation could be Type~I supernovae; normally these objects would not be found in young star-forming regions, but there is some evidence for a ``prompt'' pathway for Type~I generation \citep[e.g.,][]{Scannapieco05}.
  
How is the diffuse X-ray emission really distributed spatially?  Are we seeing mostly surface emission from hot/cold interfaces, or mostly volume-filling hot plasma?  Higher spatial resolution (e.g., finer tessellation) would help to answer these questions, but such analysis likely requires more photons in order to constrain the spectral fitting that would be necessary to study these phenomena in detail.
  
We believe that CE emission provides a plausible explanation for the many unmodeled spectral lines that are suggested by our spectral fit residuals.  CE provides a direct link between the hot plasma and the cold ISM in Carina; much future effort will be needed to understand the details of the physical processes at work.  CE also provides a link from Carina to the superwind in M82 \citep{Ranalli08,Ranalli10}.  This superwind consists primarily of old SNRs and Carina's hot plasma mimics it; apparently both plasmas have the right temperature, density, and shock speed to generate CE at their cold interfaces.    
  
Has this study provided strong evidence for supernova activity in Carina or not?  \citet{Wise89} showed that CE emission is unlikely from the fast shock of a young supernova remnant \citep{Lallement04}.  There is no radio or obvious X-ray synchrotron emission in Carina to indicate a recent supernova.  Part of the diffuse X-ray emission almost certainly comes from the energetic winds of Carina's many massive stars, but those winds apparently do not interact close to their origin because the brightest regions of diffuse emission are far from the concentrations of massive stars.
  
If CE does explain the many unmodeled lines in Carina's diffuse emission, then we have found a new example of a rarefied hot plasma striking a cold neutral medium with the right mix of physical conditions to generate CE.  This phenomenon strengthens the argument for the existence of one or more old supernova remnants that have merged in Carina's superbubble cavities to form the hot plasma, because these are the constituents of galactic superwinds (e.g., M82) where similar emission is seen \citep{Ranalli08}.  Pivotal to this argument is the recent discovery of an old neutron star in the Carina complex \citep{Hamaguchi09,Pires09}.  This neutron star and older massive star clusters such as Tr15 \citep{Wang11} are strong evidence that the Carina complex possessed massive stars several million years ago; such an older massive population is necessary to generate the cavity supernova remnants that perhaps contribute to the hot plasmas that we see today.  Thus it is quite likely that X-rays provide direct evidence of the interaction between the hot and cold denizens of Carina's ISM, where multi-million-degree plasma---echos of the last stellar generation's exploded stars and today's massive stellar winds---eats away at the shreds of molecular material that formed Carina's current stellar generation and that might be struggling, despite this caustic onslaught, to form the next.  

\bigskip


\acknowledgments
We very much appreciate the time, effort, and helpful ideas donated by our anonymous referee to improve this paper, and his/her patience with this long manuscript.  LKT appreciates enlightening conversations on Carina's diffuse emission with B-G Andersson, Svet Zhekov, and George Pavlov.  This work is supported by Chandra X-ray Observatory grant GO8-9131X (PI:  L.\ Townsley) and by the ACIS Instrument Team contract SV4-74018 (PI:  G.\ Garmire), issued by the \Chandra X-ray Center, which is operated by the Smithsonian Astrophysical Observatory for and on behalf of NASA under contract NAS8-03060.  The Digitized Sky Surveys were produced at the Space Telescope Science Institute under U.S.\ Government grant NAG W-2166.  This research made use of data products from the Midcourse Space Experiment, the NASA/IPAC Infrared Science Archive, and NASA's Astrophysics Data System.

{\em Facilities:} \facility{CXO (ACIS)}, \facility{MSX ()}.


\end{document}